\documentclass[12pt]{iopart}
\usepackage[caption=false, subrefformat=parens,labelformat=parens]{subfig}
\usepackage[backend=bibtex, style=phys, articletitle=true,biblabel=brackets,%
chaptertitle=true,pageranges=true]{biblatex}
\DeclareFieldFormat[article]{title}{\mkbibemph{#1}}
\DeclareFieldFormat[inproceedings]{title}{\mkbibemph{#1}}
\DeclareFieldFormat[Inbook]{title}{\mkbibemph{#1}}
\usepackage{graphicx}
\usepackage[percent]{overpic}
\usepackage{pict2e}
\usepackage{bbold}
\usepackage{url}
\usepackage{booktabs}
\usepackage{color}


\usepackage{float}

\usepackage{tikz}
\usetikzlibrary{decorations.shapes}
\tikzset{decorate sep/.style 2 args=
	{decorate,decoration={shape backgrounds,shape=circle,shape size=#1,shape sep=#2}}}

\usepackage{appendix}

\renewcommand{\theequation}{II.\arabic{equation}}

\usepackage{hyperref}

\begin{document}

\title[Multivariate Distributions in Non--Stationary Complex Systems II]
      {Multivariate Distributions in Non--Stationary Complex Systems II:
        Empirical Results for Correlated Stock Markets}

\author{Anton J.~Heckens, Efstratios Manolakis\footnote{Now at: Dipartimento di Fisica e Astronomia Ettore Majorana, Universit\`a degli Studi di Catania, and
		Dipartimento di Fisica e Chimica Emilio Segr\`e, Universit\`a degli Studi di Palermo, Italy}, Cedric Schuhmann and Thomas Guhr}
\address{Fakult\"at f\"ur Physik, Universit\"at Duisburg--Essen, Duisburg, Germany}
\ead{anton.heckens@uni-due.de, efstratios.manolakis@phd.unict.it, cedricschuhmann@gmail.com and thomas.guhr@uni-due.de}

\begin{abstract}
        Multivariate Distributions are needed to capture the
        correlation structure of complex systems. In previous works,
        we developed a Random Matrix Model for such correlated
        multivariate joint probability density functions that accounts
        for the non--stationarity typically found in complex
        systems. Here, we apply these results to the returns measured
        in correlated stock markets. Only the knowledge of the
        multivariate return distributions allows for a full--fledged
        risk assessment. We analyze intraday data of 479 US stocks
        included in the S\&P500 index during the trading year of
        2014. We focus particularly on the tails which are algebraic
        and heavy. The non--stationary fluctuations of the
        correlations make the tails heavier. With the few--parameter formulae of our Random
        Matrix Model we can describe and quantify how the empirical
        distributions change for varying time resolution and in the
        presence of non--stationarity.
\end{abstract}

\section{\label{sec:level1}Introduction}

Global developments and ever increasing socio--economic interactions
trigger the need to better understand and model complex systems
\cite{mantegna1999introduction,Kutner_2019}.  Large amount of
high--quality data is essential for this endeavor. A wealth of data is
nowadays available for financial markets making them particularly well
suited to develop methods of statistical analysis and new approaches
for modeling.  Rare events in the tails of the distributions are
especially sensitive for systemic risk and stability of a system.  In
financial markets, the analysis of distributions for individual stocks
is of considerable importance for a variety of reasons
\cite{Mandelbrot_1963,Gopikrishnan1998,Plerou_1999_Dist}, it is also
essential to understand the mechanisms of price formation
\cite{Bouchaud_2004,Farmer_2004,Bouchaud_2009}.  With globalization,
the interconnectedness of the considered system must be taken into
account, market--wide synchronicity and correlations of traders'
actions play a decisive role
\cite{Laloux_1999,Gopikrishnan_2001,Plerou_2002}.  Univariate
distributions of individual allow statements about the
corresponding individual risk. Thus, the shape of those distributions
is of interest
\cite{Mandelbrot_1963,Gopikrishnan1998,Plerou_1999_Dist}.  Yet, the
high correlations in financial markets imply that a univariate
assessment of risk is insufficient.  In recent years, such a
multivariate view moved in the focus, often in the context of
stochastic
processes~\cite{engle2002dynamic,tse2002multivariate,van2006modelling,ma2009pricing,GOLOSNOY2012,GOLOSNOY2012,Aielli_2013,Teng2016,Bauwens2016,Hawkes_2018,HAFNER2022}. 

Another
important aspect of complex systems is their non--stationarity
\cite{Plerou_2002,munnixIdentifyingStatesFinancial2012,Wang_2020,wang2021collective,Bette_2023}.
The standard deviations or volatilities for individual returns fluctuate seemingly erratically
over time
\cite{schwert1989,Mandelbrot1997,Bekaert2000,BEKAERT2014,MAZUR2021}.
The mutual dependencies such as Pearson correlations or copulas
\cite{Pearson_1900,Sklar1959:Fonctions,sklar1973random,nelsen2006introduction,MUNNIX2011,salinas2019high}
show
non--stationarity variations as well and play a particularly important
role in states of
crisis~\cite{munnixIdentifyingStatesFinancial2012,Chetalova_2015,Chetalova_2015_2,Stepanov_2015_MultiAsset,Stepanov_2015,Rinn_2015,Pharasi_2018,Heckens_2020,heckens2021new,heckens2022new,marti2021review,pharasi2020market,pharasi2021dynamics,James_2022,Wand_2023,Hessler_2023,Wand_2023_2}.
The multivariate distributions, \textit{i.e.}~the joint probability
density functions of several or even many stock returns are urgently
needed to assess and understand the risks of a financial market as a whole.

To carry out a thorough empirical analysis of such multivariate
distributions is our first goal.  There are various ways to look at
multivariate data. Here, we rotate the vector of returns into the
eigenbasis of the covariance or correlation matrix. These matrices
have spectra featuring a bulk as well as large eigenvalues belonging
to industrial sectors and to the entire market. We obtain individual,
\textit{i.e.}~univariate distributions for the corresponding linear
combinations of returns, which provide a full picture of the
multivariate data. By normalizing to the (square root of) the
corresponding eigenvalue and overlaying the resulting
univariate distributions we arrive at aggregated distributions
of high statistical significance. We find a strong influence
of non--stationarity.

Our second goal is the comparison of our results to our Random Matrix
Model that we discussed in depth in Ref.~\cite{Heckens_PaperI} to which we
refer as I in the sequel. It was developed in
Refs.~\cite{Schmitt_2013,Schmitt_2014,schmitt2015credit,chetalova2015portfolio,Meudt_2015,Sicking_2018,Muehlbacher_2018}
and considerably extended in Ref.~\cite{Guhr_2021}. The fluctuating correlations
in the non--stationary system are modeled by random matrices. The model
predicts heavier tails on longer time intervals.
We obtained four multivariate model distributions with heavy, mostly
algebraic tails which we fit to the data. The data analyzed are
well described by our algebraic multivariate return distributions.
The results confirm that the non--stationary fluctuations of the correlations
lift the tails.

The paper is organized as follows. In Sec.~\ref{sec:DM}, we
introduce our empirical data set and the procedure of aggregation. In
Sec.~\ref{sec:Results}, we present our empirical distributions and the
fits to the model distributions. Moreover, we discuss some caveats
relevant for such a large--scale data analysis in
Sec.~\ref{sec:Caveats}. We give our conclusions in
Sec.~\ref{sec:Conclusion}.

\section{\label{sec:DM}Data and Methods of Statistical Analysis}

We describe our data set in Sec.~\ref{sec:DataSet}. To fix the
notation and conventions, we briefly sketch the normalization of return time series
and data matrices in Sec.~\ref{sec:Returns}. In
Sec.~\ref{sec:EpochsIntervals}, we divide long time intervals into
epochs to facilitate the analysis of non--stationarity. We discuss the
issue of normalization occurring in the separation of time scales. In Sec.~\ref{sec:CorMat}, we briefly introduce
two types of Pearson correlation matrices.  Rotation and aggregation
of multivariate empirical data are explained in Sec.~\ref{sec:AggEmp}.

\subsection{\label{sec:DataSet}Data sets}

We use the Daily TAQ (Trade and Quote) of the year 2014 from the New
York stock exchange (NYSE)~\cite{NYSE_2014}. There are different
columns specified by timestamp, the bid price which is the maximum a
buyer is willing to pay and the ask price which is the minimum a
seller is willing to accept. As the number of used stocks is
comparatively large stocks across all industrial sectors according to
the Global Industry Classification Standard (GICS) \cite{GICS} are
represented, see~App.~\ref{app:ListTickerSymbols}.  Since the market
operates in the opening and closing hours differently from its main
phase, we discard the first and last 10 minutes of each
day~\cite{Wang2016_2}, \textit{i.e.}~we use data from 09:40 (UTC-5)
until 15:50 (UTC-5).  We select stocks being continually traded on
every open day while simultaneously being part of the
\textsc{S$\&$P~500} index in 2014.

There are three days which contain artifactual data for half the day.
Those three days are the 3rd of July, the 28th of November and the
24th of December, which are three public holidays on which the
\textsc{NYSE} was only open for half a day.  We remove the
corresponding data but keep the normal (non--artifactual) trading data
of (half) the mentioned day.

In addition to the aforementioned data set that we use for our
analyses in Secs.~\ref{sec:AggEmp} and~\ref{sec:Results}, we select a
second one similar to the data set from
Refs.~\cite{Schmitt_2013,YahooFinance_2013} to discuss carefully
further aspects of our analysis, see
Sec.~\ref{sec:EpochsandNumber}. The data set has a daily time
resolution with stocks that were listed in the S\&P~500 index,
see~App.~\ref{app:ListTickerSymbols_YahooFinance}. In total, it comprises
308 stocks for a time period ranging from January 23, 1992 to December
31, 2012. This data set lists daily adjusted prices.

\subsection{\label{sec:Returns}Normalization of Returns}

Our data contain stock $K$ stocks which we label as
$k=1,\ldots,K$.  
In our analysis, we include $K=479$ stocks.
For our analysis in Sec.~\ref{sec:Results}, we derive the observables from the midpoint price as
it allows analyses with higher time resolutions compared to stock
prices. Importantly, the dynamics of prices and midpoints is
comparable. With best ask $a_k(t)$ and best bid $b_k(t)$ the midpoint
price reads
\begin{equation} \label{eqn:midpoint}
	m_k(t) = \frac{a_k(t)+b_k(t)}{2} \ , \hspace{1cm} k = 1, \ldots, K \ .
\end{equation}
From the midpoint price,
we calculate the logarithmic returns
\begin{equation}\label{eqn:logreturn}
	G_k(t) = \ln\frac{m_k(t+\Delta t)}{m_{k}(t)} \ , \hspace{1cm} k = 1, \ldots, K \ .
\end{equation}
which depend on the chosen return horizon $\Delta t$.  We arrange the
return time series $G_k(t), t=1,\ldots,T$, as the rows of the $K\times
T$ data matrix
\begin{equation}\label{eqn:dataMatrixMidpointsRet}
	G = \left[
	\begin{array}{ccccc}
		G_{1}(1) & \cdots    & G_{1}(t)  &\cdots & G_{1}(T)\\
		\vdots &   \vdots & \vdots & \vdots  & \vdots\\
		G_{k}(1) & \cdots    & G_{k}(t)  &\cdots & G_{k}(T)\\
		\vdots &   \vdots & \vdots & \vdots  & \vdots\\
		G_{K}(1) & \cdots & G_{K}(t)  &\cdots & G_{K}(T)
	\end{array}   
	\right] \ .
\end{equation}
We normalize of each row in Eq.~(\ref{eqn:dataMatrixMidpointsRet}) to zero mean and unit standard deviation which yields the time series
\begin{equation}\label{eqn:NormDatMatM}
	\mathcal{M}_k(t) = \frac{G_{k}(t)-\langle G_{k}(t)\rangle_T}{\sigma_k} \ , \hspace{1cm} k = 1, \ldots, K \ .
\end{equation}
The sample average is defined as
\begin{equation}\label{samplerow}
	\langle f_k(t) \rangle_T = \frac{1}{T} \sum_{t=1}^T f_k(t) \ ,
\end{equation}
such that $\langle G_k(t) \rangle$ is the sample mean and
\begin{equation}
	\sigma_k = \sqrt{ \langle \left(G_{k}(t) -  \langle G_{k}(t) \rangle_T \right)^2 \rangle_T } \,.
\end{equation}
the sample standard deviation.  The resulting $K \times T$ data matrix $\mathcal{M}$ contains
the normalized return  time series $\mathcal{M}_{k}(t), \ t=1,\ldots,T$, as rows.

We may also normalize the columns of $G$ to zero mean and unit standard
deviation $\varrho(t)$.  This yields a different type of series in the
index $k$, referred to as position series,
\begin{equation}\label{eqn:NormDatMatE}
	\mathcal{E}_k(t) = \frac{G_{k}(t)-\langle G_{k}(t)\rangle_K}{\varrho(t)} 
\end{equation}
with the sample average
\begin{equation}\label{samplecolumn}
	\langle f_k(t) \rangle_K = \frac{1}{K} \sum_{k=1}^K f_k(t) \,.
\end{equation}
The resulting $K \times T$ data matrix $\mathcal{E}$ contains as
columns the normalized return position series $\mathcal{E}_{k}(t),
\ k=1,\ldots,K$. Time series provide information on subsequent events
in one stock or, more generally, position $k$, while position time
series collect the information on all positions at a given time $t$.

\subsection{\label{sec:EpochsIntervals}Epochs and Long Interval}

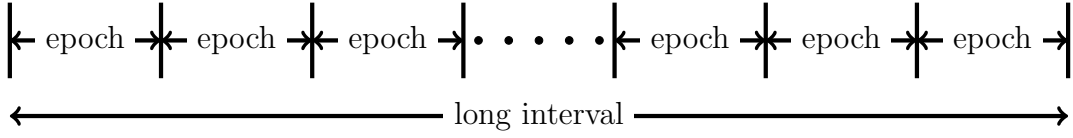
\begin{figure}[hbtp]
	\centering
	\begin{tikzpicture}
		\draw[ultra thick] (0,0.5) -- (0,-0.5);
		\draw[ultra thick] (2,0.5) -- (2,-0.5);
		\draw[ultra thick] (4,0.5) -- (4,-0.5);
		\draw[ultra thick] (6,0.5) -- (6,-0.5);
		\draw[ultra thick]  (0, 0) edge[<->] node[midway,fill=white] {epoch} (2, 0);
		\draw[ultra thick]  (2, 0) edge[<->] node[midway,fill=white] {epoch} (4, 0);
		\draw[ultra thick]  (4, 0) edge[<->] node[midway,fill=white] {epoch} (6, 0);
		\draw[decorate sep={1mm}{4mm},fill] (6.2,0) -- (7.9,0);
		\draw[ultra thick] (8,0.5) -- (8,-0.5);
		\draw[ultra thick] (10,0.5) -- (10,-0.5);
		\draw[ultra thick] (12,0.5) -- (12,-0.5);
		\draw[ultra thick] (14,0.5) -- (14,-0.5);
		\draw[ultra thick]  (8, 0) edge[<->] node[midway,fill=white] {epoch} (10, 0);
		\draw[ultra thick]  (10, 0) edge[<->] node[midway,fill=white] {epoch} (12, 0);
		\draw[ultra thick]  (12, 0) edge[<->] node[midway,fill=white] {epoch} (14, 0);
		\draw[ultra thick]  (0, -1) edge[<->] node[midway,fill=white] {long interval} (14, -1);
	\end{tikzpicture}
	\caption{Division of the whole year 2014 into epochs and long intervals.}
	\label{fig:EpochInterval}
\end{figure}

We consider the trading year 2014 with a total of $250$ trading
days. To analyze non--stationarity, we separate time scales by
dividing a long time interval into many small ones, referred to as
epochs as shown in Fig.~\ref{fig:EpochInterval}. 
Anticipating the
later discussion, one can try to choose the epoch length such that the
effects of non--stationarity are negligible or at least much smaller
than across epochs, \textit{i.e.}~on the long time
interval. Nevertheless, conceptually this is not a prerequisite for
applying our model. For the data under consideration, it turns out
reasonable to choose one trading day as epoch length.  
The
long interval can be one trading year, 25 trading days or 50 trading days. 
There are 10 or 5 long intervals for the whole trading year which have a length of 25 or
50 trading days, respectively, see~App.~\ref{app:IntervalsRange}.

Importantly, we
calculate the data matrices for the returns separately for each
epoch and concatenate those together to obtain the return data
matrix for a long interval, see Sec.~\ref{sec:EpochsandNumber}.
We carry out our analysis for different time resolutions, \textit{i.\,e.} $\Delta t = 1\,\mathrm{s}$ and $\Delta t = 10\,\mathrm{s}$.
These two considered return horizons limit the number of data points
$T$ that can be chosen for an epoch and for the long interval
in our analysis.
The number of data points are 22200 per epoch ($\Delta t = 1\,\mathrm{s}$) and
2220 per epoch ($\Delta t = 10\,\mathrm{s}$).
It is very important that we normalize our time series to the considered
epoch or long interval, see~Eq.~(\ref{eqn:NormDatMatM}).

\subsection{\label{sec:CorMat}Two Types of Correlation Matrices}

Since we introduced time and position series, there are two different
Pearson correlation matrices. Using the data matrix $\mathcal{M}$,  we have
the $K\times K$ correlation matrix of the time series
\begin{equation}\label{eq:CorrMatEst}
	C = \frac{1}{T} \mathcal M \mathcal M^{\dagger} \ ,
\end{equation}
where we employ $\dagger$ to denote the transpose of a matrix. Here,
$T$ stands for the number of data points in either the long interval or in epochs. In the sequel, the sample correlation matrices on the
long interval and in the epochs will be denoted $C$ and $C_\mathrm{ep}$,
respectively. Using the data
matrix $\mathcal{E}$, we find the $T\times T$ correlation matrix of
the position series matrix,
\begin{equation}
	D = \frac{1}{K} \mathcal E^{\dagger} \mathcal E \ .
\end{equation} 
While $C$ measures the relations between the different stocks, $D$
captures the dependencies in time, \textit{i.e.}~the non--Markovian
features. Contrary to some confusing remarks in the literature,
$C$ and $D$ are not equivalent. Due to the different normalizations
of time and position series, eigenvalues and eigenvectors differ.
Financial markets are known to have small non--Markovian effects \cite{Bouchaud_2004,LilloFarmer+2004,Wang2016,Wang2016_2,WSG2015preprint,schuhmann2025newtradersgame}, see Sec.~I.2.5. Thus, $D$ is not needed in the present case, but will be relevant for other data.

\subsection{\label{sec:AggEmp}Rotation and Aggregation of Empirical Data}

In Eq.~(\ref{eqn:NormDatMatM}) we introduced the notation
$\mathcal{M}_k(t)$ for the returns that is then used to calculate the
correlation matrices $C$ or $C_\mathrm{ep}$. As we wish to analyze
multivariate return distributions, it is advised to employ another
notation for the returns if they appear as arguments of the distributions,
we choose the notation $r_k(t)$ such that
$r(t)=(r_1(t),\ldots,r_K(t))$ is the $K$ component vector of returns
at a given time $t$. In the sequel, we will simply write $r$, because
the steps taken are the same for all times $t$.
Importantly, we use the correlation matrices $C$ or $C_\mathrm{ep}$ and the returns $r$ from the same long interval or epoch for the multivariate return distributions when comparing the model distributions with data.
We recall that the returns $r$ are always normalized on the
considered epoch or on the long time interval. As the correlation
matrices are real symmetric, all eigenvalues are real. Due to the
specific form (\ref{eq:CorrMatEst}), the eigenvalues of a correlation matrix
are positive semidefinite, in our case positive definite, since we always
work with correlation matrices of full rank. As in the theoretical
discussion in I, we diagonalize
\begin{eqnarray}
	C=U \Lambda U^{\dagger} \qquad \mathrm{with} \qquad \Lambda = \mathrm{diag}(\Lambda_1, \ldots, \Lambda_K) \ , \quad \Lambda_k>0 
\end{eqnarray}
with an an orthogonal matrix $U$. The same applies to
$C_{\mathrm{ep}}$ with eigenvalues $\Lambda_{\mathrm{ep},k}$. For the
inverse correlation matrices, we then have $C^{-1} = U \Lambda^{-1}
U^{\dagger}$. As we work with full--rank
correlation matrices, the existence of their
inverse is warranted. For the squared Mahalanobis distance
\cite{MahalanobisReprint2018} we have
\begin{eqnarray}\label{Maha}
r^{\dagger} C^{-1} r = r^{\dagger} U \Lambda^{-1} U^{\dagger} r = \sum_{k=1}^K \frac{\bar{r}_k^2}{\Lambda_k}
        \qquad \mathrm{with} \qquad \bar{r}_k = \sum_{l=1}^K  U_{lk} r_l \ ,
\end{eqnarray}
see I, and similarly for $C_{\mathrm{ep}}$.
The linear combinations
$\bar{r}_k$, are the returns rotated into the eigenbasis of the
correlation matrix.
In the case we use the covariance instead of the correlation matrix, we only normalize the return time series to zero mean but not to unit standard deviation. To calculate the corresponding rotated returns, we use the eigenvectors of the covariance matrix.
All our formulae can be adjusted accordingly and remain valid.

By sampling, we then work out the empirical univariate distributions
$p^{(\mathrm{rot},k)}(\bar{r}_k)$ of the rotated returns
$\bar{r}_k$. These $K$ distributions provide full information on the
multivariate system, because all linear combinations differ.  To
accumulate statistics, we can normalize to the square root of the
corresponding eigenvalue,
\begin{equation}\label{eq:widetilde}
	\widetilde{r}_k = \frac{\bar{r}_k}{\sqrt{\Lambda_{k}}}  ,
\end{equation}
or $\Lambda_{\mathrm{ep},k}$, respectively, and lump together all $K$
distributions. We refer to this averaging procedure as aggregation.
It yields a statistically highly significant univariate empirical
distribution which facilitates a careful study of the tail behavior.

\begin{figure}[htbp]
	\centering
	{\begin{overpic}[width=.85\linewidth]{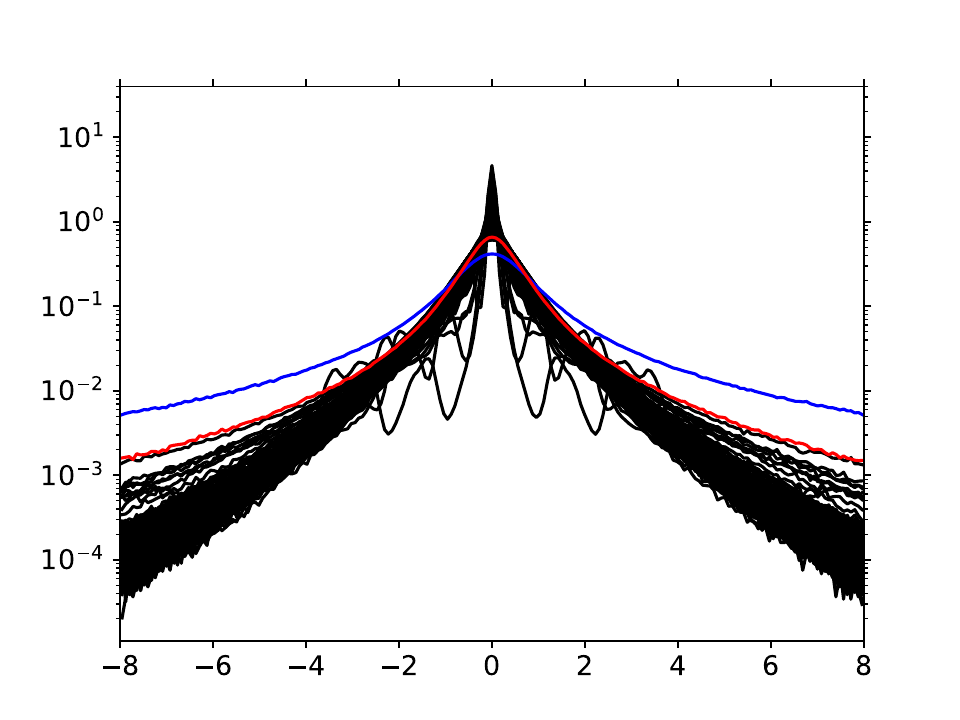}
			\put(50,2){\makebox(0,0){\large\sffamily  rotated return}}
			\put(1,35){\makebox(0,0){\rotatebox{90}{\large\sffamily pdf}}}
		\end{overpic}
	}
	\caption{\label{fig:DistYearly2014_RotRet}Empirical distributions of the rotated returns $p^{(\mathrm{rot},k)}(\bar{r}_k)$ for the whole year of 2014 with $\Delta t = 1\,\mathrm{s}$. Distributions corresponding to the largest eigenvalue (blue) and second largest eigenvalue (red) are highlighted.}
\end{figure}

Figure \ref{fig:DistYearly2014_RotRet} shows the univariate
distributions of the rotated returns $p^{(\mathrm{rot},k)}(\bar{r}_k)$
derived from the Daily TAQ data set with
$\Delta t = 1\,\mathrm{s}$, see~Sec.~\ref{sec:DataSet}.
\begin{figure}[htbp]
	\centering
	{\begin{overpic}[width=.85\linewidth]{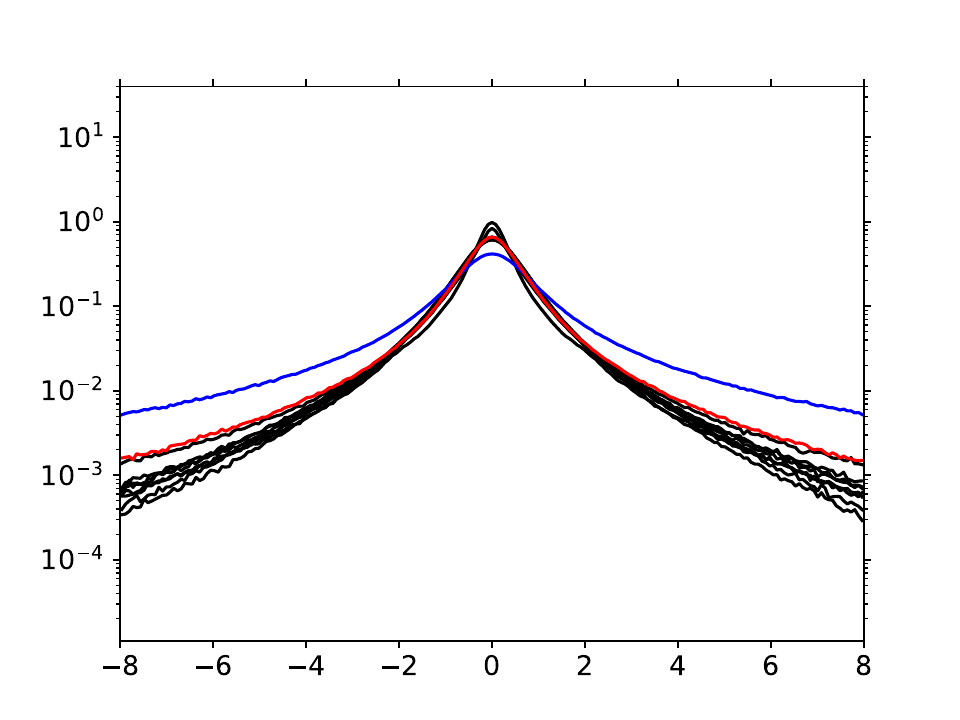}
			\put(50,2){\makebox(0,0){\large\sffamily  rotated return}}
			\put(1,35){\makebox(0,0){\rotatebox{90}{\large\sffamily pdf}}}
		\end{overpic}
	}%
	\caption{\label{fig:DistYearly2014_RotRet_10largestEigenval} Empirical distributions of the rotated returns $p^{(\mathrm{rot})}(\bar{r}_k)$ corresponding to the ten largest eigenvalues for the whole year of 2014 with $\Delta t = 1\,\mathrm{s}$. Distributions corresponding to the largest eigenvalue (blue) and second largest eigenvalue (red) are highlighted.}
\end{figure}%
\begin{figure}[htbp]
	\centering
	{\begin{overpic}[width=.85\linewidth]{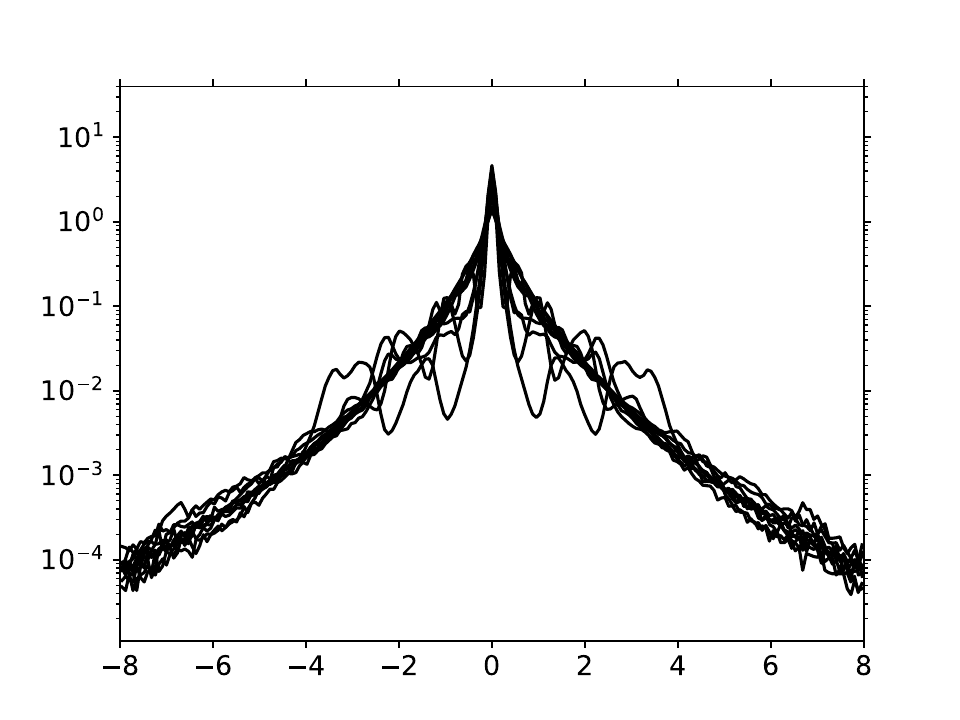}
			\put(50,2){\makebox(0,0){\large\sffamily  rotated return}}
			\put(1,35){\makebox(0,0){\rotatebox{90}{\large\sffamily pdf}}}
		\end{overpic}
	}%
	\caption{\label{fig:DistYearly2014_RotRet_10SmallestEigenval} Empirical distributions of the rotated returns $p^{(\mathrm{rot})}(\bar{r}_k)$ distributions corresponding to the ten smallest eigenvalues for the whole year of 2014 with $\Delta t = 1\,\mathrm{s}$.}
\end{figure}%
\begin{figure}[htbp]
	\centering
	{\begin{overpic}[width=.85\linewidth]{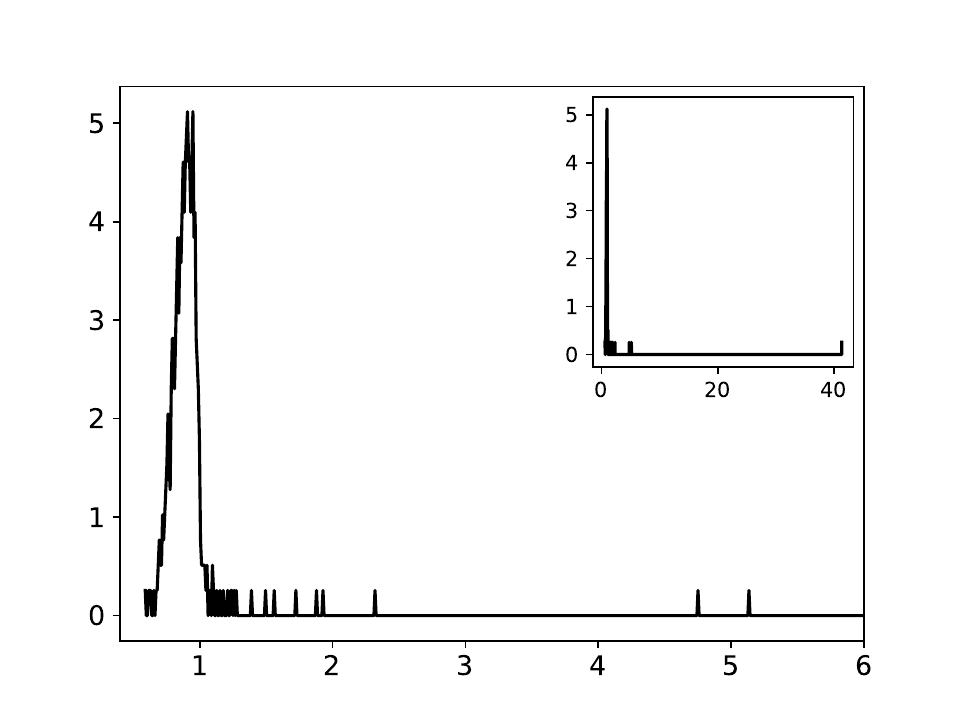}
			\put (70,51) {\Large \sffamily market}
			\put(76,50){\color{blue}\vector(1,-1){10}}
			\put (36,17) {\Large \sffamily industrial sectors}
			\put(26,12){\makebox(52,5){{\color{blue}\downbracefill}}}
			\put (30,55) {\Large \sffamily bulk}
			\put(30,54){\color{blue}\vector(-1,-1){9.1}}
			\put (25,36) {\Large \sffamily small eigenvalues}
			\put(30,35){\color{blue}\vector(-0.68,-1){14.5}}
			\put(50,2){\makebox(0,0){\large\sffamily eigenvalue}}
			\put(5,35){\makebox(0,0){\rotatebox{90}{\large\sffamily pdf}}}
		\end{overpic}
	}%
	\caption{\label{fig:SpectrumCorrMat2014}Spectrum of correlation matrix for the whole year of 2014 with $K = 479$ stocks with $\Delta t = 1\,\mathrm{s}$.}
\end{figure}%
To zoom into the details, the distributions of the rotated returns corresponding to the ten largest and ten smallest eigenvalues are depicted in Figs.~\ref{fig:DistYearly2014_RotRet_10largestEigenval} and~\ref{fig:DistYearly2014_RotRet_10SmallestEigenval}, respectively.
Anticipating the later
discussion, an important remark is in order. It is well--known that
the spectra of large financial correlation or covariance matrices consist
of a rather universal bulk and of outliers which are due to the collective behavior in the industrial sectors, the largest one captures the collective motion of the market as a whole~\cite{Plerou_2002},  see Fig.~\ref{fig:SpectrumCorrMat2014}. 
The second, third, etc. largest eigenvalues correspond to the industrial sectors. Hence, the large eigenvalues capture collective motion in the entire system or in parts of it. The eigenvectors also contain this information and carry it over to the linear combinations~(\ref{Maha}) and their distributions.
As expected
the univariate distributions of the rotated returns in Fig.~\ref{fig:DistYearly2014_RotRet_10largestEigenval} corresponding to the ten largest outliers differ
from those corresponding to the bulk eigenvalues and thus carry important
additional information. 
Moreover, the distributions corresponding to the largest and second largest eigenvalues are heavier-tailed than the distributions corresponding to the bulk eigenvalues.
For the distributions of the rotated returns corresponding to the smallest eigenvalues in Fig.~\ref{fig:DistYearly2014_RotRet_10SmallestEigenval}, we observe stronger oscillations presumably caused by the discrete nature of the prices due to the tick size.
We notice that the Epps effect~\cite{Epps01061979} has an impact on our measured correlation matrices. In particular, tick size and asynchronicity in the trading play an important role~\cite{MUNNIX2010767,MUENNIX2010,Muennix_2011}. Hence, the entries of our empirical correlation matrices have slightly smaller absolute values than they should. Since the focus here is on the multivariate distributions and not primarily on the correlations, we decided to not correct the Epps effect in our empirical study as this will always have the same small impact for data with the same return horizon $\Delta t$. Moreover, it would require additional parameters whose influence on the data analysis will be small but difficult to tell apart from the one of our model parameters.

For comparison, it is also instructive to work out the univariate
distributions of the normalized, original (unrotated) returns $p^{(\mathrm{orig},k)} (r_k)$. 
\begin{figure}[htbp]
	\centering
	{\begin{overpic}[width=.85\linewidth]{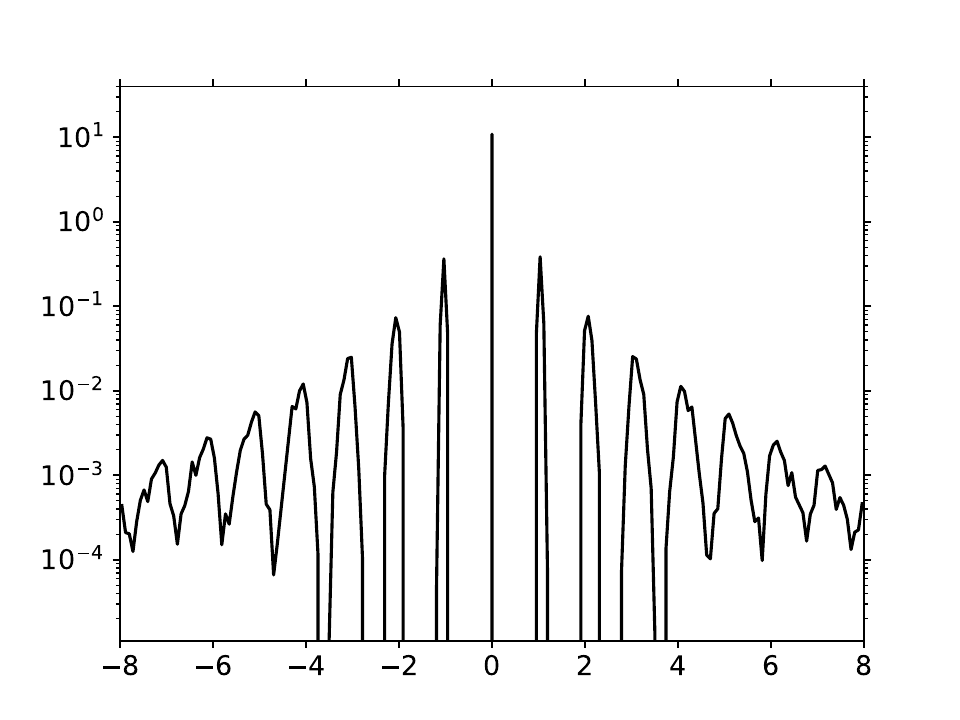}
			\put(50,2){\makebox(0,0){\large\sffamily  original return}}
			\put(1,35){\makebox(0,0){\rotatebox{90}{\large\sffamily pdf}}}
		\end{overpic}
	}
	\caption{\label{fig:Year2014_NormalizedOriginalRet_TROW}Empirical distribution of the normalized, original returns $p^{(\mathrm{orig},k)} (r_k)$ for the whole year 2014 with $\Delta t = 1\,\mathrm{s}$ for stock \textsc{T. Rowe Price} (TROW).}
\end{figure}%
\begin{figure}[htbp]
	\centering
	{\begin{overpic}[width=.85\linewidth]{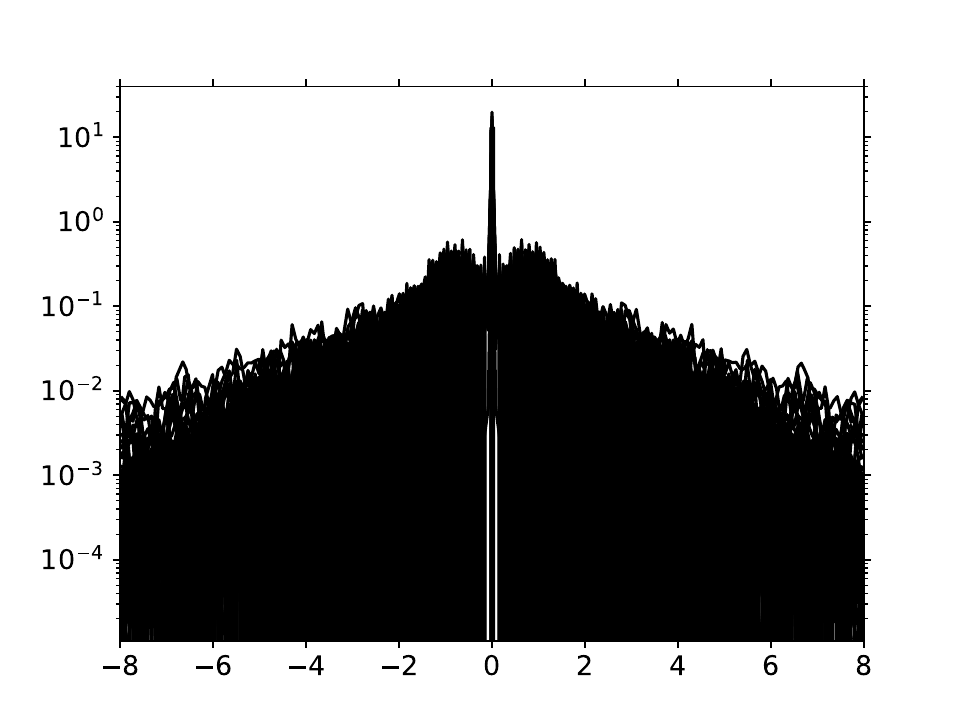}
			\put(50,2){\makebox(0,0){\large\sffamily original return}}
			\put(1,35){\makebox(0,0){\rotatebox{90}{\large\sffamily pdf}}}
		\end{overpic}
	}
	\caption{\label{fig:Year2014_NormalizedOriginalRet}Empirical distributions of the normalized, original returns $p^{(\mathrm{orig},k)} (r_k)$ for all stocks $k=1,\ldots,K$ with $K=479$ for the whole year 2014 with $\Delta t = 1\,\mathrm{s}$.}
\end{figure}%
We show a single typical return distribution for a time resolution of $\Delta t = 1\,\mathrm{s}$ in Fig.~\ref{fig:Year2014_NormalizedOriginalRet_TROW}.
Surprisingly, deep dips give it a fence--like appearance.
This shape of the distribution can also be traced back to the tick size as smallest trading unit~\cite{MUENNIX2010}. 
Analogous to Fig.~\ref{fig:DistYearly2014_RotRet}, we display all $k=1,\ldots,K$, univariate distributions of the normalized, original returns together in Fig.~\ref{fig:Year2014_NormalizedOriginalRet}.
Apart from this very peculiar feature of the univariate distributions for the original returns, it becomes obvious that the univariate distributions for the rotated returns carry much more information, namely on the correlation structure. They depend on the corresponding eigenvalue with a strong influence on the shapes, as see in Figs.~\ref{fig:DistYearly2014_RotRet}--\ref{fig:DistYearly2014_RotRet_10SmallestEigenval}.

Empirical correlation or covariance matrices are known to be noise dressed, there is a variety of noise reduction methods, see for example~\cite{Laloux_1999} and \cite{Guhr_2003,Schäfer_2010}. As the emphasis in the present contribution is on the construction of multivariate return distributions, we study the influence of noise dressing only by means of an example.
We use Ledoit--Wolf shrinkage~\cite{LedoitWolf:2004}, which reduces noise in covariance matrices, on the distribution of the aggregated returns, see~App.~\ref{app:LedoitWolfShrinkage}. The distributions of the aggregated returns are almost not affected.

\section{\label{sec:Results}Comparison of the Multivariate Model Distributions With the Data}

In Sec.~\ref{sec:Aggregating}, we briefly review the process of aggregation and discuss the model distributions for the aggregated empirical ones.
In Sec.~\ref{sec:EmpiricalEpochDistributions}, we determine the fit parameters for the epoch distributions of the aggregated returns. 
Based on the determination of these values, we fit the distributions of the aggregated returns on the long intervals in Sec.~\ref{sec:EmpiricalLongIntervalDistributions}. In Sec.~\ref{sec:ComparingShapesDdistributionsEpochsLongInterval}, we show that these distributions on the long intervals indeed have a stronger tail behavior caused by the fluctuations of the correlation matrices. Furthermore, we take a closer look at the tails in the distributions of the aggregated returns in Sec.~\ref{sec:TailBehaviorComparison}.

\subsection{\label{sec:Aggregating}Aggregating the Return Distributions}

\begin{figure}[htbp]
	\centering {\begin{overpic}[width=.85\linewidth]{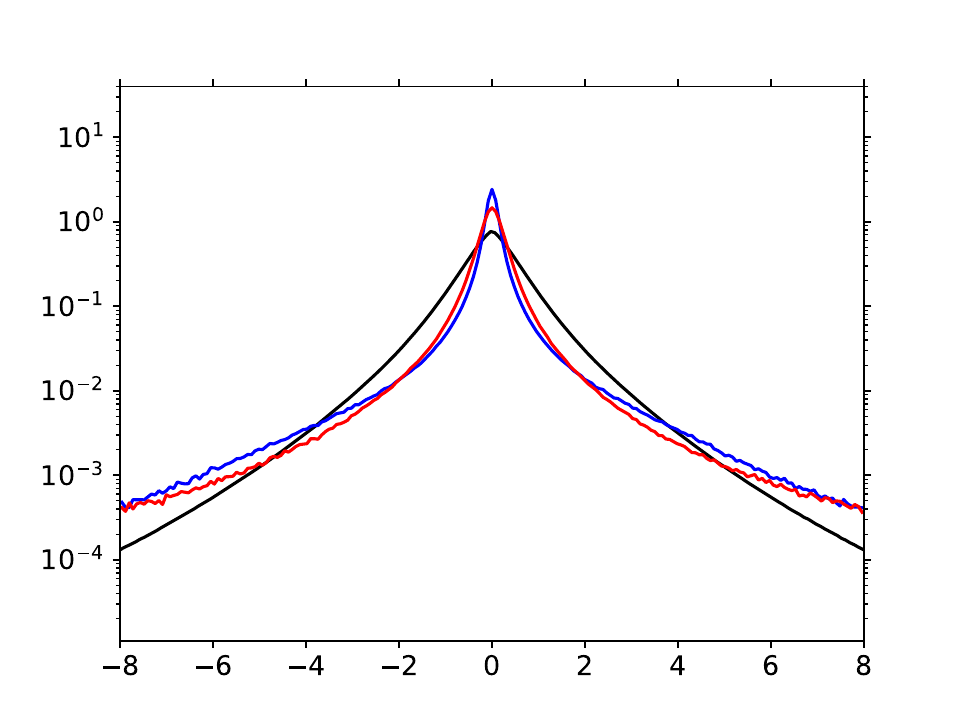}
			\put(52,0){\makebox(0,0){\large\sffamily aggregated return}}
			\put(2,35){\makebox(0,0){\rotatebox{90}{\large\sffamily pdf}}}
			\put(75,55){\makebox(0,0)}
		\end{overpic}
	}
	\caption{Empirical distribution of the aggregated returns $p^{(\mathrm{aggr})}(\widetilde{r})$ for the whole year 2014 with $\Delta t = 1\,\mathrm{s}$ (black). Empirical distribution $p^{(\mathrm{rot,scal},k)}(\widetilde{r}_k)$ corresponding to the largest eigenvalue for $k=K$ is displayed in blue color and for second largest eigenvalue for $k=K-1$ in red color.}
	\label{fig:AggRet_2014_EigenVal}
\end{figure}

In Sec.~I.2.2, we introduced the univariate distributions of the rotated returns.
To obtain better statistics for larger return horizon $\Delta t$ we go over to the distributions $p^{(\mathrm{aggr})}(\widetilde{r})$ of the aggregated returns~(\ref{eq:widetilde}).
By using the aggregated returns we facilitate the data analysis, but our results are not restricted to the eigenbasis, they are in particular also valid  in the original basis of the individual stocks.
The rescaling with the corresponding eigenvalues moves the empirical, univariate distributions of the rotated, unrescaled returns, see Fig.~\ref{fig:DistYearly2014_RotRet}, corresponding to different eigenvalues closer together, see~Fig.~\ref{fig:AggRet_2014_EigenVal}. 
However, the empirical distributions $p^{(\mathrm{rot,scal},k)}(\widetilde{r}_k)$ of the rotated and rescaled returns corresponding to the largest eigenvalue for $k=K$ and second largest eigenvalues for $k=K-1$ are still heavier--tailed than the empirical aggregated distribution of all returns $p^{(\mathrm{aggr})} (\widetilde{r})$.
The situation is reminiscent of other statistical situations where one has a null hypothesis as for example in the case of the Marchenko–-Pastur distribution~\cite{MP_distribution_1967}. Here, the distribution of the aggregated returns is the null hypothesis.

In the sequel, we analogously transform the model distribution ~(I.18) for the epochs and those for the long interval (I.31), (I.32), (I.33) and (I.34) of the rotated returns into distributions of the rotated and rescaled returns.
In the formulae, this is done by replacing the eigenvalues $\Lambda_k$ with one, as the functional forms of the distributions is the same for all $k$ in the respective model.
In general, we use the notation $Y, Y^\prime = A,G$ (algebraic or Gaussian), where $Y$ refers to the epoch distribution and $Y^\prime$ to the random matrix distribution. For further details, see I.

The non--stationarity of distributions for the aggregated returns was already studied by looking at the Gaussian--Gaussian case ($Y, Y^\prime = G,G$)~\cite{Chetalova_2015}. For daily data, the long intervals were chosen by the length of market states which are based on the quasi--stationarity of correlations matrices. In particular,
it was shown that during crises times the distributions of aggregated returns become more heavy-tailed.

\subsection{\label{sec:EmpiricalEpochDistributions}Fits of the Aggregated Empirical Distributions on Epochs}

When looking at heavy--tailed distributions, one is usually interested in their shapes on two scales, the linear one that emphasizes the region around zero, and the logarithmic one that allows an assessment of the tail--behavior. Thus, we carry out two least square fits for each distributions, a linear and a logarithmic one. We use the normalized $\chi^2_{\mathrm{lin}}$ and $\chi^2_{\mathrm{ln}}$ \cite{Bevington2003} as measures for the goodness of the fit.

\begin{figure}[htbp]
	\captionsetup[subfigure]{labelformat=empty}
	\centering
	\begin{minipage}{.5\textwidth}
		\subfloat[]{\begin{overpic}[width=1.\linewidth]{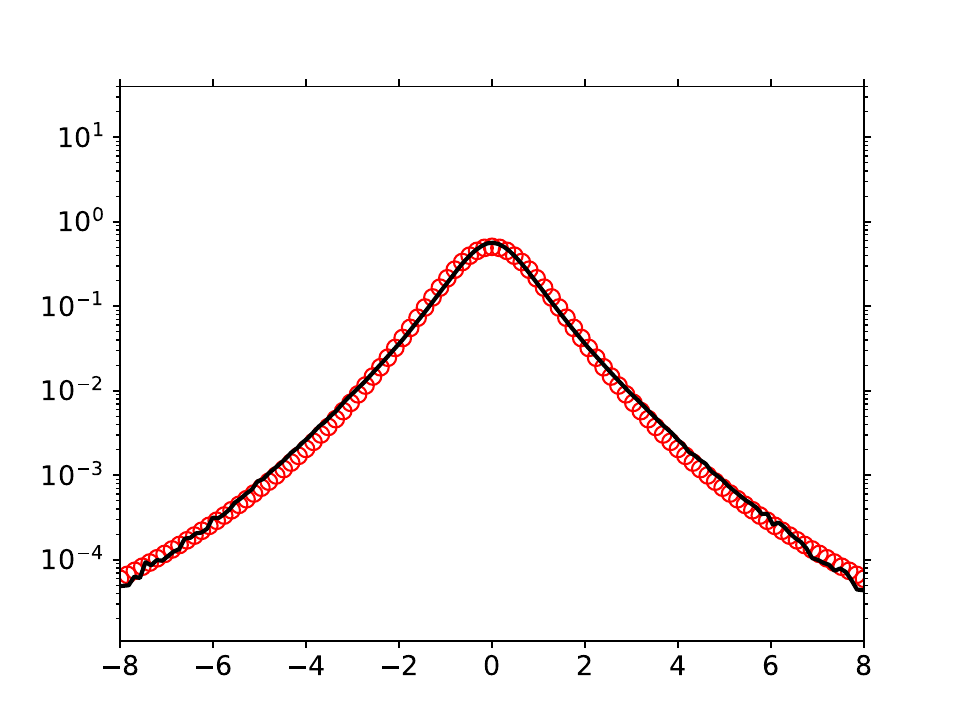}
				\put(18,55){\noindent\fbox{\parbox{1.7cm}{\sffamily Oct 23\\$\mathsf{\Delta t = 1\,s}$}}}
				\put(50,0){\makebox(0,0){\small\sffamily aggregated return}}
				\put(2,35){\makebox(0,0){\rotatebox{90}{\small\sffamily pdf}}}
				\put(78,55){\makebox(0,0){\sffamily\Large A}}
			\end{overpic}
		}
	\end{minipage}%
	\begin{minipage}{.5\textwidth}
		\centering
		\subfloat[]{\begin{overpic}[width=1.\linewidth]{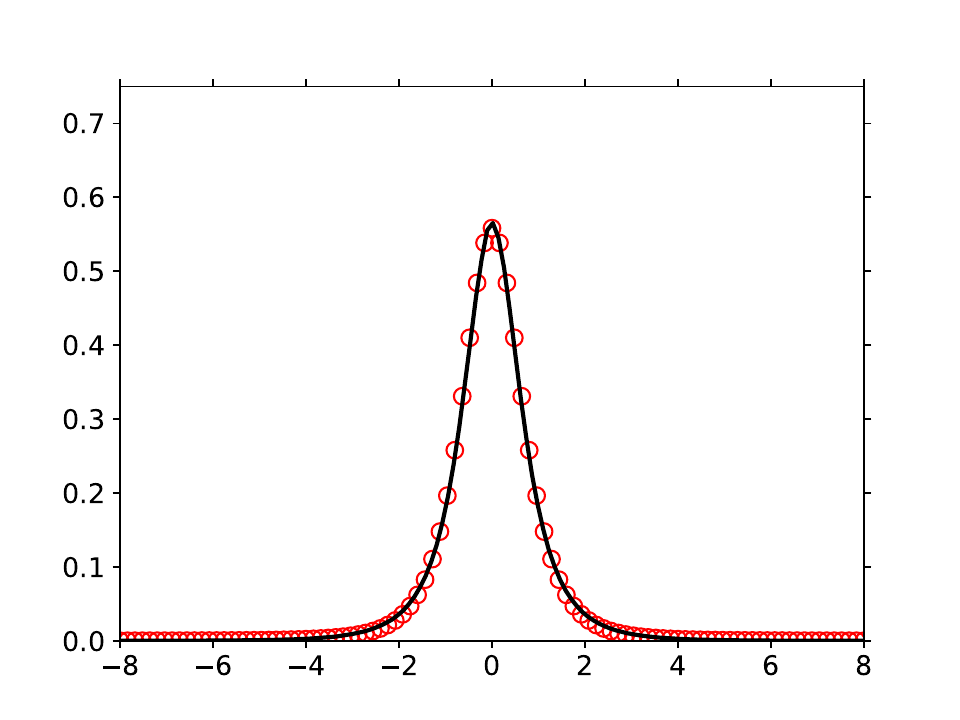}
				\put(18,55){\noindent\fbox{\parbox{1.7cm}{\sffamily Oct 23\\$\mathsf{\Delta t = 1\,s}$}}}
				\put(50,0){\makebox(0,0){\small aggregated return}}
				\put(2,35){\makebox(0,0){\rotatebox{90}{\small pdf}}}
				\put(78,55){\makebox(0,0){\sffamily\Large A}}
			\end{overpic}
		}
	\end{minipage}
	\caption{Empirical distributions of aggregated returns with $\Delta t = 1\,\mathrm{s}$ (black) for epoch October 23, 2014 with $p_\mathrm{A}^{(\mathrm{aggr})}(\widetilde{r})$ (red circles), left: for $l_{\mathrm{rot}} = 2.769$ on a logarithmic scale, right: for $l_{\mathrm{rot}} = 2.303$ on a linear scale.}
	\label{fig:2220DataPoints_AggretRetDist_Epoch_20141023}
\end{figure}

\begin{figure}[htbp]
	\captionsetup[subfigure]{labelformat=empty}
	\centering
	\begin{minipage}{.5\textwidth}
		\centering
		\subfloat[]{\begin{overpic}[width=1.\linewidth]{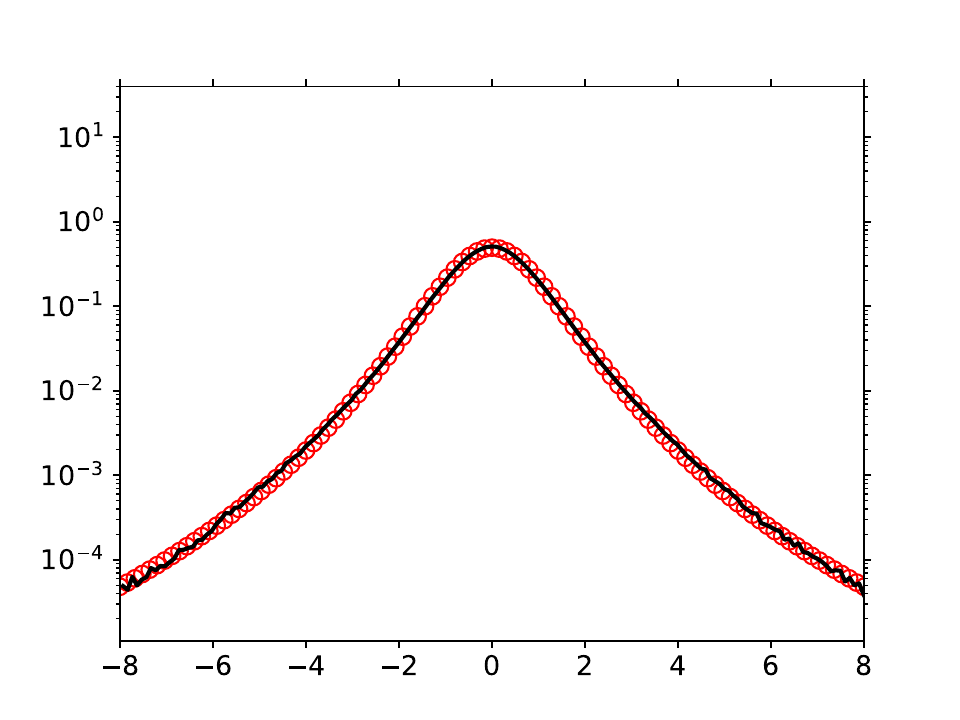}
				\put(18,55){\noindent\fbox{\parbox{1.5cm}{\sffamily Dec 8\\$\mathsf{\Delta t = 1\,s}$}}}
				\put(50,0){\makebox(0,0){\small\sffamily aggregated return}}
				\put(2,35){\makebox(0,0){\rotatebox{90}{\small\sffamily pdf}}}
				\put(78,55){\makebox(0,0){\sffamily\Large A}}
			\end{overpic}
		}
	\end{minipage}%
	\begin{minipage}{.5\textwidth}
		\centering
		\subfloat[]{\begin{overpic}[width=1.\linewidth]{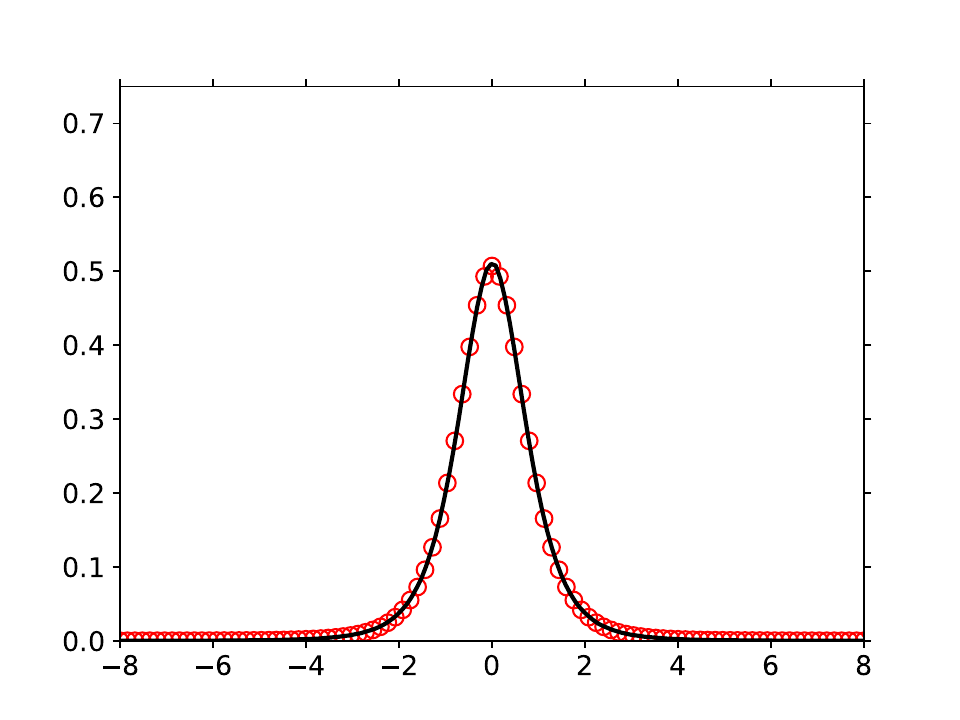}
				\put(18,55){\noindent\fbox{\parbox{1.5cm}{\sffamily Dec 8\\$\mathsf{\Delta t = 1\,s}$}}}
				\put(50,0){\makebox(0,0){\small\sffamily aggregated return}}
				\put(2,35){\makebox(0,0){\rotatebox{90}{\small\sffamily pdf}}}
				\put(78,55){\makebox(0,0){\sffamily\Large A}}
			\end{overpic}
		}
	\end{minipage}
	\caption{Empirical distributions of aggregated returns with $\Delta t = 1\,\mathrm{s}$ (black) for epoch December 8, 2014 with $p_\mathrm{A}^{(\mathrm{aggr})}(\widetilde{r})$ (red circles), left:  for $l_{\mathrm{rot}} = 2.933$ on a logarithmic scale, right: for $l_{\mathrm{rot}} = 2.744$ on a linear scale.}
	\label{fig:22200DataPoints_AggretRetDist_Epoch_20141208}
\end{figure}

\begin{figure}[htbp]
	\captionsetup[subfigure]{labelformat=empty}
	\centering
	\begin{minipage}{.5\textwidth}
		\centering
		\subfloat[]{\begin{overpic}[width=1\linewidth]{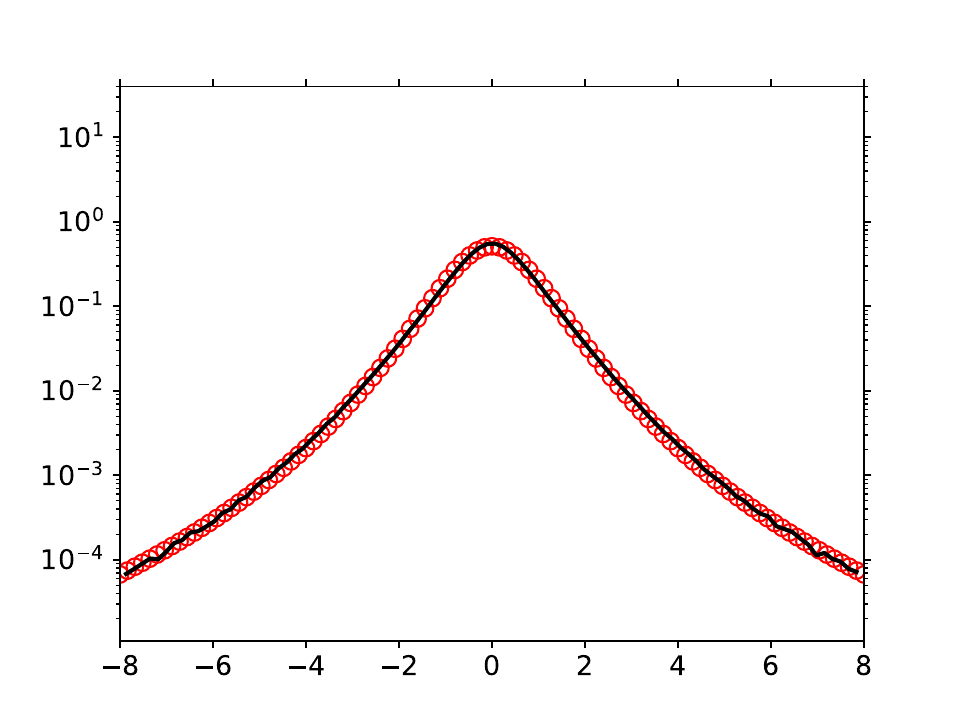}
			\put(18,55){\noindent\fbox{\parbox{1.5cm}{\sffamily Dec 17\\$\mathsf{\Delta t = 1\,s}$}}}
			\put(50,0){\makebox(0,0){\small\sffamily aggregated return}}
			\put(2,35){\makebox(0,0){\rotatebox{90}{\small\sffamily pdf}}}
			\put(78,55){\makebox(0,0){\sffamily\Large A}}
		\end{overpic}
	}%
	\end{minipage}%
	\begin{minipage}{.5\textwidth}
		\centering
		\subfloat[]{\begin{overpic}[width=1\linewidth]{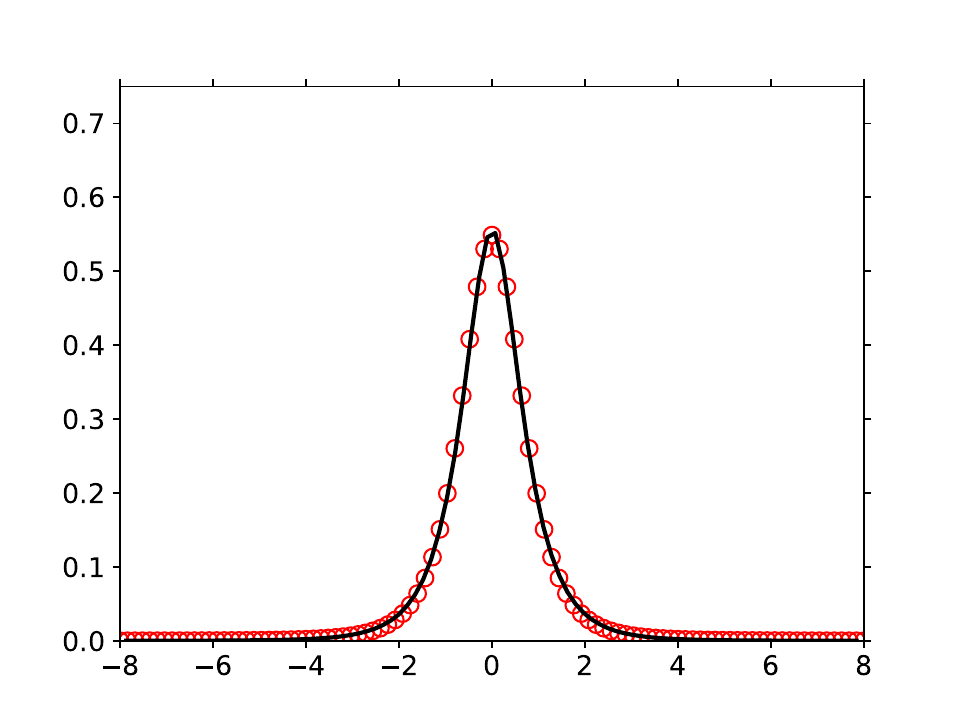}
				\put(18,55){\noindent\fbox{\parbox{1.5cm}{\sffamily Dec 17\\$\mathsf{\Delta t = 1\,s}$}}}
				\put(50,0){\makebox(0,0){\small\sffamily aggregated return}}
				\put(2,35){\makebox(0,0){\rotatebox{90}{\small\sffamily pdf}}}
				\put(78,55){\makebox(0,0){\sffamily\Large A}}
			\end{overpic}
		}
	\end{minipage}
	\caption{Empirical distributions of aggregated returns with $\Delta t = 1\,\mathrm{s}$ (black) for epoch December 17, 2014 with $p_\mathrm{A}^{(\mathrm{aggr})}(\widetilde{r})$ (red circles), left: for $ l_{\mathrm{rot}} = 2.679$  on a logarithmic scale, right: for $ l_{\mathrm{rot}} = 2.361$ on a linear scale.}
	\label{fig:22200DataPoints_AggretRetDist_Epoch_20141217}
\end{figure}

\begin{figure}[htbp]
	\captionsetup[subfigure]{labelformat=empty}
	\centering
	\begin{minipage}{.5\textwidth}
		\centering
		\subfloat[]{\begin{overpic}[width=.9\linewidth]{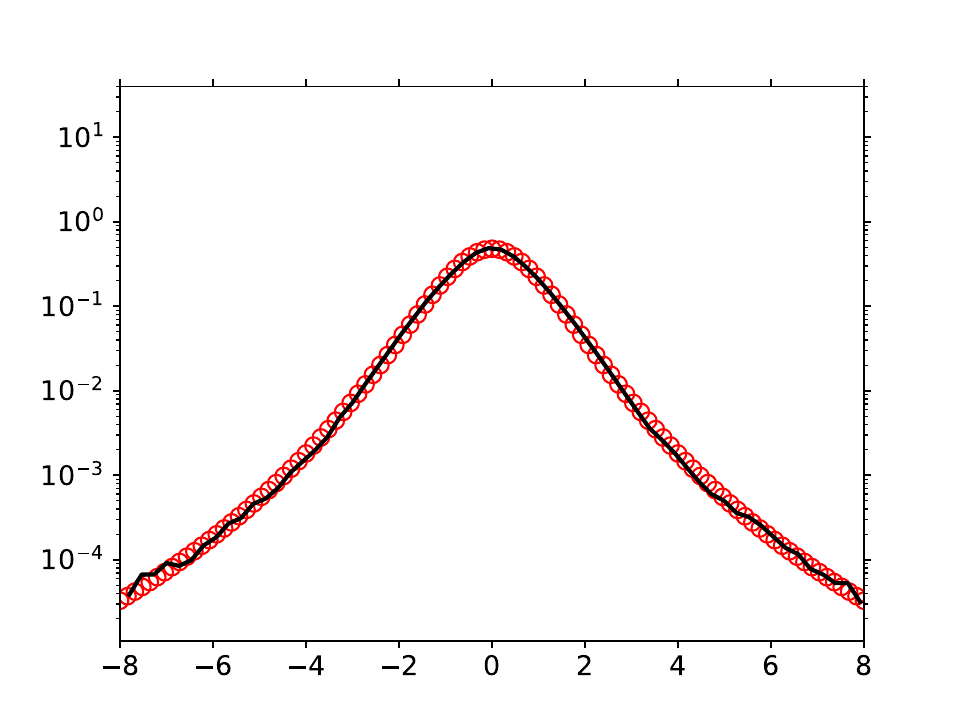}
				\put(18,55){\noindent\fbox{\parbox{1.7cm}{\sffamily Feb 20\\$\mathsf{\Delta t = 10\,s}$}}}
				\put(50,0){\makebox(0,0){\small\sffamily aggregated return}}
				\put(2,35){\makebox(0,0){\rotatebox{90}{\small\sffamily pdf}}}
				\put(78,55){\makebox(0,0){\sffamily\Large A}}
			\end{overpic}
		}
	\end{minipage}%
	\begin{minipage}{.5\textwidth}
		\centering
		\subfloat[]{\begin{overpic}[width=.9\linewidth]{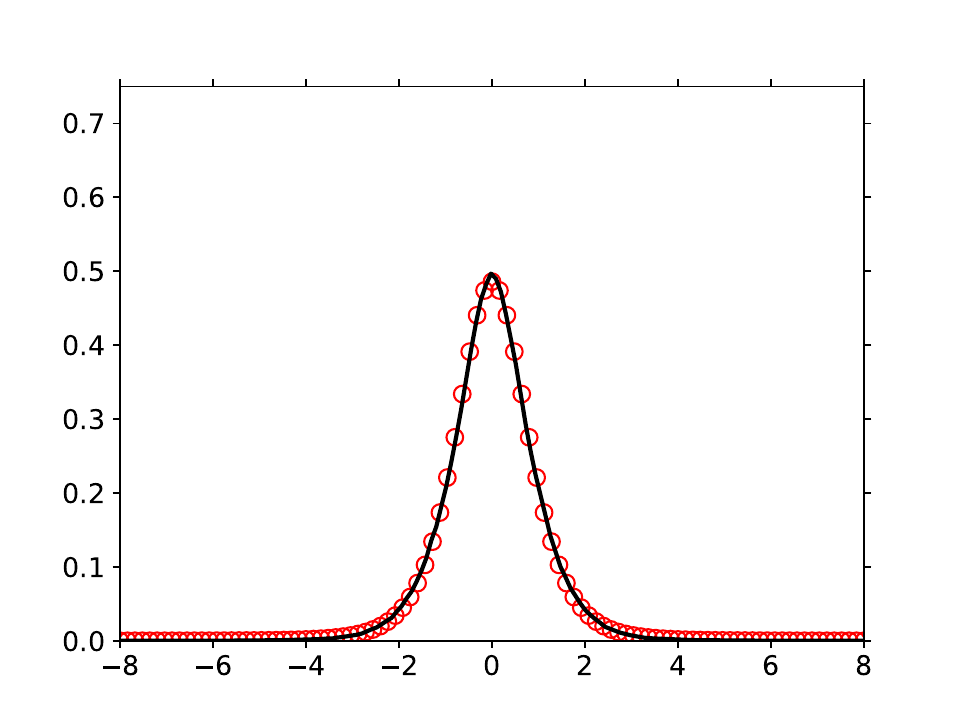}
				\put(18,55){\noindent\fbox{\parbox{1.7cm}{\sffamily Feb 20\\$\mathsf{\Delta t = 10\,s}$}}}
				\put(50,0){\makebox(0,0){\small\sffamily aggregated return}}
				\put(2,35){\makebox(0,0){\rotatebox{90}{\small\sffamily pdf}}}
				\put(78,55){\makebox(0,0){\sffamily\Large A}}
			\end{overpic}
		}
	\end{minipage}
	\caption{Empirical distributions of aggregated returns with $\Delta t = 10\,\mathrm{s}$ (black) for epoch February 20, 2014 with $p_\mathrm{A}^{(\mathrm{aggr})}(\widetilde{r})$ (red circles), left: for $l_{\mathrm{rot}} = 3.227$ on a logarithmic scale, right: for $l_{\mathrm{rot}} = 3.079$ on a linear scale.}
	\label{fig:2220DataPoints_AggretRetDist_Epoch_20140220}
\end{figure}

\begin{figure}[htbp]
	\captionsetup[subfigure]{labelformat=empty}
	\centering
	\begin{minipage}{.5\textwidth}
		\subfloat[]{\begin{overpic}[width=1.\linewidth]{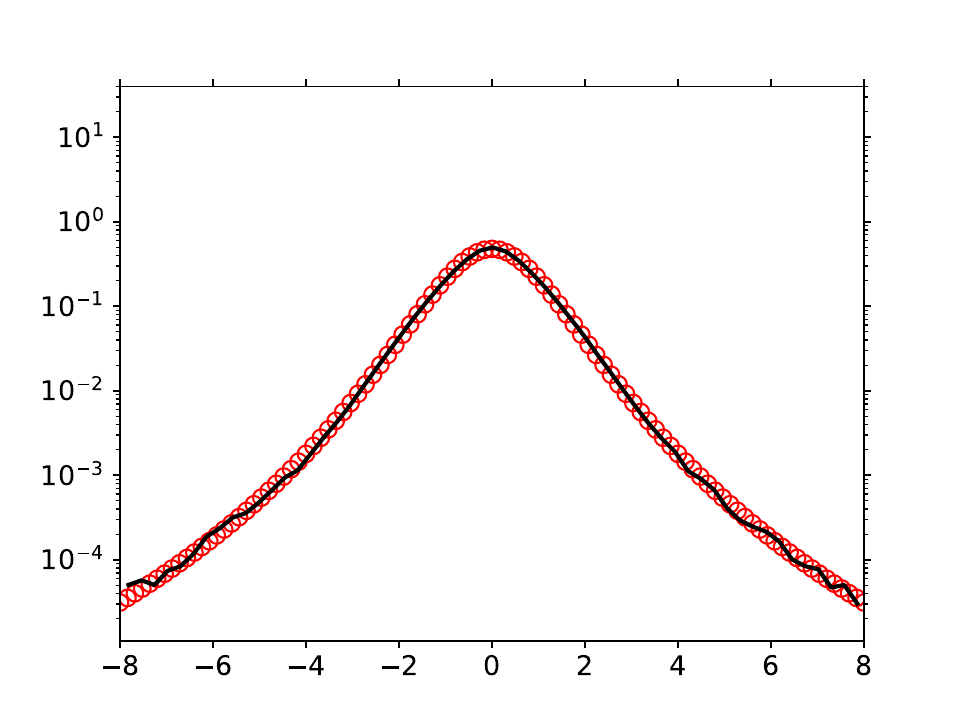}
				\put(18,55){\noindent\fbox{\parbox{1.7cm}{\sffamily Jun 2\\$\mathsf{\Delta t = 10\,s}$}}}
				\put(50,0){\makebox(0,0){\small\sffamily aggregated return}}
				\put(2,35){\makebox(0,0){\rotatebox{90}{\small\sffamily pdf}}}
				\put(78,55){\makebox(0,0){\sffamily\Large A}}
			\end{overpic}
		}
	\end{minipage}%
	\begin{minipage}{.5\textwidth}
		\centering
		\subfloat[]{\begin{overpic}[width=1.\linewidth]{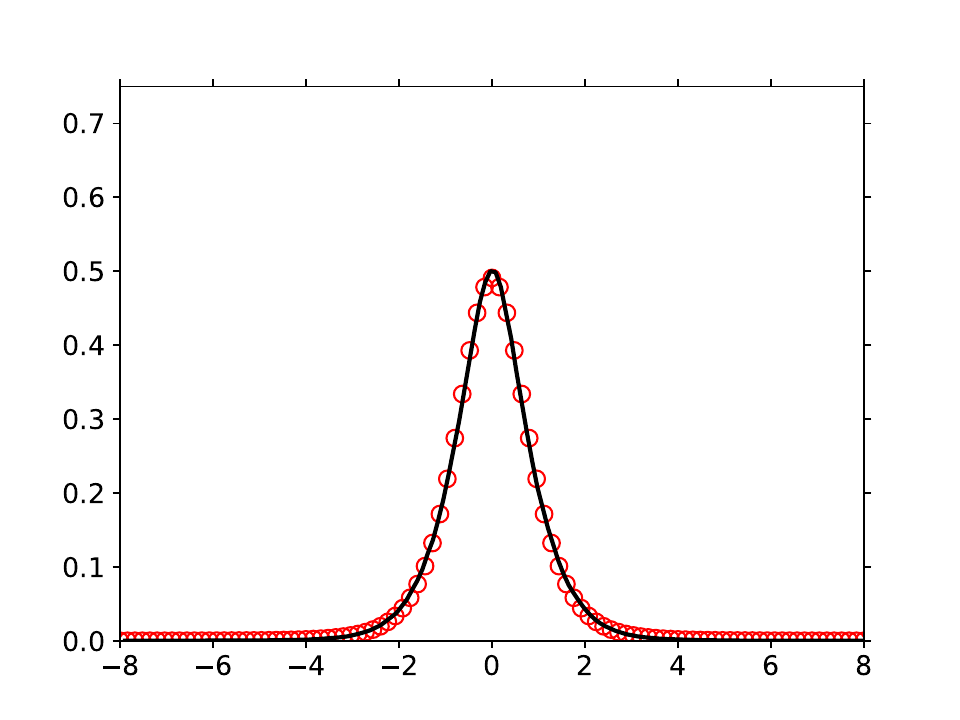}
				\put(18,55){\noindent\fbox{\parbox{1.7cm}{\sffamily Jun 2\\$\mathsf{\Delta t = 10\,s}$}}}
				\put(50,0){\makebox(0,0){\small aggregated return}}
				\put(2,35){\makebox(0,0){\rotatebox{90}{\small pdf}}}
				\put(78,55){\makebox(0,0){\sffamily\Large A}}
			\end{overpic}
		}
	\end{minipage}
	\caption{Empirical distributions of aggregated returns with $\Delta t = 10\,\mathrm{s}$ (black) for epoch June 2, 2014 with $p_\mathrm{A}^{(\mathrm{aggr})}(\widetilde{r})$ (red circles), left: for $l_{\mathrm{rot}} = 3.264$ on a logarithmic scale, right: for $l_{\mathrm{rot}} = 2.985$ on a linear scale.}
	\label{fig:2220DataPoints_AggretRetDist_Epoch_20140602}
\end{figure}

For three selected epochs, 23 October, 8 December and 17 December, we show in Figs.~\ref{fig:2220DataPoints_AggretRetDist_Epoch_20141023}--\ref{fig:22200DataPoints_AggretRetDist_Epoch_20141217} empirical distributions $p^{(\mathrm{aggr})} (\widetilde{r})$ of the aggregated returns with fits on a logarithmic and a linear scale for $\Delta t = 1\,\mathrm{s}$. For $\Delta t = 10\,\mathrm{s}$, the same comparisons are depicted in Figs.~\ref{fig:2220DataPoints_AggretRetDist_Epoch_20140220} and \ref{fig:2220DataPoints_AggretRetDist_Epoch_20140602} for the 20th of December and the 2nd of June, respectively.
As expected for intraday data the empirical distributions have strong heavy tails. 
The quality of the fits for distributions depends on the non--stationarity.
For a smaller return horizon $\Delta t$, the distributions are heavier--tailed and the distributions shows a higher probability near their centers~\cite{Plerou_1999_Dist,Gopikrishnan_1999}.
For these five different epochs, we obtain very good fits for the distributions of the aggregated returns $ p_{\mathrm{A}}^{(\mathrm{aggr})} (\widetilde{r})$. The model parameter $l_\mathrm{rot}$ and the fit parameter are listed in~App.~\ref{app:FitParameterslaggr_Single}.  
Furthermore, we notice that the fit parameters change due to the non-stationarity of the epoch distributions.

By averaging over all fit parameters $l_{\mathrm{rot}}$ for all epoch distributions in 2014,
we determine the parameter $\langle l_{\mathrm{rot}} \rangle$ as input for the upcoming discussion of the distributions on the long interval, see~App.~\ref{app:FitParameterslaggr_Averaged}. The parameters vary for different return horizons $\Delta t$ and also for linear or logarithmic fits.


\subsection{\label{sec:EmpiricalLongIntervalDistributions}Fits of the Aggregated Distributions on the Long Interval}

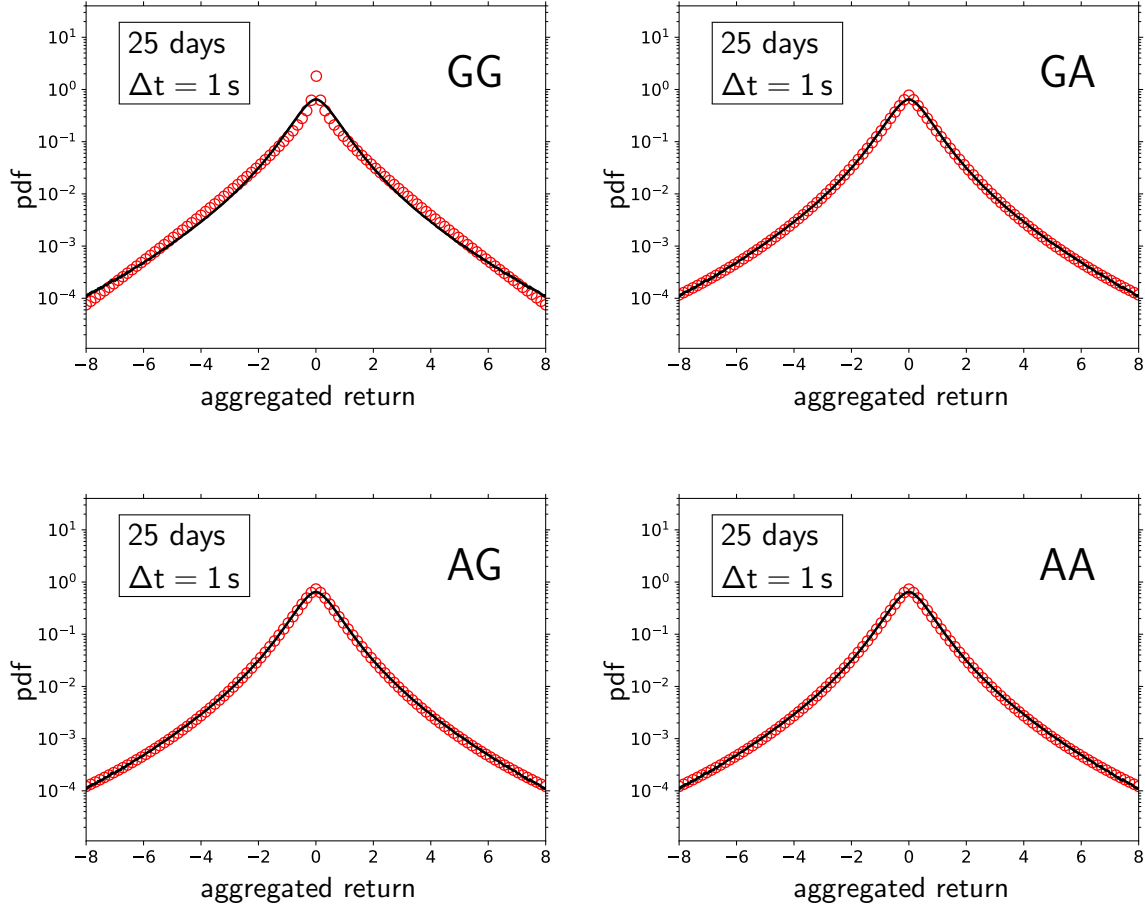
\begin{figure}[htbp]
	\captionsetup[subfigure]{labelformat=empty}
	\centering
	\begin{minipage}{.5\textwidth}
		\centering
		\subfloat[]{\begin{overpic}[width=1.\linewidth]{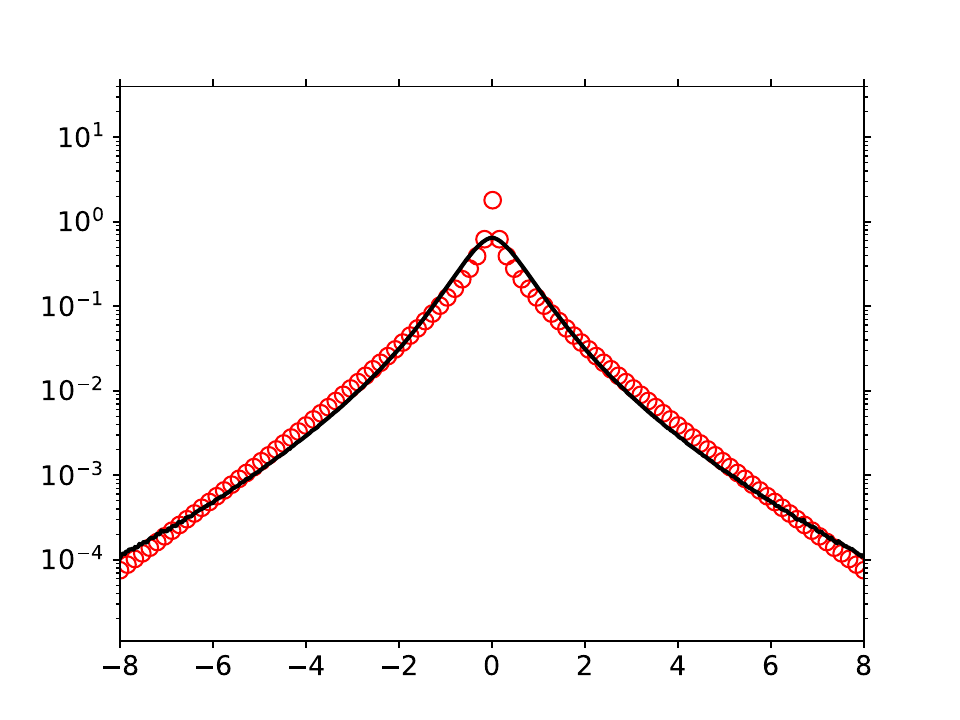}
				\put(18,55){\noindent\fbox{\parbox{1.5cm}{\sffamily 25 days\\$\mathsf{\Delta t = 1\,s}$}}}
				\put(50,0){\makebox(0,0){\small\sffamily aggregated return}}
				\put(2,35){\makebox(0,0){\rotatebox{90}{\small\sffamily pdf}}}
				\put(78,55){\makebox(0,0){\sffamily\Large GG}}
			\end{overpic}
		}
	\end{minipage}%
	\begin{minipage}{.5\textwidth}
		\centering
		\subfloat[]{\begin{overpic}[width=1.\linewidth]{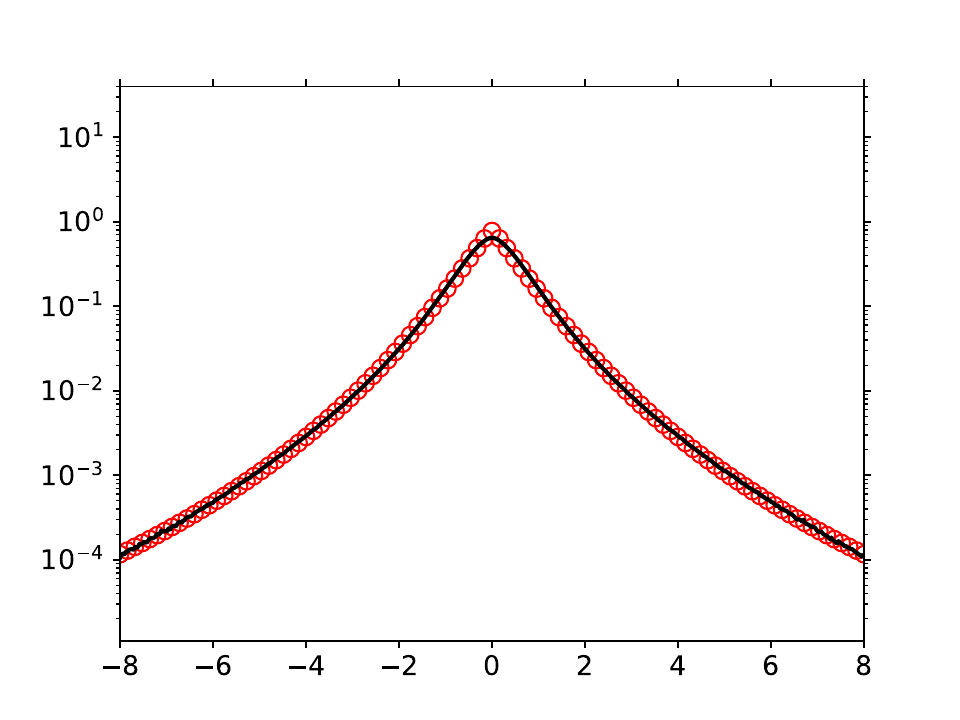}
				\put(18,55){\noindent\fbox{\parbox{1.5cm}{\sffamily 25 days\\$\mathsf{\Delta t = 1\,s}$}}}
				\put(50,0){\makebox(0,0){\small\sffamily aggregated return}}
				\put(2,35){\makebox(0,0){\rotatebox{90}{\small\sffamily pdf}}}
				\put(78,55){\makebox(0,0){\sffamily\Large GA}}
			\end{overpic}
		}
	\end{minipage}\\%
	\begin{minipage}{.5\textwidth}
		\centering
		\subfloat[]{\begin{overpic}[width=1.\linewidth]{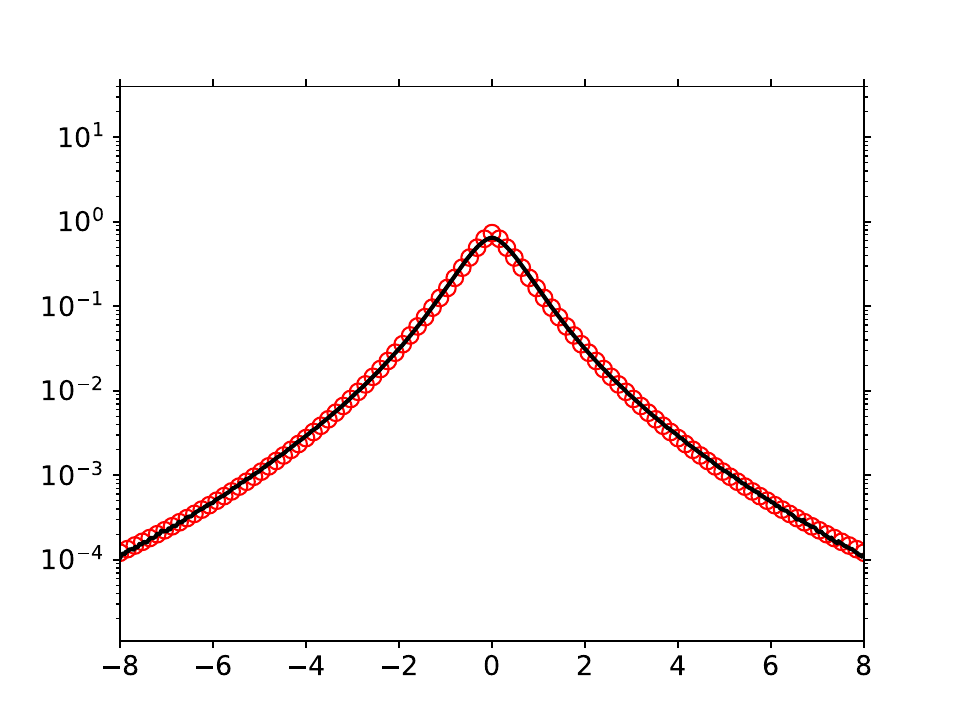}
				\put(18,55){\noindent\fbox{\parbox{1.5cm}{\sffamily 25 days\\$\mathsf{\Delta t = 1\,s}$}}}
				\put(50,0){\makebox(0,0){\small\sffamily aggregated return}}
				\put(2,35){\makebox(0,0){\rotatebox{90}{\small\sffamily pdf}}}
				\put(78,55){\makebox(0,0){\sffamily\Large AG}}
			\end{overpic}
		}
	\end{minipage}%
	\begin{minipage}{.5\textwidth}
		\centering
		\subfloat[]{\begin{overpic}[width=1.\linewidth]{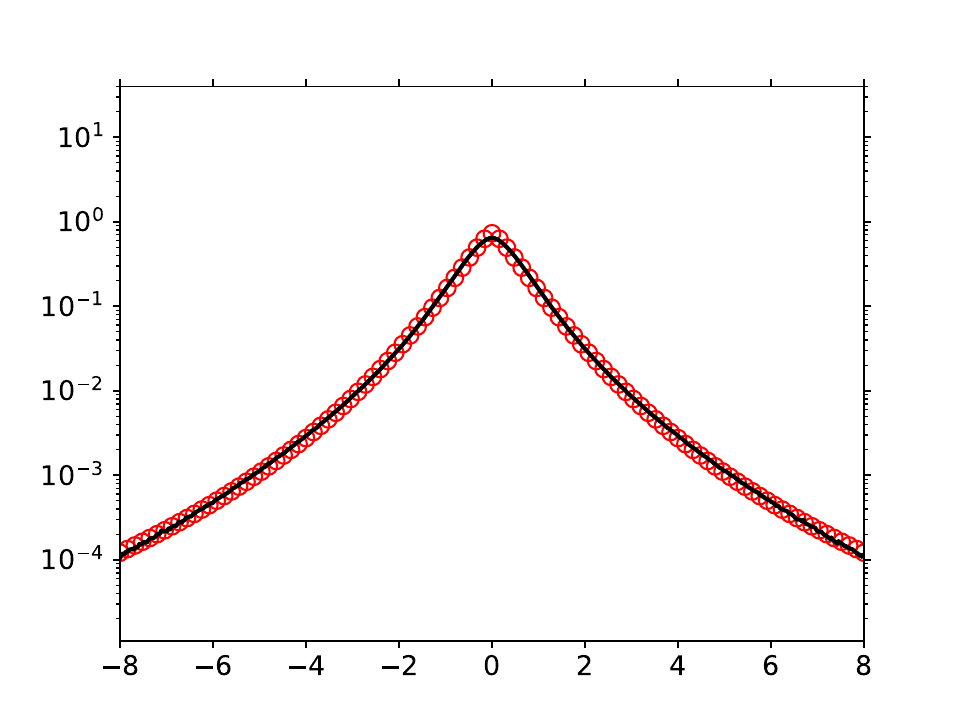}
				\put(18,55){\noindent\fbox{\parbox{1.5cm}{\sffamily 25 days\\$\mathsf{\Delta t = 1\,s}$}}}
				\put(50,0){\makebox(0,0){\small\sffamily aggregated return}}
				\put(2,35){\makebox(0,0){\rotatebox{90}{\small\sffamily pdf}}}
				\put(78,55){\makebox(0,0){\sffamily\Large AA}}
			\end{overpic}
		}
	\end{minipage}
	\caption{Empirical distributions of aggregated returns with $\Delta t = 1\,\mathrm{s}$ (black) for the 9th long interval (25 trading days) on a logarithmic scale. Model distributions in red color with Gaussian--Gaussian: $N = 0.782$, Gaussian--Algebraic: $L_{\mathrm{rot}}= 3.467$, $N = 2.438$, Algebraic--Gaussian: $N = 2.852$, Algebraic--Algebraic: $L_{\mathrm{rot}}= 99.554$, $N = 2.910$.}
	\label{fig:22200DataPoints_AggretRetDist_Int_Log}
\end{figure}

\begin{figure}[htbp]
	\captionsetup[subfigure]{labelformat=empty}
	\centering
	\begin{minipage}{.5\textwidth}
	\centering
	\subfloat[]{\begin{overpic}[width=1.\linewidth]{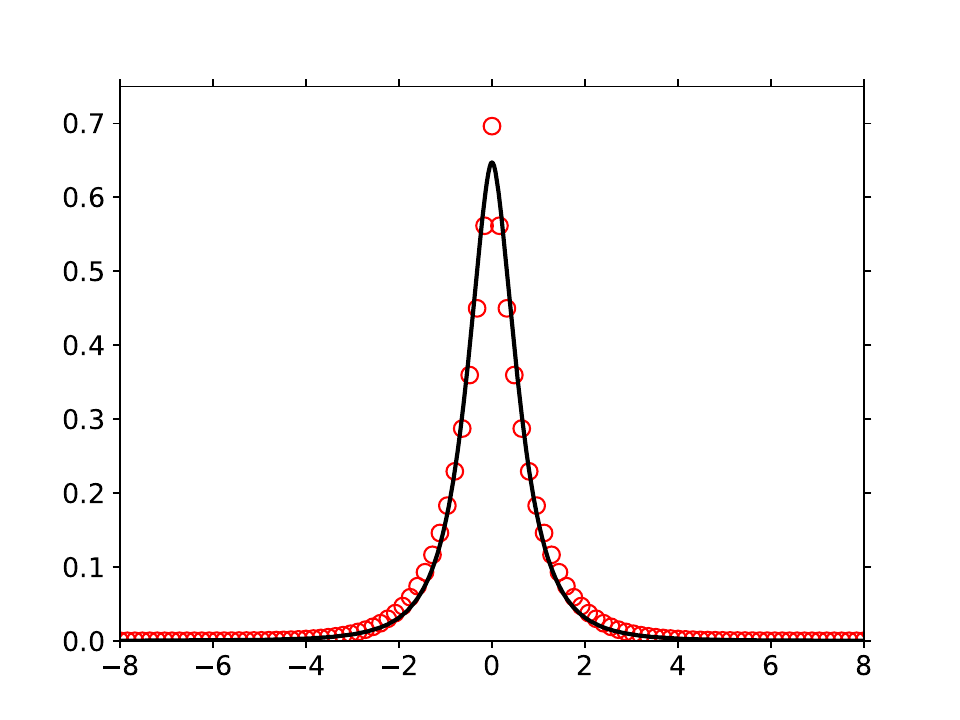}
			\put(18,55){\noindent\fbox{\parbox{1.5cm}{\sffamily 25 days\\$\mathsf{\Delta t = 1\,s}$}}}
			\put(50,0){\makebox(0,0){\small\sffamily aggregated return}}
			\put(2,35){\makebox(0,0){\rotatebox{90}{\small\sffamily pdf}}}
			\put(78,55){\makebox(0,0){\sffamily\Large GG}}
		\end{overpic}
	}
\end{minipage}%
\begin{minipage}{.5\textwidth}
	\centering
	\subfloat[]{\begin{overpic}[width=1.\linewidth]{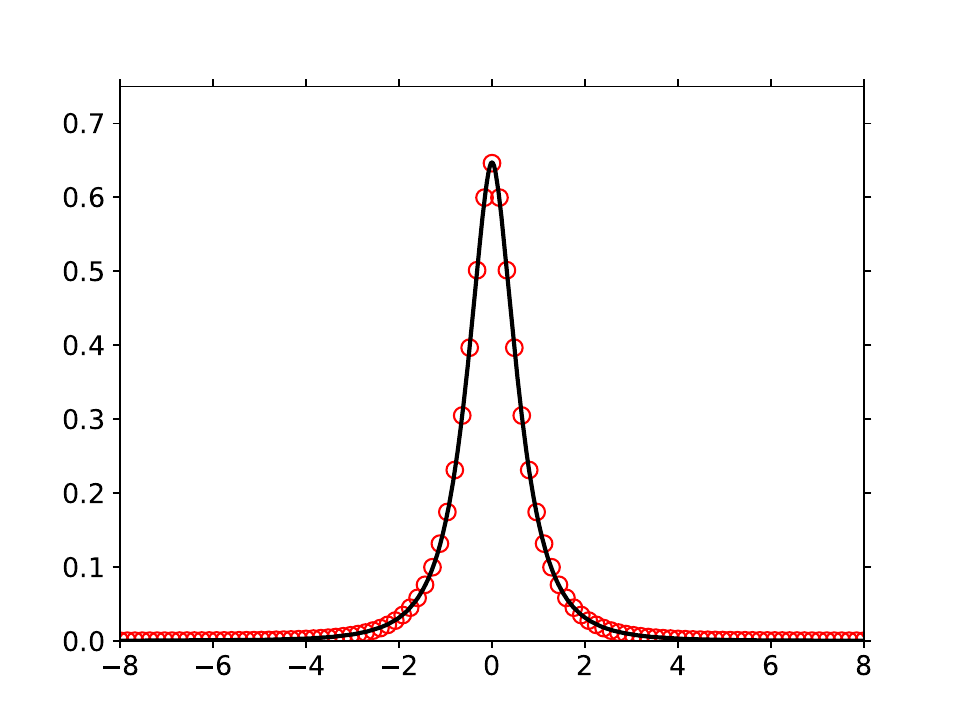}
			\put(18,55){\noindent\fbox{\parbox{1.5cm}{\sffamily 25 days\\$\mathsf{\Delta t = 1\,s}$}}}
			\put(50,0){\makebox(0,0){\small\sffamily aggregated return}}
			\put(2,35){\makebox(0,0){\rotatebox{90}{\small\sffamily pdf}}}
			\put(78,55){\makebox(0,0){\sffamily\Large GA}}
		\end{overpic}
	}
\end{minipage}\\%
\begin{minipage}{.5\textwidth}
	\centering
	\subfloat[]{\begin{overpic}[width=1.\linewidth]{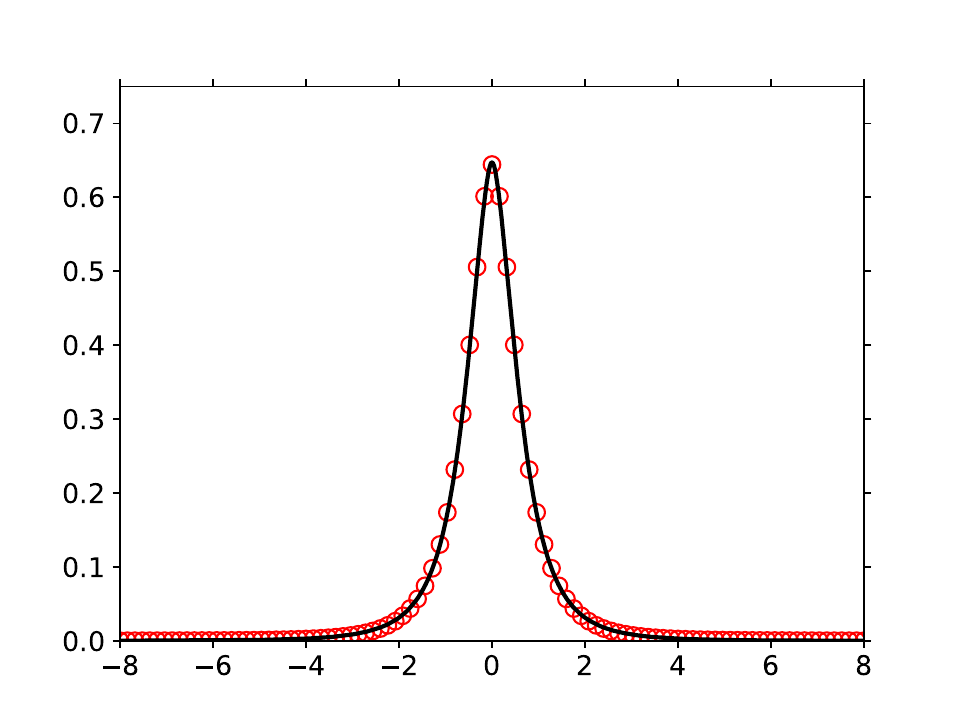}
			\put(18,55){\noindent\fbox{\parbox{1.5cm}{\sffamily 25 days\\$\mathsf{\Delta t = 1\,s}$}}}
			\put(50,0){\makebox(0,0){\small\sffamily aggregated return}}
			\put(2,35){\makebox(0,0){\rotatebox{90}{\small\sffamily pdf}}}
			\put(78,55){\makebox(0,0){\sffamily\Large AG}}
		\end{overpic}
	}
\end{minipage}%
\begin{minipage}{.5\textwidth}
	\centering
	\subfloat[]{\begin{overpic}[width=1.\linewidth]{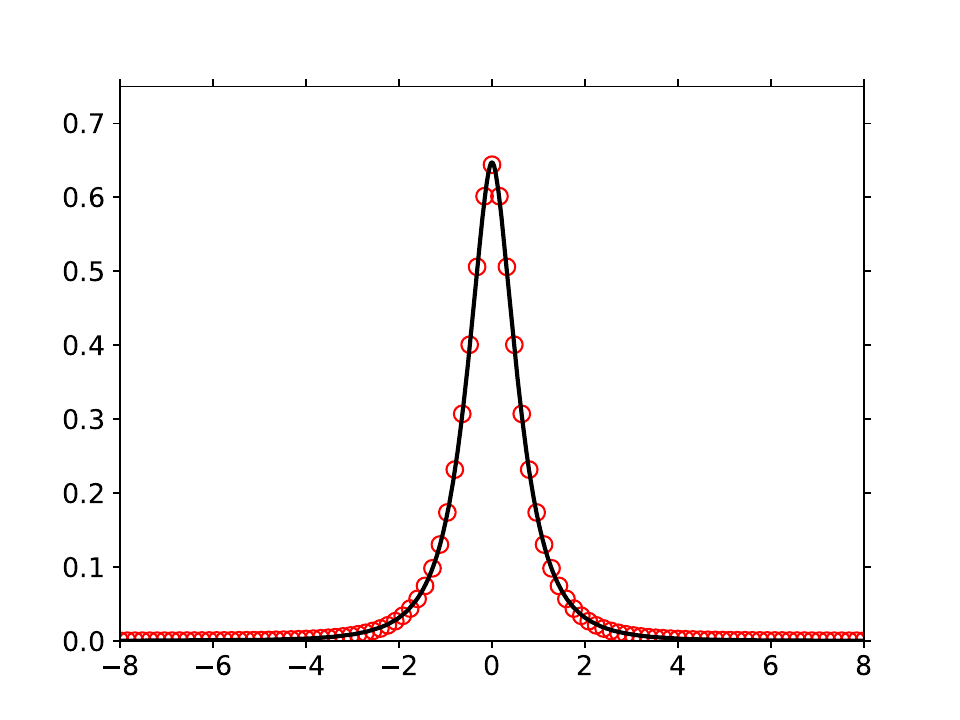}
			\put(18,55){\noindent\fbox{\parbox{1.5cm}{\sffamily 25 days\\$\mathsf{\Delta t = 1\,s}$}}}
			\put(50,0){\makebox(0,0){\small\sffamily aggregated return}}
			\put(2,35){\makebox(0,0){\rotatebox{90}{\small\sffamily pdf}}}
			\put(78,55){\makebox(0,0){\sffamily\Large AA}}
		\end{overpic}
	}
\end{minipage}
	\caption{Empirical distributions of aggregated returns with $\Delta t = 1\,\mathrm{s}$ (black) for the 9th long interval (25 trading days) on a linear scale. Model distributions in red color with Gaussian--Gaussian: $N = 2.036$, Gaussian--Algebraic: $L_{\mathrm{rot}}= 4.419$, $N = 5.058$, Algebraic--Gaussian: $N = 5.926$, Algebraic--Algebraic: $L_{\mathrm{rot}}= 100.346$, $N = 6.085$.}
	\label{fig:22200DataPoints_AggretRetDist_Int_Lin}
\end{figure}

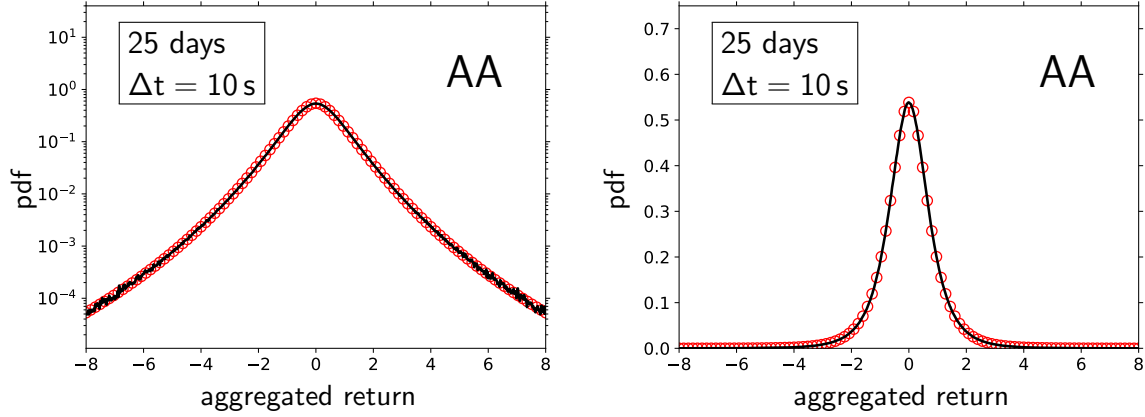
\begin{figure}[htbp]
	\captionsetup[subfigure]{labelformat=empty}
	\centering
\begin{minipage}{.5\textwidth}
	\centering
	\subfloat[]{\begin{overpic}[width=1.\linewidth]{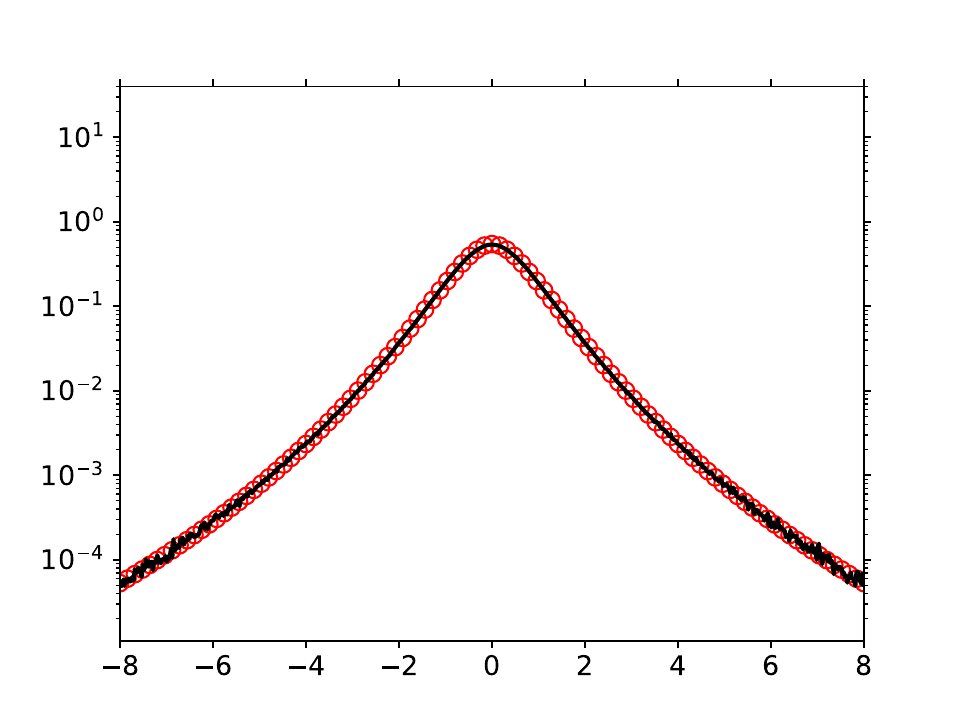}
			\put(18,55){\noindent\fbox{\parbox{1.7cm}{\sffamily 25 days\\$\mathsf{\Delta t = 10\,s}$}}}
			\put(50,0){\makebox(0,0){\small\sffamily aggregated return}}
			\put(2,35){\makebox(0,0){\rotatebox{90}{\small\sffamily pdf}}}
			\put(78,55){\makebox(0,0){\sffamily\Large AA}}
		\end{overpic}
	}
\end{minipage}%
\begin{minipage}{.5\textwidth}
	\centering
	\subfloat[]{\begin{overpic}[width=1.0\linewidth]{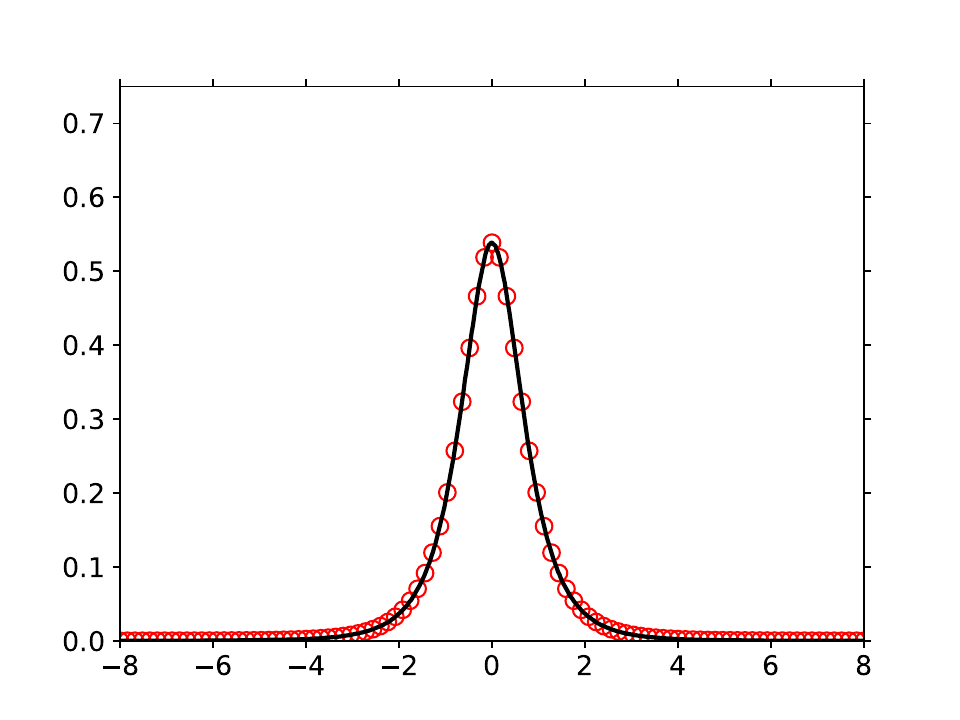}
			\put(18,55){\noindent\fbox{\parbox{1.7cm}{\sffamily 25 days\\$\mathsf{\Delta t = 10\,s}$}}}
			\put(50,0){\makebox(0,0){\small\sffamily aggregated return}}
			\put(2,35){\makebox(0,0){\rotatebox{90}{\small\sffamily pdf}}}
			\put(78,55){\makebox(0,0){\sffamily\Large AA}}
		\end{overpic}
	}
\end{minipage}
	\caption{Empirical distributions of aggregated returns with $\Delta t = 10\,\mathrm{s}$ (black) for the 8th long interval (25 trading days) with $\langle p \rangle_{\mathrm{AA}}^{(\mathrm{aggr})} (\widetilde{r})$ (red circles), 
	left: for $L_{\mathrm{rot}}= 10.990$, $N = 9.935$ on a logarithmic scale, right: for $L_{\mathrm{rot}}= 13.328$, $N = 18.328$ on linear scale.}
	\label{fig:2220DataPoints_AggretRetDist_Int_Lin}
\end{figure}

For all four model distributions on the long interval, we insert the averaged values $\langle l_\mathrm{rot} \rangle$ which are between 2 and 4, see~App.~\ref{app:FitParameterslaggr_Averaged}. By fitting, we determine the remaining parameters $N$ and $L_{\mathrm{rot}}$.
 Figures~\ref{fig:22200DataPoints_AggretRetDist_Int_Log}--\ref{fig:2220DataPoints_AggretRetDist_Int_Lin} display the empirical distributions of the aggregated returns and the corresponding fits on a long interval of 25 trading days. 
 We show the fits for all four model distributions in Figs.~\ref{fig:22200DataPoints_AggretRetDist_Int_Log} and \ref{fig:22200DataPoints_AggretRetDist_Int_Lin} for a long interval ranging from the 17th of October to the 20th of November.
 For both return horizons $\Delta t = 1 \, \mathrm{s}$ and $\Delta t = 10 \, \mathrm{s}$, the fits for $\langle p \rangle_{\mathrm{GA}}^{(\mathrm{aggr})} (\widetilde{r})$, $\langle p \rangle_{\mathrm{AG}}^{(\mathrm{aggr})} (\widetilde{r})$ and $\langle p \rangle_{\mathrm{AA}}^{(\mathrm{aggr})} (\widetilde{r})$ outperform  the one for $\langle p \rangle_{\mathrm{GG}}^{(\mathrm{aggr})} (\widetilde{r})$. The corresponding fit parameters are listed in App.~\ref{app:FitParametersCapital_Single}. Visually, there are hardly any differences between the three better fits, Gaussian-Algebraic performs slightly better than Algebraic-Gaussian and Algebraic-Algebraic on logarithmic and linear scale, see the parameters in App.~\ref{app:FitParametersCapital_Single_ChiSquared}. 
Analogous to App.~\ref{sec:EmpiricalEpochDistributions}, the distributions are heavier-tailed and the distributions show a higher probability near their centers for a smaller return horizon.
To be consistent in our modeling, we must exclude the Gaussian--Gaussian and the Gaussian--Algebraic cases as we confirmed the validity of the algebraic distribution in the epochs. The Algebraic--Gaussian and Algebraic--Algebaic model distributions perform almost equally well. 
In the sequel, we consistently choose the Algebraic--Algebraic case to present our results on equal footing. We could have chosen the Algebraic--Gaussian instead, the results are hard to distinguish. A presently not possible (see above) direct comparison theory--data of the really existing ensemble of empirical correlation matrices $C_{\mathrm{ep}}$ in the epochs could help determine which model provides a better fit to the data.
The fit parameters tend to be larger for a larger return horizon~$\Delta t$, see App.~\ref{app:FitParametersCapital_Single}.
 This observation is consistent with Ref.~\cite{Schmitt_2013} where the Gaussian-Gaussian case was discussed for the fit parameter $N$ which increases for a larger return horizon $\Delta t$.
 An exception is $L_\mathrm{rot}$ for the Algebraic-Algebraic case which decreases for a larger return horizon.
 
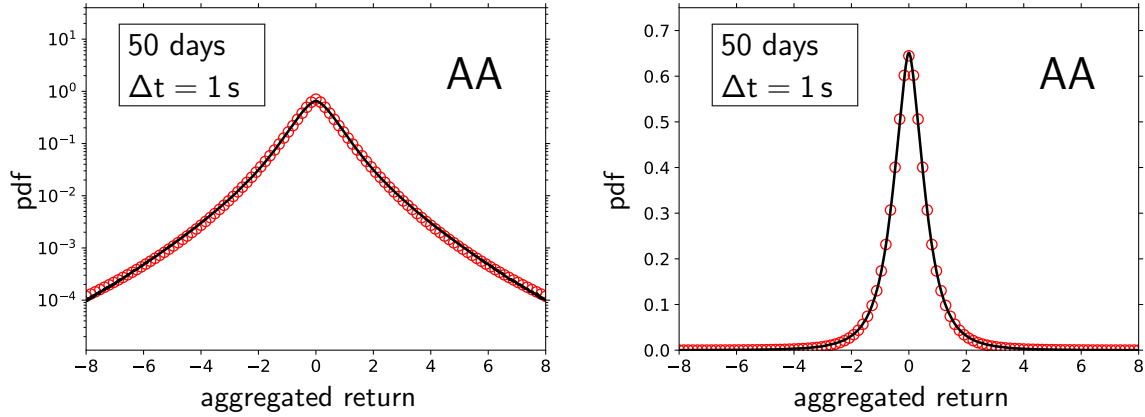
\begin{figure}[htbp]
	\captionsetup[subfigure]{labelformat=empty}
	\centering
	\begin{minipage}{.5\textwidth}
		\centering
		\subfloat[]{\begin{overpic}[width=1.\linewidth]{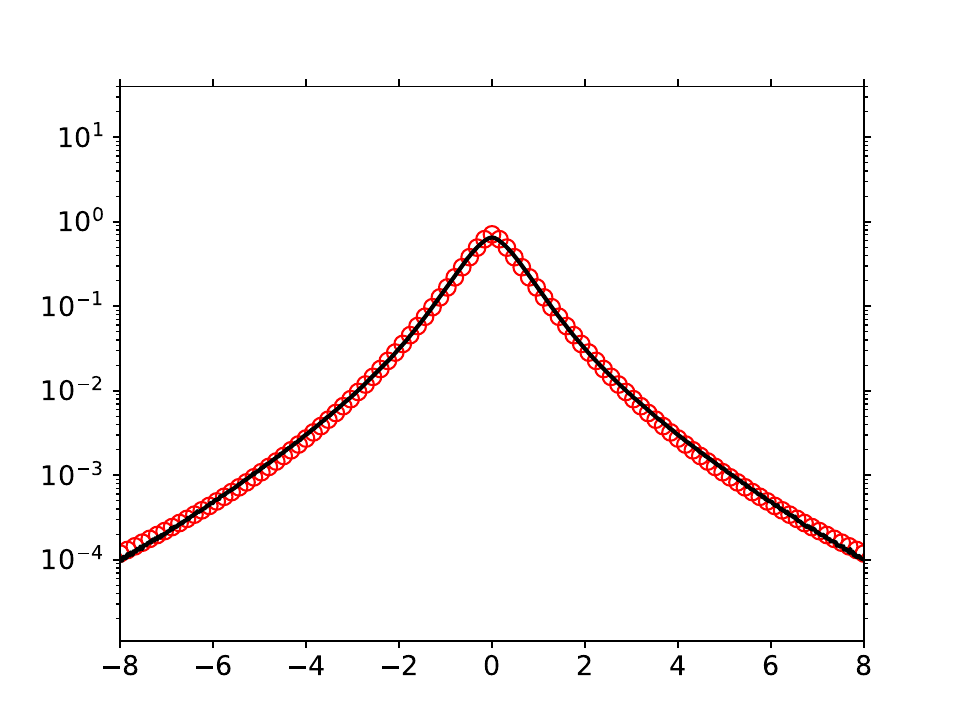}
				\put(18,55){\noindent\fbox{\parbox{1.7cm}{\sffamily 50 days\\$\mathsf{\Delta t = 1\,s}$}}}
				\put(50,0){\makebox(0,0){\small\sffamily aggregated return}}
				\put(2,35){\makebox(0,0){\rotatebox{90}{\small\sffamily pdf}}}
				\put(78,55){\makebox(0,0){\sffamily\Large AA}}
			\end{overpic}
		}
	\end{minipage}%
\begin{minipage}{.5\textwidth}
	\centering
	\subfloat[]{\begin{overpic}[width=1.\linewidth]{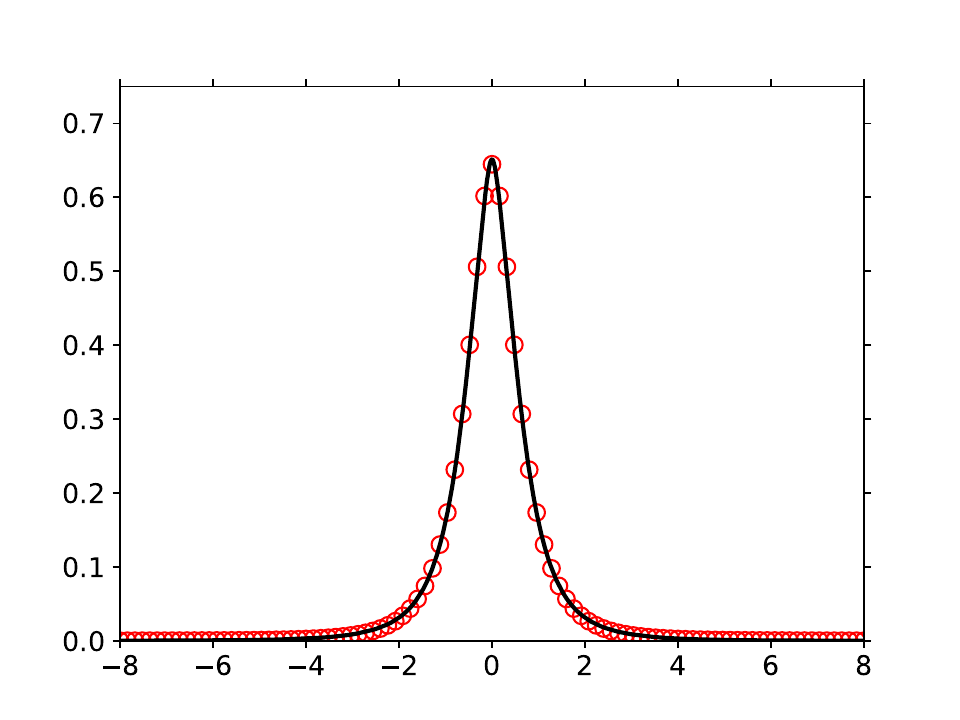}
			\put(18,55){\noindent\fbox{\parbox{1.7cm}{\sffamily 50 days\\$\mathsf{\Delta t = 1\,s}$}}}
			\put(50,0){\makebox(0,0){\small\sffamily aggregated return}}
			\put(2,35){\makebox(0,0){\rotatebox{90}{\small\sffamily pdf}}}
			\put(78,55){\makebox(0,0){\sffamily\Large AA}}
		\end{overpic}
	}
\end{minipage}
	\caption{Empirical distributions of aggregated returns with $\Delta t = 1\,\mathrm{s}$ (black) for the 1st long interval (50 trading days) with $\langle p \rangle_{\mathrm{AA}}^{(\mathrm{aggr})} (\widetilde{r})$ (red circles), 
	left: for $L_{\mathrm{rot}}= 99.607$, $N = 3.123$ on a logarithmic scale, right: for $L_{\mathrm{rot}}= 100.334$, $N = 6.051 $ on a linear scale.}
	\label{fig:50TD_22200DataPoints_AggretRetDist_Int_Log}
\end{figure}

\begin{figure}[htbp]
	\captionsetup[subfigure]{labelformat=empty}
	\centering
	\begin{minipage}{.5\textwidth}
		\centering
		\subfloat[]{\begin{overpic}[width=1.\linewidth]{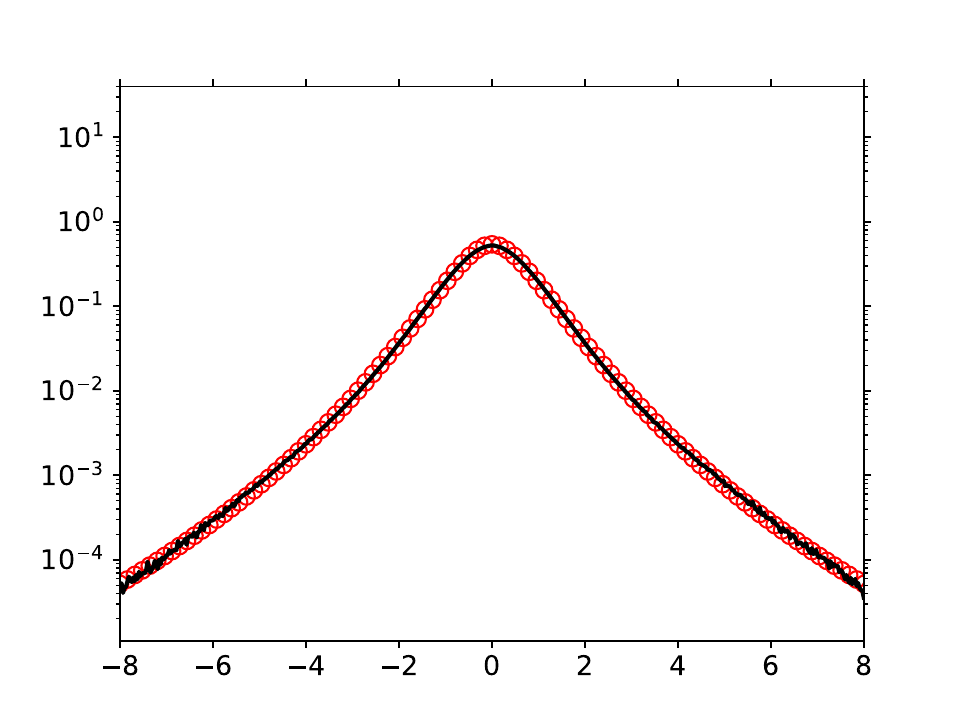}
				\put(18,55){\noindent\fbox{\parbox{1.7cm}{\sffamily 50 days\\$\mathsf{\Delta t = 10\,s}$}}}
				\put(50,0){\makebox(0,0){\small\sffamily aggregated return}}
				\put(2,35){\makebox(0,0){\rotatebox{90}{\small\sffamily pdf}}}
				\put(78,55){\makebox(0,0){\sffamily\Large AA}}
			\end{overpic}
		}
	\end{minipage}%
	\begin{minipage}{.5\textwidth}
	\centering
	\subfloat[]{\begin{overpic}[width=1.0\linewidth]{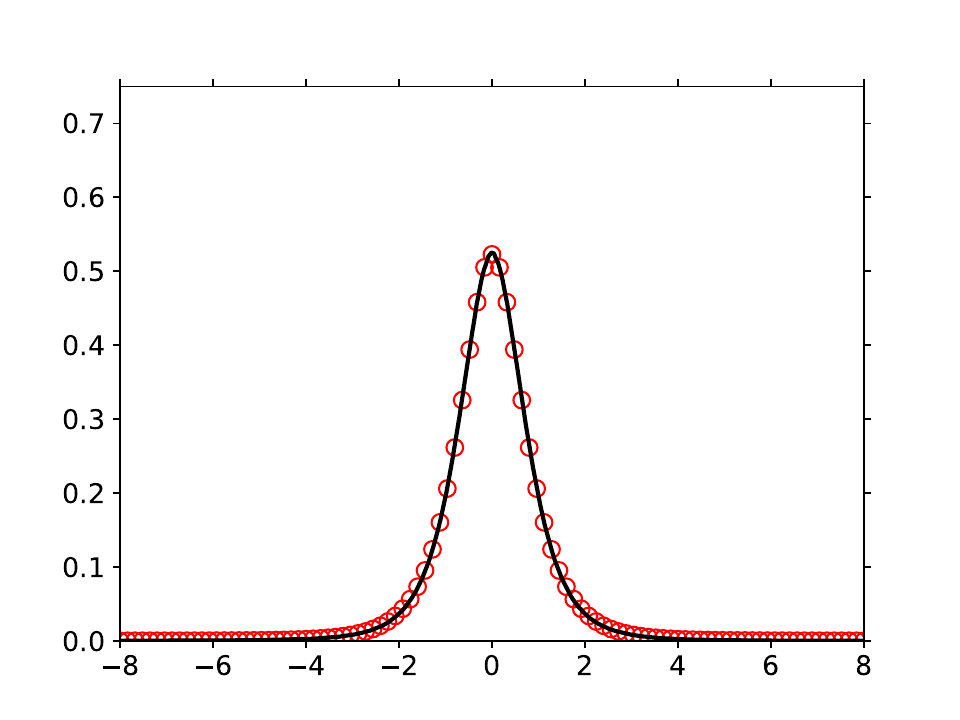}
			\put(18,55){\noindent\fbox{\parbox{1.7cm}{\sffamily 50 days\\$\mathsf{\Delta t = 10\,s}$}}}
			\put(50,0){\makebox(0,0){\small\sffamily aggregated return}}
			\put(2,35){\makebox(0,0){\rotatebox{90}{\small\sffamily pdf}}}
			\put(78,55){\makebox(0,0){\sffamily\Large AA}}
		\end{overpic}
	}
\end{minipage}
	\caption{Empirical distributions of aggregated returns with $\Delta t = 10\,\mathrm{s}$ (black) for the 2nd long interval (50 trading days) with $\langle p \rangle_{\mathrm{AA}}^{(\mathrm{aggr})} (\widetilde{r})$ (red circles), 
		left: for $L_{\mathrm{rot}}= 11.407$, $N = 10.768$ on a logarithmic scale, right: for $L_{\mathrm{rot}}= 14.449$, $N = 17.661$ on a linear scale.}
	\label{fig:50TD_2220DataPoints_AggretRetDist_Int_Log}
\end{figure}

 Analogous to the fits on the long interval of 25 trading days, we also study the empirical distributions of the aggregated returns on a long interval of 50 trading days, as displayed Figs.~\ref{fig:50TD_22200DataPoints_AggretRetDist_Int_Log}--\ref{fig:50TD_2220DataPoints_AggretRetDist_Int_Log}. Qualitatively, the results for this interval length are similar to those for 25 trading days. By averaging all values for a specified fit scale, a fixed return horizon $\Delta t$ and a fixed interval length, we notice a trend. A larger long interval results in a smaller fit parameters for $N$ and $L_{\mathrm{rot}}$.
However, the smaller $N$ and $L_{\mathrm{rot}}$ are, the stronger are the fluctuations of correlations and the heavier are the tails.
The fit parameters are listed in App.~\ref{app:FittingParameters}.
 
 \subsection{\label{sec:ComparingShapesDdistributionsEpochsLongInterval}Comparing the Shapes of the Distributions on the Epochs and on the Long Interval}

\begin{figure}[htbp]
	\centering {\begin{overpic}[width=.85\linewidth]{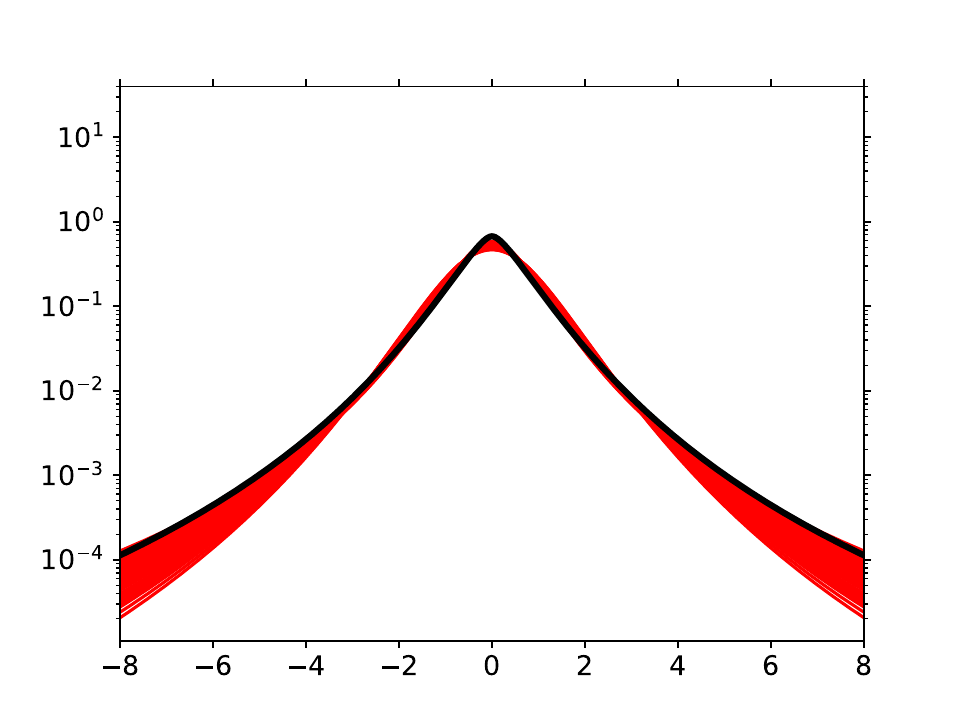}
			\put(50,0){\makebox(0,0){\large\sffamily aggregated return}}
			\put(2,35){\makebox(0,0){\rotatebox{90}{\large\sffamily pdf}}}
			\put(75,55){\makebox(0,0)}
			\put(18,55){\noindent\fbox{\parbox{2.0cm}{\Large$\mathsf{\Delta t = 1\,s}$}}}
			\put(78,55){\makebox(0,0){\sffamily\huge AA}}
		\end{overpic}
	}
	\caption{Model distribution $p_\mathrm{AA}^{(\mathrm{aggr})}(\widetilde{r})$ of the aggregated returns on long interval 1 (black) for 25 trading days and all 250 model distributions $\langle p \rangle_\mathrm{A}^{(\mathrm{aggr})}(\widetilde{r})$ on the epochs (red) with $\Delta t = 1\,\mathrm{s}$, determined with parameters from logarithmic fit.}
	\label{fig:AggRet_DistInt1_Epoch}
\end{figure}
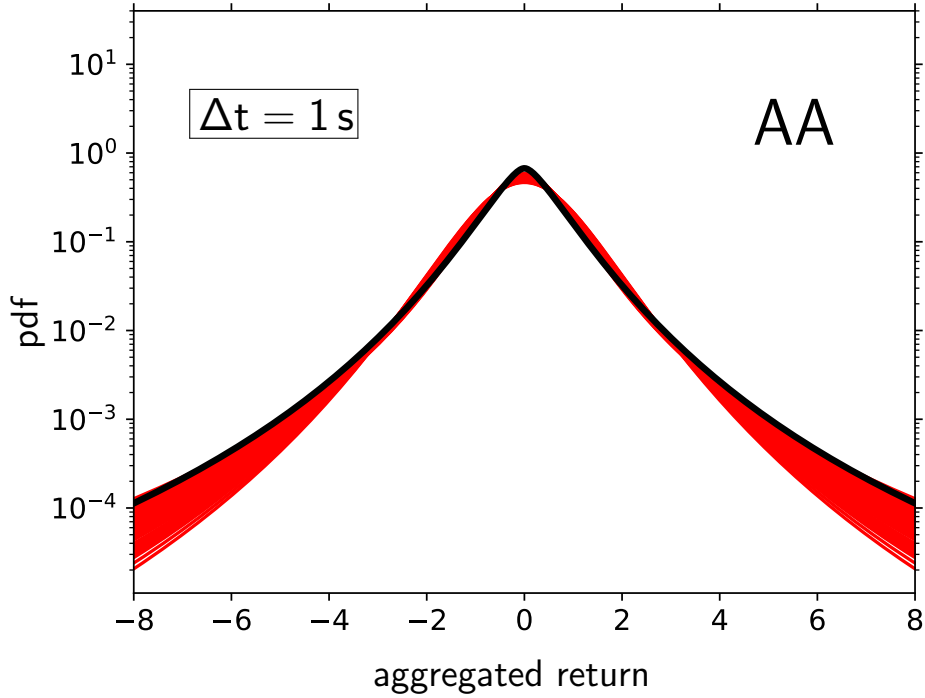

We want to demonstrate that distributions of the aggregated returns on long intervals are heavier--tailed than on the epochs.
To this end, we overlay in Fig.~\ref{fig:AggRet_DistInt1_Epoch} the 250 model distributions $p_\mathrm{A}^{(\mathrm{aggr})}(\widetilde{r})$ calculated with the fit parameters for each epoch
with the model distribution $\langle p\rangle_\mathrm{AA}^{(\mathrm{aggr})}(\widetilde{r})$ from the first long interval for 25 trading days with $\Delta t = 1\,\mathrm{s}$.
Indeed, the latter one is heavier-tailed than almost all 250 model distributions on the epochs.

\begin{figure}[htbp]
	\captionsetup[subfigure]{labelformat=empty}
	\centering
	\begin{minipage}{.5\textwidth}
		\centering
		\subfloat[]{\begin{overpic}[width=1.\linewidth]{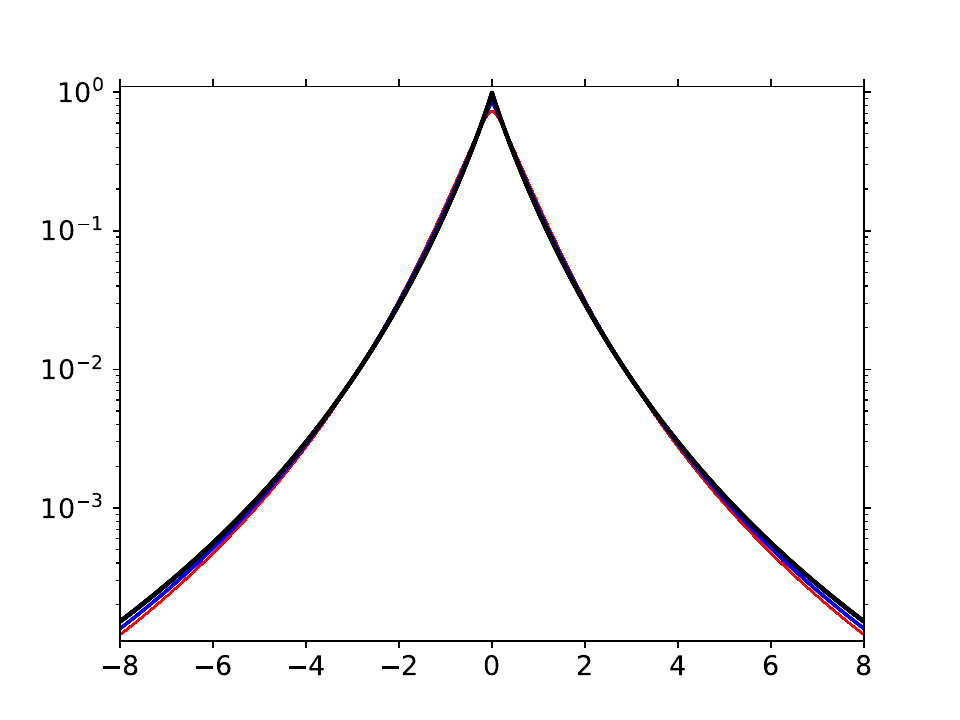}
				\put(18,55){\noindent\fbox{\parbox{1.6cm}{$\mathsf{\Delta t = 1\,\mathrm{s}}$}}}
				\put(50,0){\makebox(0,0){\small\sffamily aggregated return}}
				\put(2,35){\makebox(0,0){\rotatebox{90}{\small\sffamily pdf}}}
				\put(78,55){\makebox(0,0){\sffamily\Large AA}}
			\end{overpic}
		}
	\end{minipage}%
	\begin{minipage}{.5\textwidth}
		\centering
		\subfloat[]{\begin{overpic}[width=1.\linewidth]{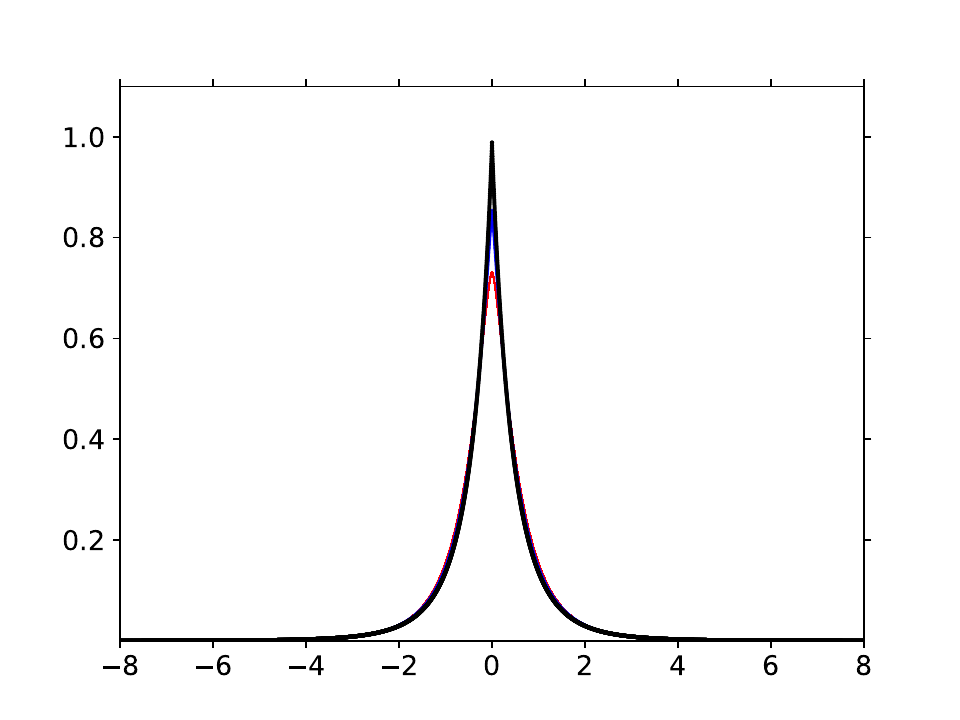}
				\put(18,55){\noindent\fbox{\parbox{1.6cm}{$\mathsf{\Delta t = 1\,\mathrm{s}}$}}}
				\put(50,0){\makebox(0,0){\small\sffamily aggregated return}}
				\put(2,35){\makebox(0,0){\rotatebox{90}{\small\sffamily pdf}}}
				\put(78,55){\makebox(0,0){\sffamily\Large AA}}
			\end{overpic}
		}
	\end{minipage}
	\begin{minipage}{.5\textwidth}
		\centering
		\subfloat[]{\begin{overpic}[width=1.\linewidth]{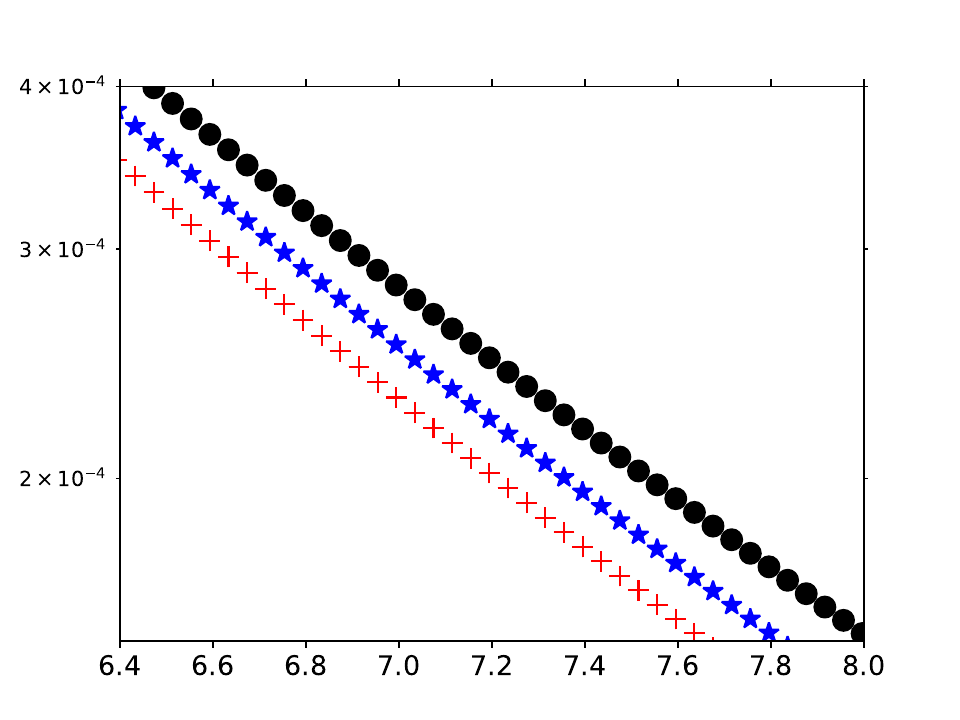}
				\put(16,27){\noindent\fbox{\parbox{1.6cm}{$\mathsf{\Delta t = 1\,\mathrm{s}}$}}}
				\put(50,0){\makebox(0,0){\small\sffamily aggregated return}}
				\put(2,35){\makebox(0,0){\rotatebox{90}{\small\sffamily pdf}}}
				\put(78,55){\makebox(0,0){\sffamily\Large AA}}
			\end{overpic}
		}
	\end{minipage}%
	\begin{minipage}{.5\textwidth}
		\centering
		\subfloat[]{\begin{overpic}[width=1.\linewidth]{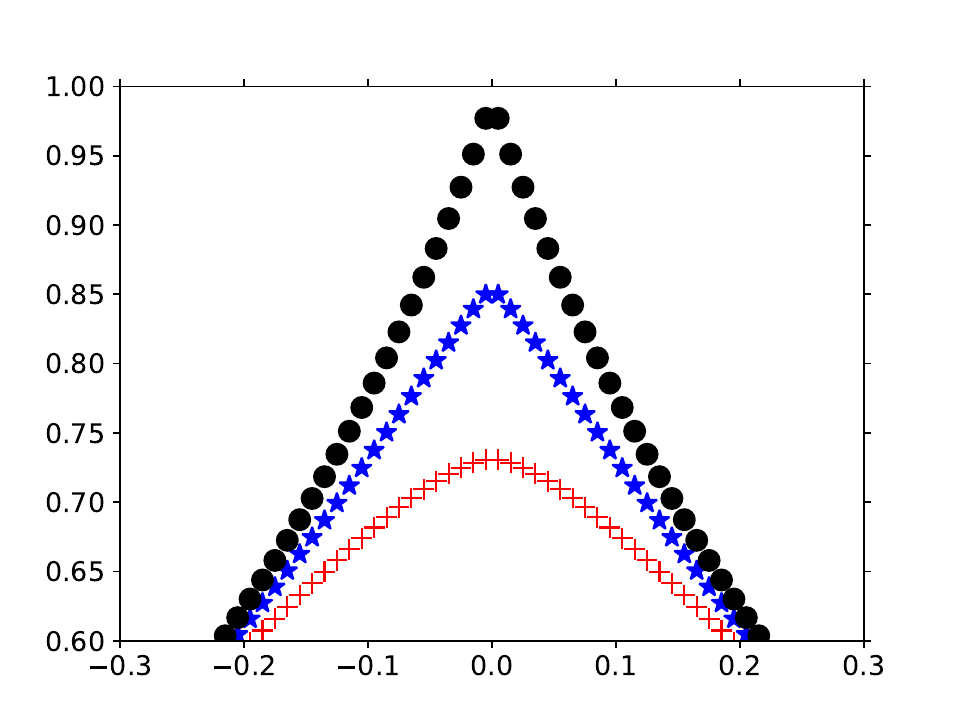}
				\put(18,55){\noindent\fbox{\parbox{1.6cm}{$\mathsf{\Delta t = 1\,\mathrm{s}}$}}}
				\put(50,0){\makebox(0,0){\small\sffamily aggregated return}}
				\put(2,35){\makebox(0,0){\rotatebox{90}{\small\sffamily pdf}}}
				\put(78,55){\makebox(0,0){\sffamily\Large AA}}
			\end{overpic}
		}
	\end{minipage}
	\caption{Fitted distributions $p_\mathrm{AA}^{(\mathrm{aggr})}(\widetilde{r})$ of the aggregated returns on long intervals of 50 trading days from August 7 2014 to October 16, 2014 (black, circles) and of 25 trading days from August 7, 2014 to September 11, 2014 (red, +) and September 12, 2014 to October 16, 2014 (blue, asterisk) with $\Delta t = 1\,\mathrm{s}$, determined with parameters from logarithmic fit, on a logarithmic scale (left, top and bottom), on a linear scale (right, top and bottom).}
	\label{fig:AggRet_DistInt1_25_50_TradingDays}
\end{figure}

Similarly, we compare the model distributions for the Algebraic--Algebraic case on long intervals of 25 and 50 trading days in Fig.~\ref{fig:AggRet_DistInt1_25_50_TradingDays}.
We notice that the model distributions on 50 trading day intervals are heavier--tailed.
When going from the one--day epochs to the long interval of 25 days, the differences in the distributions are larger as there is a factor of 25 between the lengths of the considered intervals. Here, there is only a factor of two.
These results strongly corroborate our model assumption that the fluctuations of the correlation matrices make the tails heavier the longer the considered interval.

\subsection{\label{sec:TailBehaviorComparison}Tail Behavior of Distributions for the Original Returns and Aggregated Returns}

In the literature, the tail behavior of the original returns was studied in great detail, in particular for very large returns~\cite{Gopikrishnan1998,Plerou_1999_Dist}.
The distribution of the original returns show for smaller values a power law with Lévy exponent of about 2 that changes for very large returns to a value of about 3.
This is also referred to as ``inverse cubic law''.
It was suggested that this is caused by the investment strategies of large mutual funds~\cite{Gabaix2003}.

\begin{figure}[htbp]
	\captionsetup[subfigure]{labelformat=empty}
	\centering
	\subfloat[]{
		\begin{overpic}[width=0.5\textwidth]{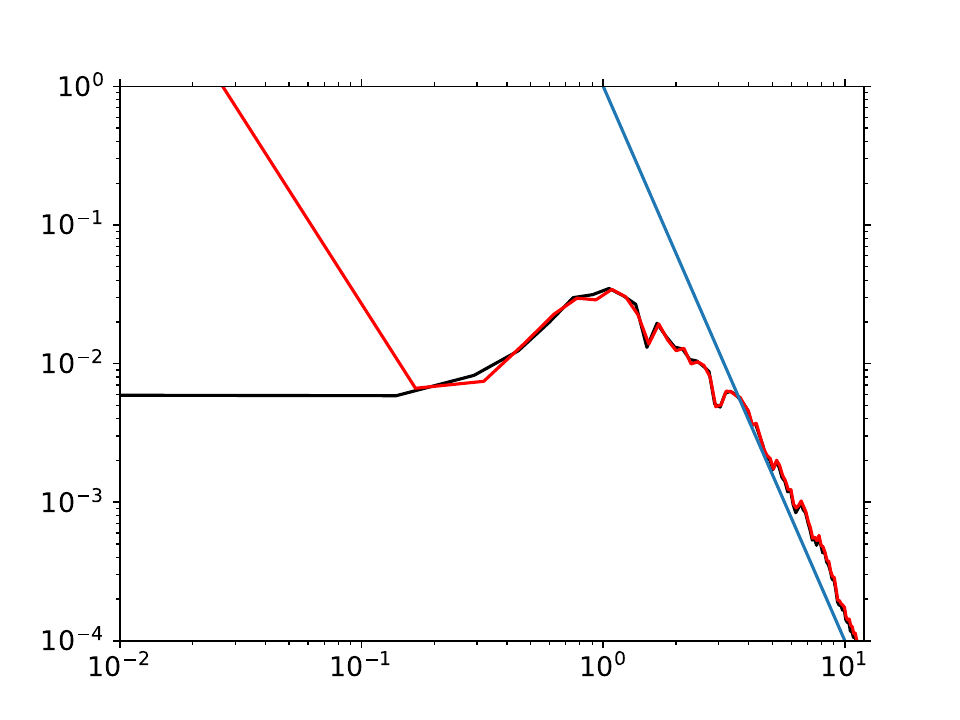}
			\put(50,0){\makebox(0,0){\sffamily original return}}
			\put(0,35){\makebox(0,0){\rotatebox{90}{\sffamily pdf}}}
			\put(75,55){\makebox(0,0)}
		\end{overpic}
	\label{fig:OrigRet_25TradDaysInt1}}
	\subfloat[]{
		\begin{overpic}[width=0.5\textwidth]{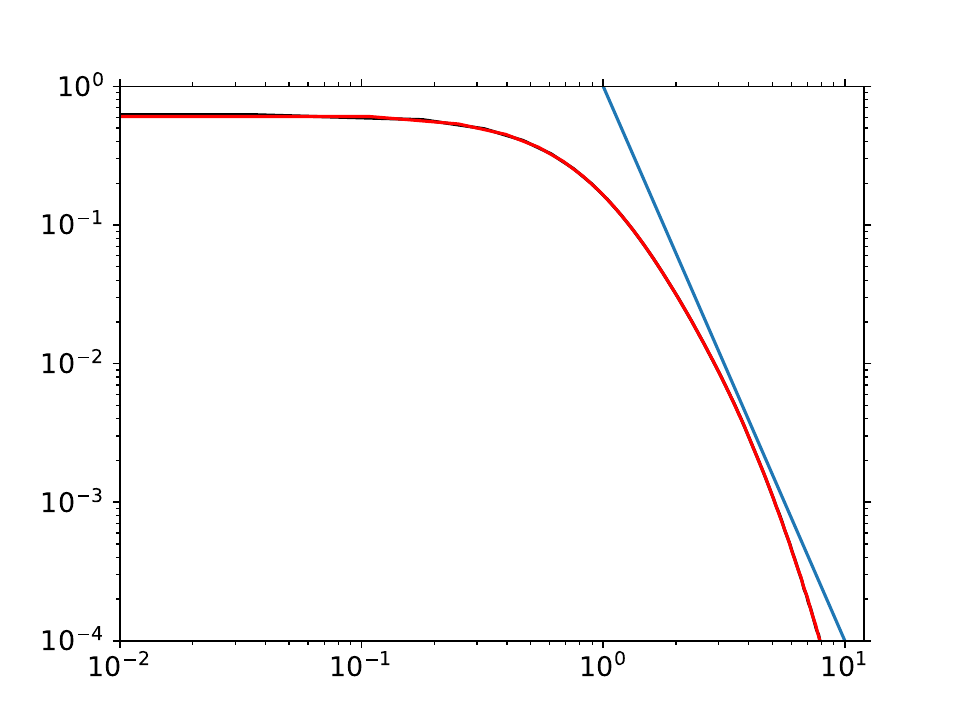}
			\put(50,0){\makebox(0,0){\sffamily aggregated return}}
			\put(0,35){\makebox(0,0){\rotatebox{90}{\sffamily pdf}}}
			\put(75,55){\makebox(0,0)}
		\end{overpic}
	\label{fig:AggregRet_25TradDaysInt1}}
	\caption{\label{fig:TailBehavDoubleLog}Distributions of original returns (left) and aggregated returns (right) for the first 25 interval in 2014 with $\Delta t = 1\,\mathrm{s}$ compared to $1/|x|^4$ behavior (blue). Positive values of the distributions are displayed in black and negative ones are highlighted in red.}
\end{figure}

In Fig.~\ref{fig:TailBehavDoubleLog}, we show for the first 25 trading day interval in 2014 the original returns which are lumped together for all 479 stocks after each individual return time series was normalized.
For the aggregated returns
the Lévy exponents is slightly larger than 3.
We notice that the tail behavior of the original returns carries over to the distributions of the aggregated returns.

\section{\label{sec:Caveats}Caveats}

A large--scale empirical analysis as carried out here requires
special care, as the data set is divided into
epochs. In Sec.~\ref{sec:ExtremeHeavyTails} we show, how an improper
calculation of returns can produce artifacts, in particular extremely
heavy tails. We demonstrate in Sec.~\ref{sec:EpochsandNumber} that the
normalization with a limited number of data points in the epochs
leads to results that need careful interpretation.

\subsection{\label{sec:ExtremeHeavyTails}Occurrence of Extremely Heavy Tails}

\begin{figure}[htbp]
	\captionsetup[subfigure]{labelformat=empty}
	\centering
	\subfloat[]{
	\begin{overpic}[width=0.5\textwidth]{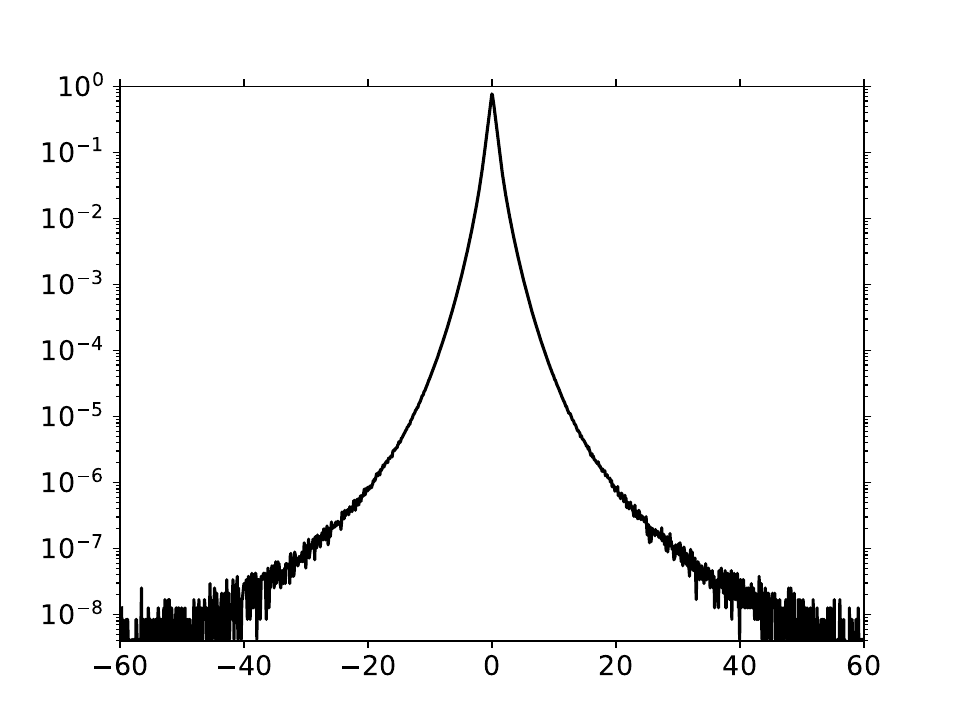}
	\put(50,0){\makebox(0,0){\sffamily aggregated return}}
	\put(0,35){\makebox(0,0){\rotatebox{90}{\sffamily pdf}}}
	\put(75,55){\makebox(0,0)}
	\end{overpic}
	}
	\subfloat[]{
		\begin{overpic}[width=0.5\textwidth]{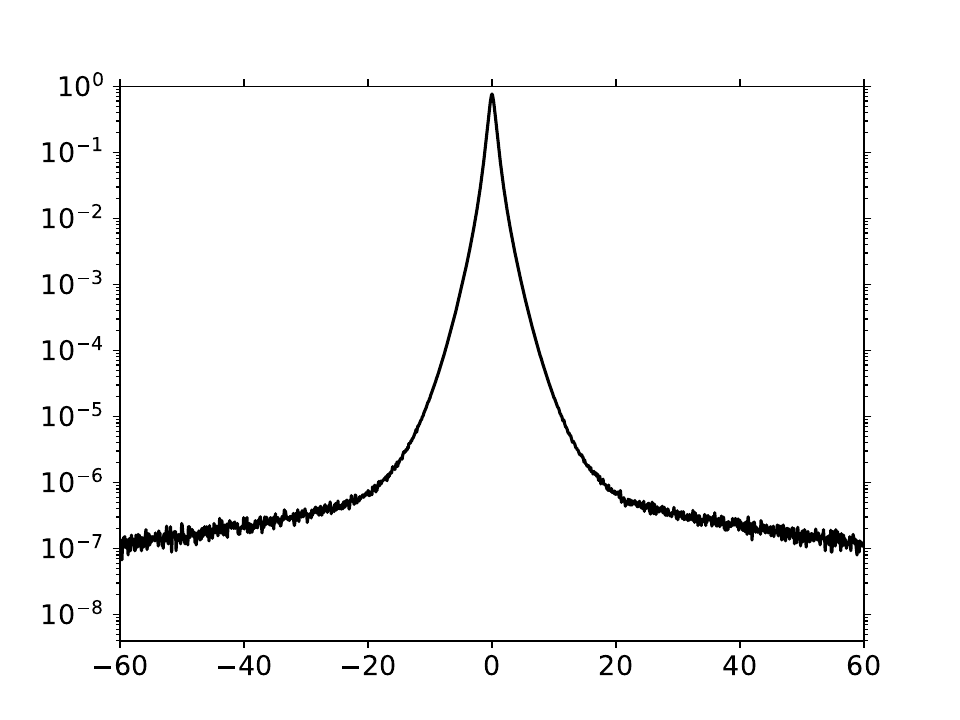}
			\put(50,0){\makebox(0,0){\sffamily aggregated return}}
			\put(0,35){\makebox(0,0){\rotatebox{90}{\sffamily pdf}}}
			\put(75,55){\makebox(0,0)}
		\end{overpic}
	}
	\caption{\label{fig:DistYearly2014_WithWithoutOvernight}Distributions of aggregated returns for 2014 without overnight returns (left) and with overnight returns (right).}
\end{figure}

In the analysis of Sec.~\ref{sec:Results}, we work out the
corresponding multivariate return distributions in the epochs and on
the long interval. We now demonstrate how strongly the tail behavior
depends on a consistent empirical analysis of the returns. For the
epochs and long intervals, we only use intraday data, there are no overnight
returns. What happens if we include overnight returns? In
Fig.~\ref{fig:DistYearly2014_WithWithoutOvernight} we display the
distributions of the aggregated returns for the whole year 2014 excluding and
including the overnight returns.

\begin{figure}[htbp]
	\captionsetup[subfigure]{labelformat=empty}
	\centering
	\subfloat[]{
	\begin{overpic}[width=0.5\textwidth]{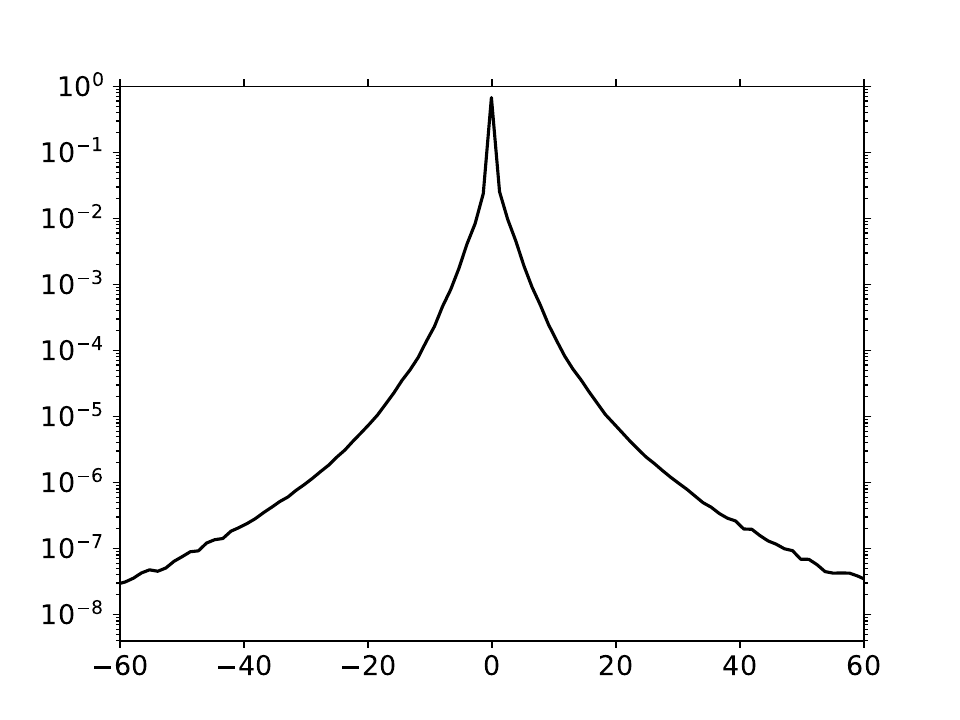}
		\put(50,0){\makebox(0,0){\sffamily original return}}
		\put(0,35){\makebox(0,0){\rotatebox{90}{\sffamily pdf}}}
		\put(75,55){\makebox(0,0)}
	\end{overpic}
	}
	\subfloat[]{
	\begin{overpic}[width=0.5\textwidth]{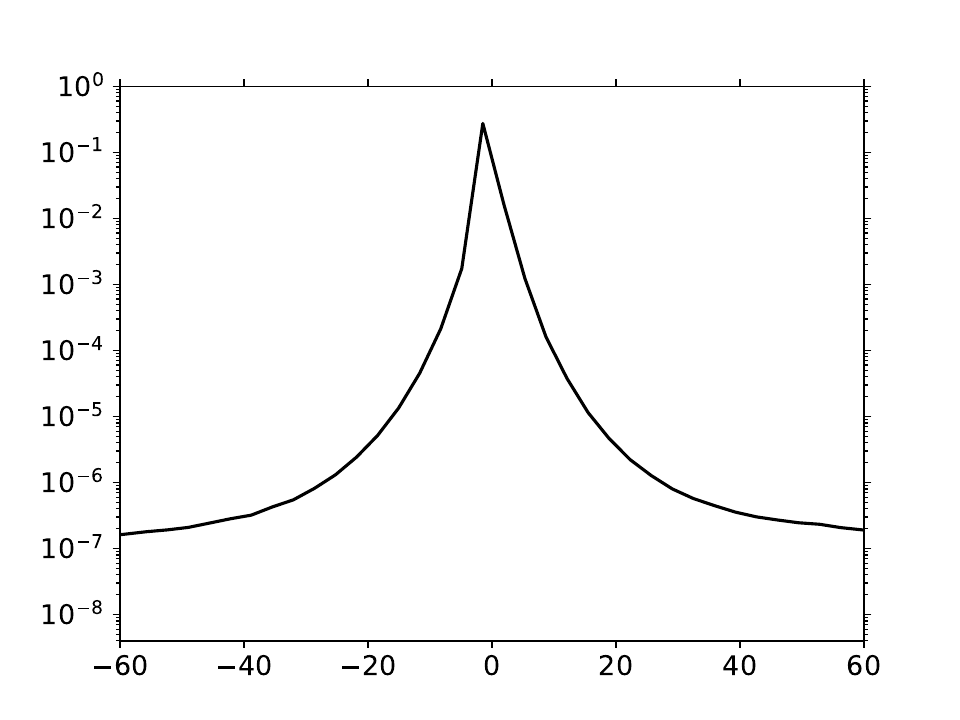}
		\put(50,0){\makebox(0,0){\sffamily original return}}
		\put(0,35){\makebox(0,0){\rotatebox{90}{\sffamily pdf}}}
		\put(75,55){\makebox(0,0)}
	\end{overpic}
	}
	\caption{\label{fig:DistNORMORIGINALRET_Yearly2014_WithWithoutOvernight}Distributions of normalized, original returns for 2014 without overnight returns (left) and with overnight returns (right), return horizon $\Delta t = 1\,\mathrm{s}$.}
\end{figure}

Obviously, the inclusion produces extremely
heavy tails. To understand them, we show in Fig.~\ref{fig:DistNORMORIGINALRET_Yearly2014_WithWithoutOvernight} the distribution of the normalized, original returns lumping together all returns for all 479 stocks, once more including and excluding overnight returns.
The distribution with overnight returns has extremely heavy tails as well.
Overnight returns tend to be larger than intraday returns
because the return horizon $\Delta t$ is much larger for overnight than for intraday returns.
Thus, two different statistics of returns are mixed together leading to extremely heavy tails.
The tail behavior in the distribution of the normalized, original returns is carried over to the distributions of the aggregated returns.
This is why the exclusion of overnight returns is advised in our analysis. To avoid misunderstandings, we emphasize that the long intervals we consider consist of the concatenated epochs, there are no overnight return either.

\subsection{\label{sec:EpochsandNumber}{Epochs and the Normalization With a Limited Number of Data Points}}

The normalization of time series in epochs with a limited number of data points can strongly
influence the tails of a distribution, cf.~Sec.~\ref{sec:EpochsIntervals}. Here we use the daily data set with $K = 308$ stocks and $T=5220$ data points, corresponding to the total length of the daily price time series, see~Sec.~\ref{sec:DataSet}.
After calculating the time series of the logarithmic returns
we divide them into epochs.
In the sequel we do not work out the normalized, original return distributions of the individual stocks as in Sec.~\ref{sec:AggEmp}.
Here, we lump together (or aggregate) the normalized, original returns for all stocks and all epochs and work out the overall univariate distribution of the normalized, original returns.

Using the aggregation method, we also determine the univariate distribution of the
aggregated returns lumping together the aggregated returns for all
epochs with a specified number of data points, see~Sec.~\ref{sec:AggEmp}. However, we must be careful if the
number of data points is smaller than that of the stocks $T < K$. The correlation
matrices then do not have full rank. To circumvent this problem, we use an
approach introduced in Ref.~\cite{Schmitt_2013} where correlation
matrices of dimension $2\times2$ were calculated.  Since the $2 \times
2$ correlation matrices have full rank, we can apply the
aggregation method to the pairs of all stocks and lump together these
aggregated returns for all epochs with a given number of data points.  

\begin{figure}[htbp]
	\captionsetup[subfigure]{labelformat=empty}
	\centering
	\begin{minipage}{.5\textwidth}
		\centering
		\subfloat[]{\begin{overpic}[width=1.\linewidth]{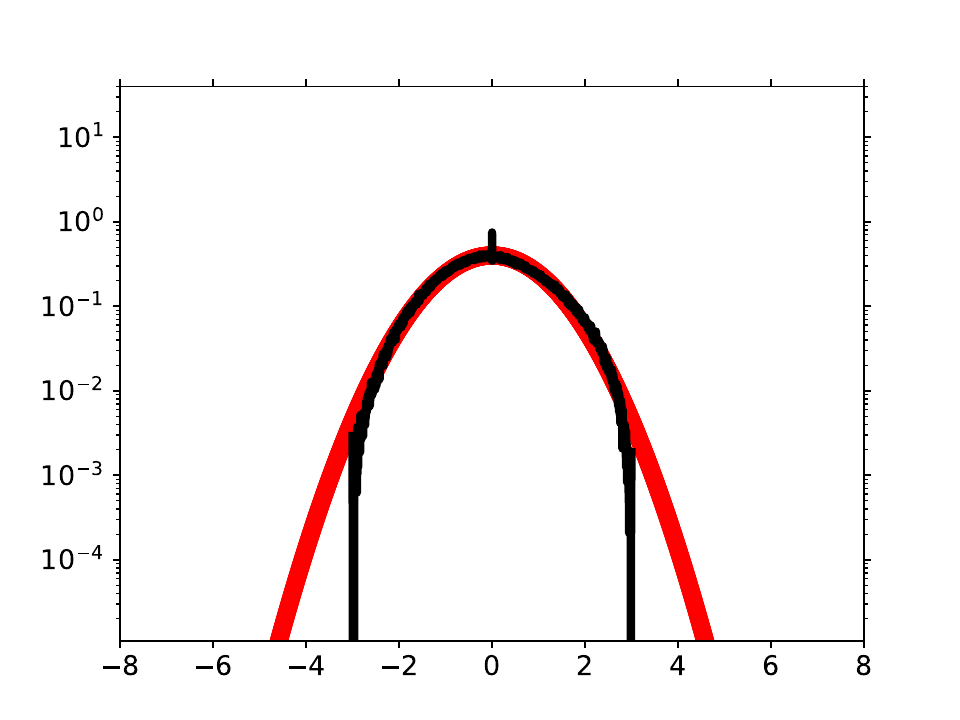}
				\put(50,0){\makebox(0,0){\small\sffamily original return}}
				\put(1,35){\makebox(0,0){\rotatebox{90}{\small\sffamily pdf}}}
			\end{overpic}
		}
	\end{minipage}%
	\begin{minipage}{.5\textwidth}
		\centering
		\subfloat[]{\begin{overpic}[width=1.\linewidth]{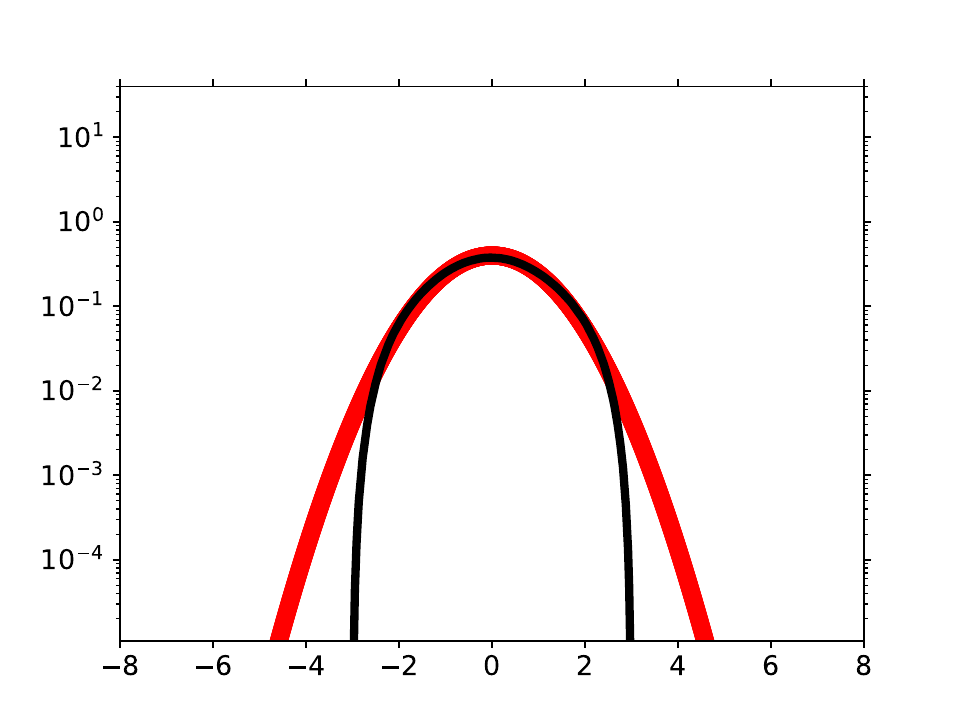}
				\put(50,0){\makebox(0,0){\small\sffamily aggregated return}}
				\put(1,35){\makebox(0,0){\rotatebox{90}{\small\sffamily pdf}}}
			\end{overpic}
		}
	\end{minipage}\\
	\begin{minipage}{.5\textwidth}
		\centering
		\subfloat[]{\begin{overpic}[width=1.\linewidth]{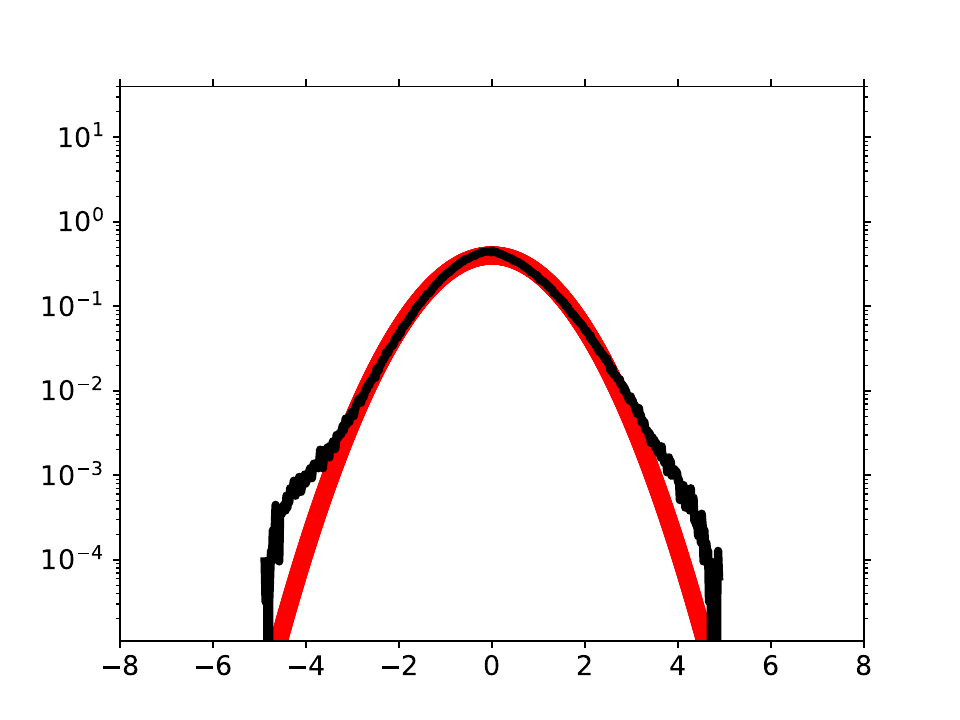}
				\put(50,0){\makebox(0,0){\small\sffamily original return}}
				\put(1,35){\makebox(0,0){\rotatebox{90}{\small\sffamily pdf}}}
			\end{overpic}
		}
	\end{minipage}%
	\begin{minipage}{.5\textwidth}
		\centering
		\subfloat[\label{subfig:Schmitt_Reproduced}]{\begin{overpic}[width=1.\linewidth]{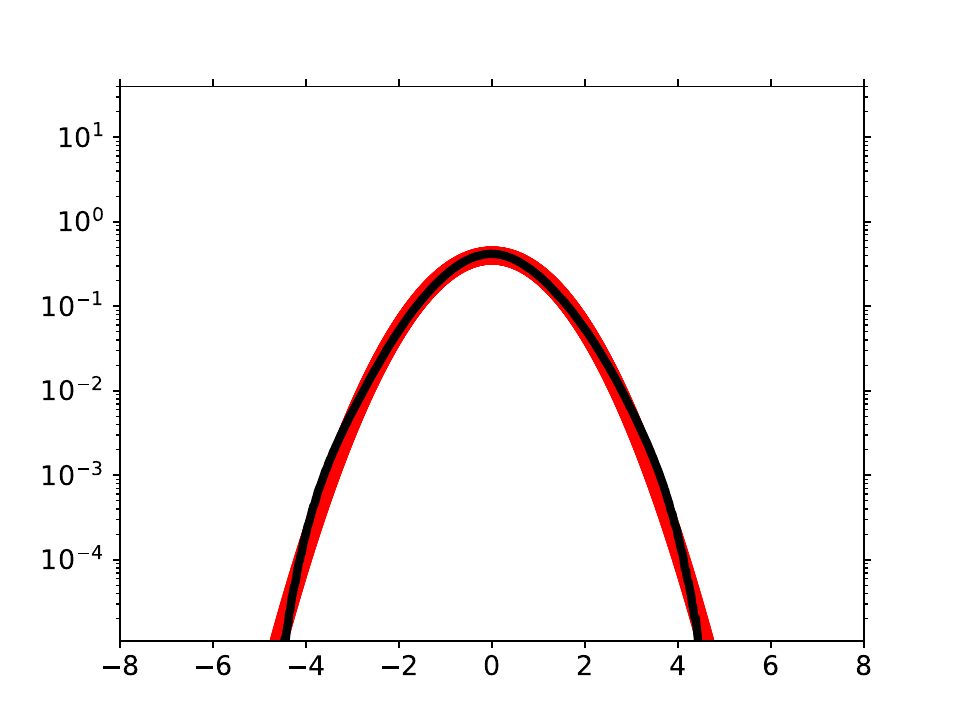}
				\put(50,0){\makebox(0,0){\small\sffamily aggregated return}}
				\put(1,35){\makebox(0,0){\rotatebox{90}{\small\sffamily pdf}}}
			\end{overpic}
		}
	\end{minipage}\\
	\begin{minipage}{.5\textwidth}
		\centering
		\subfloat[]{\begin{overpic}[width=1.\linewidth]{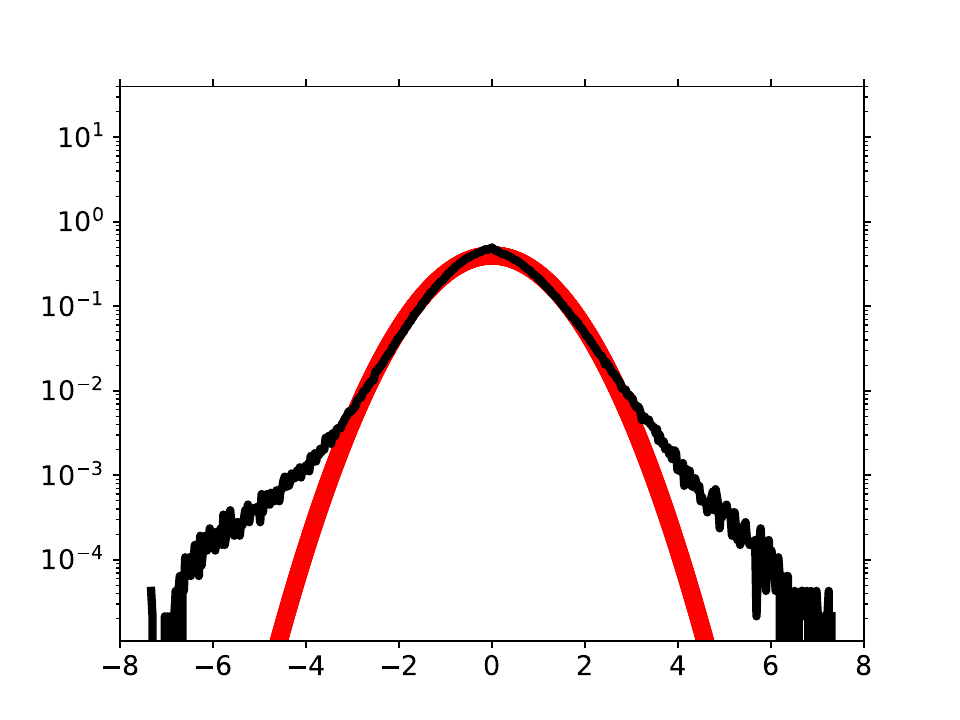}
				\put(50,0){\makebox(0,0){\small\sffamily original return}}
				\put(1,35){\makebox(0,0){\rotatebox{90}{\small\sffamily pdf}}}
			\end{overpic}
		}
	\end{minipage}%
	\begin{minipage}{.5\textwidth}
		\centering
		\subfloat[]{\begin{overpic}[width=1.\linewidth]{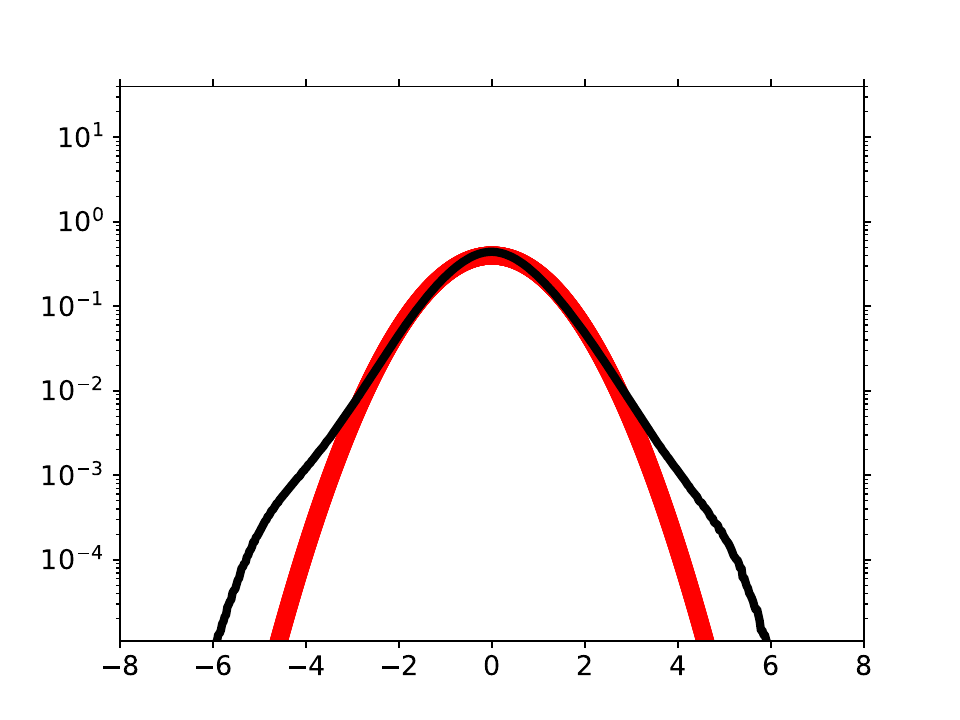}
				\put(50,0){\makebox(0,0){\small\sffamily aggregated return}}
				\put(1,35){\makebox(0,0){\rotatebox{90}{\small\sffamily pdf}}}
			\end{overpic}
		}
	\end{minipage}%
	\caption{\label{fig:DataPoints_DistNormalizedOriginalAggregated}Univariate distribution of the original returns for 10 (top, left), 25 (middle, left) and 55 data points (bottom, left) per epoch and univariate distributions of the rotated and aggregated returns for 10 (top, right), 25 (middle, right) and 55 data points (bottom, right) per epoch. Normal distribution is shown in red color. Calculated from daily data set with $\Delta t$ = 1 trading day, see~Sec.~\ref{sec:DataSet}.}
\end{figure}

In contrast to our choice of $T= 22200$ and $T =2220$ data points per epoch in our main analysis in Sec.~\ref{sec:Results}, we now
choose $T = 10$, $25$ and $55$ data points \textit{i.e.} we deal with epochs with a rather small number of data points as studied in Ref.~\cite{HenanoLondono_2021}. Corresponding univariate distributions of the normalized, original returns and aggregated ones are depicted in
Figs.~\ref{fig:DataPoints_DistNormalizedOriginalAggregated}.
In comparison to a Gaussian distribution, we see that for increasing epoch lengths both
type of distributions change their platykurtic behavior to a
leptokurtic one.  For a fixed number of data points, distributions of normalized, original returns and aggregated
returns look very similar.  Thus, the normalization itself
strongly affects the tail behavior of the distributions of the aggregated returns.

To gain a better understanding of this effect, we discuss the normalization procedure itself.
We work out the mean value and standard deviation of the return time series  over rather short epochs.
However, for such a small number of data points, mean values and standard deviations for different epochs do strongly deviate which has to be distinguished from the non--stationarity present in financial time series. It is only caused by the small number of data points.
The normalization leads to a broadening in the center of the distribution, \textit{i.\,e.} a higher probability density near the center.
Consequently, the probability density in the tails must decrease. 
The more data points per epoch, the smaller the influence on the tails.
This explains the change from a platykurtic to a leptokurtic behavior of the tails of the distributions.

\begin{figure}[htbp]
	\captionsetup[subfigure]{labelformat=empty}
	\begin{minipage}{.5\textwidth}
		\subfloat[]{\begin{overpic}[width=1.\linewidth]{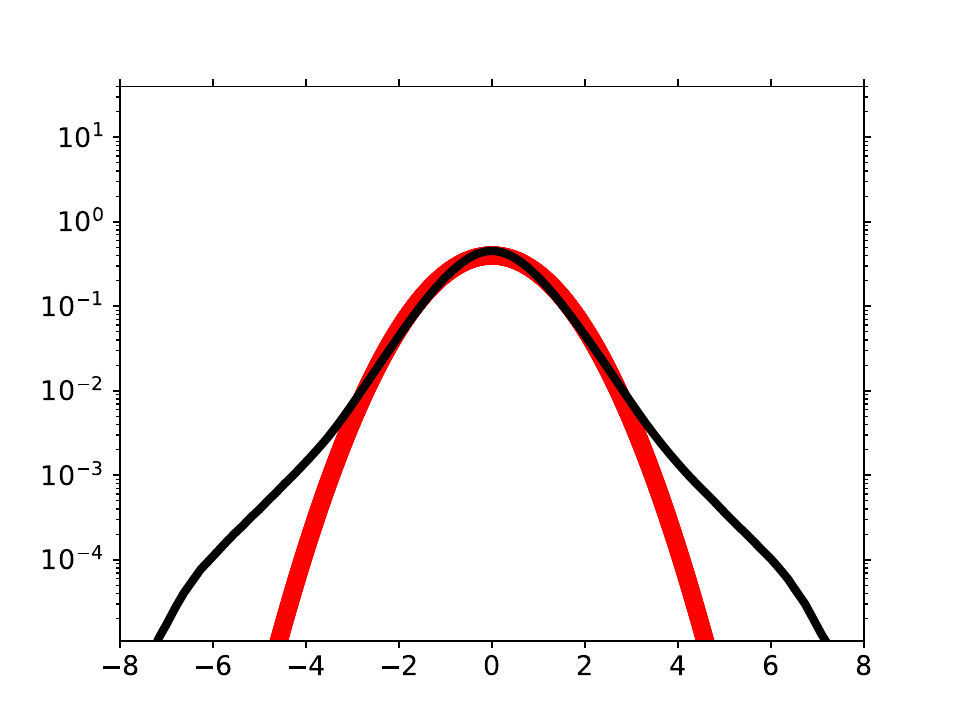}
				\put(50,0){\makebox(0,0){\small\sffamily aggregated return}}
				\put(1,35){\makebox(0,0){\rotatebox{90}{\small\sffamily pdf}}}
			\end{overpic}
		}
	\end{minipage}%
	\hfill
	\begin{minipage}{.5\textwidth}
		\subfloat[]{\begin{overpic}[width=1.\linewidth]{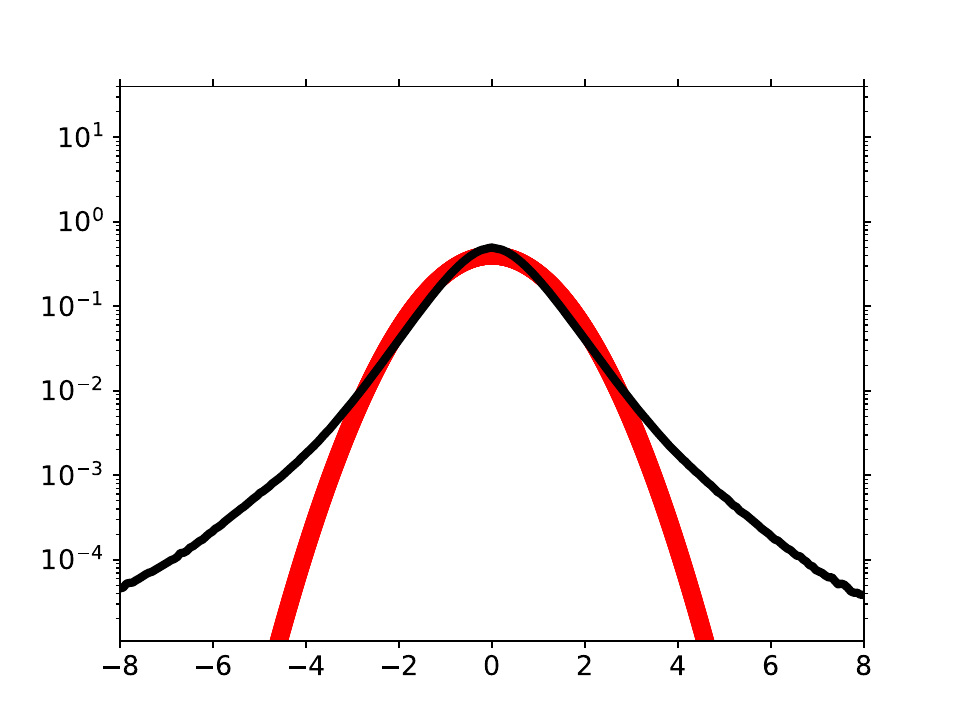}
				\put(50,0){\makebox(0,0){\small\sffamily aggregated return}}
				\put(1,35){\makebox(0,0){\rotatebox{90}{\small\sffamily pdf}}}
			\end{overpic}
		}
	\end{minipage}
	\begin{minipage}{.5\textwidth}
		\subfloat[]{\begin{overpic}[width=1.\linewidth]{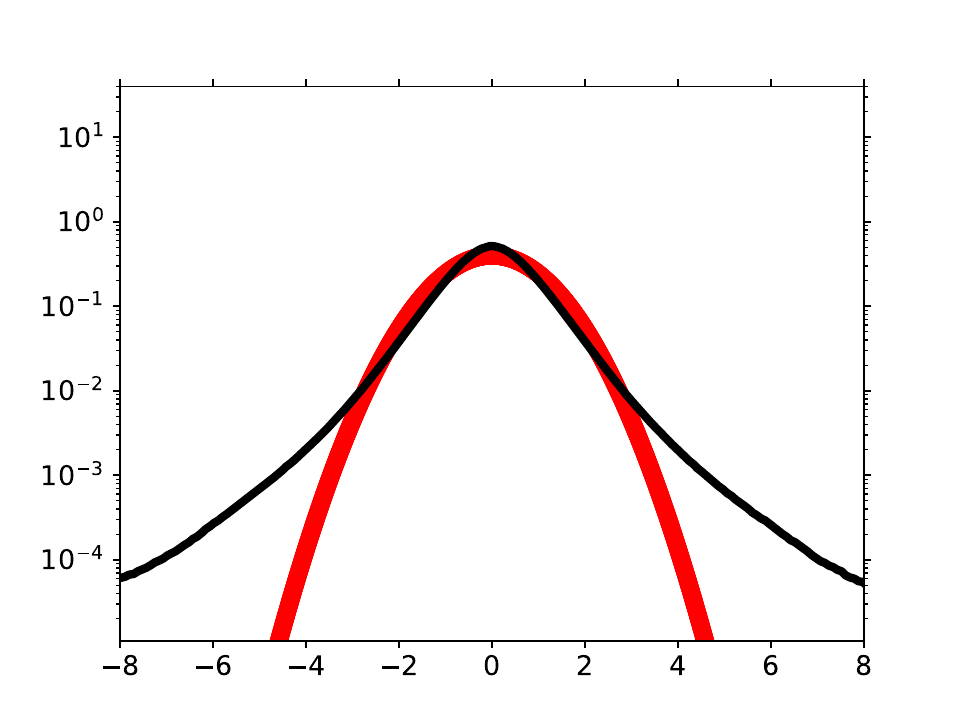}
				\put(50,0){\makebox(0,0){\small\sffamily aggregated return}}
				\put(1,35){\makebox(0,0){\rotatebox{90}{\small\sffamily pdf}}}
			\end{overpic}
		}
	\end{minipage}%
	\hfill
	\begin{minipage}{.5\textwidth}
		\subfloat[]{\begin{overpic}[width=1.\linewidth]{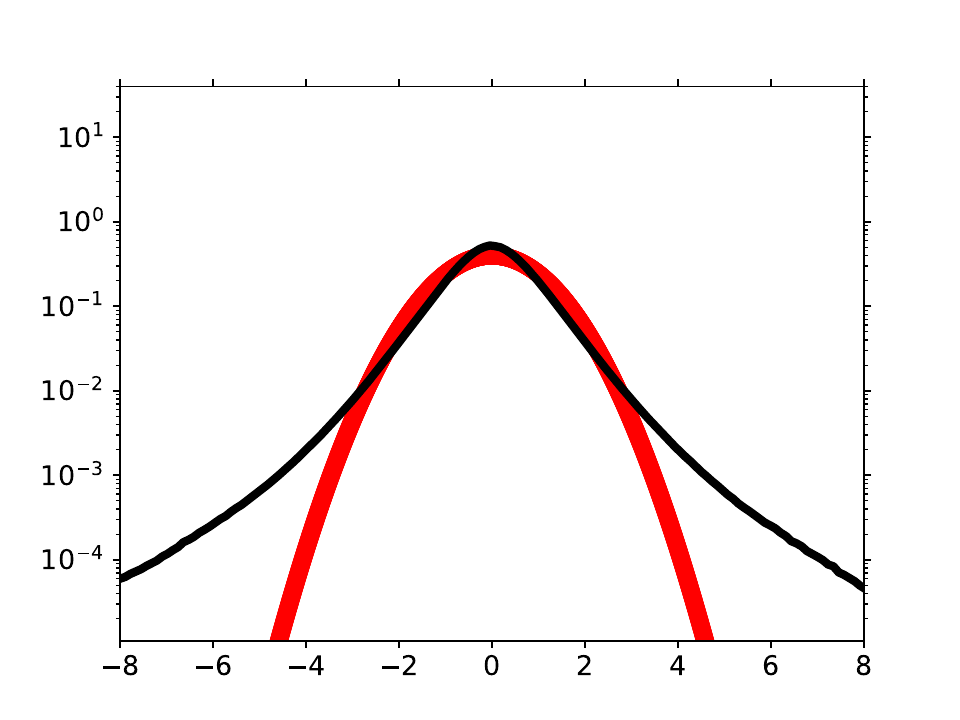}
				\put(50,0){\makebox(0,0){\small\sffamily aggregated return}}
				\put(1,35){\makebox(0,0){\rotatebox{90}{\small\sffamily pdf}}}
			\end{overpic}
		}
	\end{minipage}
	\caption{Aggregated return distribution for 100 (top, left), 500 (top, right), 1000 (bottom, left) and 2000 data points (bottom, right) per epoch. Normal distribution is shown in red color. Calculated from daily data set with $\Delta t$ = 1 trading day, see~Sec.~\ref{sec:DataSet}.}
	\label{fig:AggregReturnDist_DifferentDataPoints}
\end{figure}

As stated above, we use $T= 22200$ and $T =2220$ data points for our main analysis. To demonstrate that this choice for the number of data points has a negligible effect on all distributions of our main analysis, we also work out the return distributions of the aggregated returns in Fig.~\ref{fig:AggregReturnDist_DifferentDataPoints} for $T = 100, 500, 1000$ and 2000 data points. For 100 data points the tails are still suppressed.
For 500 data points, the distribution of the aggregated returns appears to be free from the suppression in the tails and more influenced by the non-stationarity in the correlation matrices.
Differences in the tail behavior for 1000 and 2000 data points are almost not discernible.

\begin{figure}[htbp]
	\centering 
	\includegraphics[width=.80\linewidth]{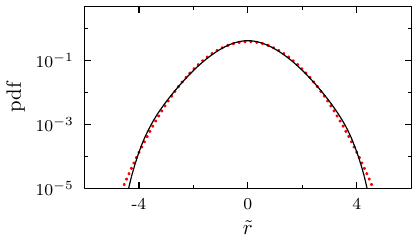}
	\caption{Distribution of the aggregated returns, here
		denoted by $\widetilde{r}$, for fixed covariances from S\&P~500
		daily data set, $\Delta t$ = 1 trading day and 25 data points per epoch. The circles show a normal distribution. Taken from Ref.~\cite{Schmitt_2013}.}
	\label{fig:Schmitt_fig4.pdf}
\end{figure}

In Ref.~\cite{Schmitt_2013} it was argued that the epoch distribution  of the aggregated returns shown in Fig.~\ref{fig:Schmitt_fig4.pdf} and reproduced in Fig~\ref{fig:DataPoints_DistNormalizedOriginalAggregated} indicated stationarity. However, as just shown this tail behavior can be traced back to an artifact due to the small number of data points. For $T=25$, the univariate distribution of the aggregated returns happens to be a Gaussian-like distribution. In this context, we recall that stationarity is not a necessary prerequisite for the model construction as outlined in I.

\begin{figure}[htbp]
	\centering 
	\includegraphics[width=.80\linewidth]{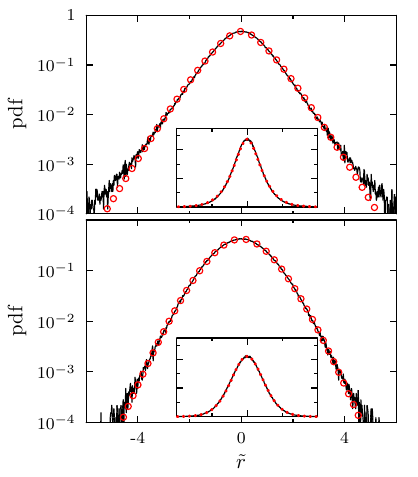}
	\caption{Empirical distributions of the aggregated returns (black) for $\Delta t =1$ (top) and $\Delta t =20$ (bottom) from S\&P~500
		daily data set.
	Model distribution $\langle p \rangle_{\mathrm{GG}}^{(\mathrm{aggr})} (\widetilde{r})$ is indicated by red circles. Taken from Ref.~\cite{Schmitt_2013}.}
	\label{fig:Schmitt_fig6.pdf}
\end{figure}

Furthermore, in Ref.~\cite{Schmitt_2013}, the empirical, univariate distributions of the aggregated returns were compared to the model distribution $\langle p \rangle_{\mathrm{GG}}^{(\mathrm{aggr})} (\widetilde{r})$ on a long interval from 1992 to 2012, see~Fig.~\ref{fig:Schmitt_fig6.pdf}. 
A daily data set was used, very similar to our daily data set, \textit{i.e.} the study was performed for a data set with several thousand data points. Hence, the analysis for the long interval from 1992 to 2012 does not suffer from problematic estimation of mean value and standard deviation for the long intervals. In contrast to $\langle p \rangle_{\mathrm{AG}}^{(\mathrm{aggr})} (\widetilde{r})$ and $\langle p \rangle_{\mathrm{AA}}^{(\mathrm{aggr})} (\widetilde{r})$ no parameter estimated from the epoch distribution of the aggregated returns goes into $\langle p \rangle_{\mathrm{GG}}^{(\mathrm{aggr})} (\widetilde{r})$.  Although $\langle p \rangle_{\mathrm{GG}}^{(\mathrm{aggr})} (\widetilde{r})$ performs poorly for intraday data with a resolution of 22200 and 2220 data points as shown in Sec.~\ref{sec:Results},  
$\langle p \rangle_{\mathrm{GG}}^{(\mathrm{aggr})} (\widetilde{r})$ agrees very well with the empirical distribution of the aggregated returns using daily data.
Hence, even though the data on the epochs is problematic in view of the above discussion, the data analysis and model comparison on the long interval in Ref.~\cite{Schmitt_2013} remains valid without restrictions.


\section{\label{sec:Conclusion}Conclusions}

We empirically analyzed multivariate return distributions of the US stock markets
in the year 2014. Strong correlations are present which fluctuate in time because the company performances, the business relations
and the traders' market expectations change. To assess this non--stationarity 
and to quantitatively describe the multivariate distributions, we apply a random matrix model recently put forward and presented
with formulae for the data analyses in I.

We carry out the empirical analysis on (short) epochs and long intervals.
To this end, we rotate the data into the eigenbasis of the correlation matrix.
As opposed to the univariate distributions of the original, unrotated returns, the univariate distributions of the rotated returns 
depend on the eigenvalues of the correlation matrix and thus contain the full information on the correlated system.
To accumulate statistics we also resort to aggregation.

We emphasize that the rotation into the eigenbasis facilitates the data analysis, but does not restrict our results to this basis! Once all parameters are determined, we have a multivariate distribution which may be further used in any arbitrary basis: in the original one of the individual stocks, in the eigenbasis of the correlation matrix or any other basis defined by an arbitrary rotation of the original returns.

We find heavy tails which are very well described by an algebraic distributions on the epochs and the Algebraic--Algebraic model on the long interval.
Thus, the distributions on the epochs are characterized by only one fit parameter. 
Having that fixed, the distributions on the long interval depend on just two fit parameters which can readily be determined.

Importantly, the distributions on the epochs on the one hand and on the long interval on the other hand
differ in the empirical analysis and their functional forms of the model are different.
The tails on the long interval become heavier.
Moreover, comparing two long intervals demonstrates that the fluctuations of the correlation further accumulated.
This clearly confirms the model of I.
The non--stationary fluctuations of the correlations lift the tails.

\section*{Acknowledgment}

We thank Henrik M. Bette and Shanshan Wang for fruitful discussions. We are particularly grateful to Holger Kantz for helpful remarks on the epoch distributions.
We thank Thilo Schmitt for providing the daily data set.
		
\clearpage
\markboth{References}{References}

\newcommand{\newblock}{}

\printbibliography[heading=none]											   

\clearpage

\begin{appendices}
\markboth{Appendix}{Appendix}

\section{\label{app:TickerSymbols}Ticker Symbols}

\subsection{\label{app:ListTickerSymbols}List of Ticker Symbols for Intraday Data set from NYSE}

A, AA, AAPL, ABBV, ABC, ABT, ACE, ACN, ACT, ADBE, ADI, ADM, ADP, ADS, ADSK, ADT, AEE, AEP, AES, AET, AFL, AGN, AIG, AIV, AIZ, AKAM, ALL, ALLE, ALTR, ALXN, AMAT, AME, AMGN, AMP, AMT, AMZN, AN, AON, APA, APC, APD, APH, ARG, ATI, AVB, AVP, AVY, AXP, AZO, BA, BAC, BAX, BBBY, BBT, BBY, BCR, BDX, BEN, BHI, BIIB, BK, BLK, BLL, BMY, BRCM, BSX, BWA, BXP, C, CA, CAG, CAH, CAM, CAT, CB, CBG, CBS, CCE, CCI, CCL, CELG, CERN, CF, CFN, CHK, CHRW, CI, CINF, CL, CLX, CMA, CMCSA, CME, CMG, CMI, CMS, CNP, CNX, COF, COG, COH, COL, COP, COST, COV, CPB, CRM, CSC, CSCO, CSX, CTAS, CTL, CTSH, CTXS, CVC, CVS, CVX, D, DAL, DD, DE, DFS, DG, DGX, DHI, DHR, DIS, DISCA, DLPH, DLTR, DNB, DNR, DO, DOV, DOW, DPS, DRI, DTE, DTV, DUK, DVA, DVN, EA, EBAY, ECL, ED, EFX, EIX, EL, EMC, EMN, EMR, EOG, EQR, EQT, ESRX, ESV, ETFC, ETN, ETR, EW, EXC, EXPD, EXPE, F, FAST, FB, FCX, FDO, FDX, FE, FFIV, FIS, FISV, FITB, FLIR, FLR, FLS, FMC, FOSL, FOXA, FSLR, FTI, FTR, GAS, GCI, GD, GE, GGP, GILD, GIS, GLW, GM, GME, GNW, GOOG, GPC, GPS, GRMN, GS, GT, GWW, HAL, HAR, HAS, HBAN, HCBK, HCN, HCP, HD, HES, HIG, HOG, HON, HOT, HP, HPQ, HRB, HRL, HRS, HSP, HST, HSY, HUM, IBM, ICE, IFF, INTC, INTU, IP, IPG, IR, IRM, ISRG, ITW, IVZ, JCI, JEC, JNJ, JNPR, JOY, JPM, JWN, K, KEY, KIM, KLAC, KMB, KMI, KMX, KO, KR, KRFT, KSS, KSU, L, LB, LEG, LEN, LH, LIFE, LLL, LLTC, LLY, LM, LMT, LNC, LO, LOW, LRCX, LUK, LUV, LYB, M, MA, MAC, MAR, MAS, MAT, MCD, MCHP, MCK, MCO, MDLZ, MDT, MET, MHFI, MHK, MJN, MKC, MMC, MMM, MNST, MO, MON, MOS, MPC, MRK, MRO, MS, MSFT, MSI, MTB, MU, MUR, MWV, MYL, NBR, NDAQ, NE, NEE, NEM, NFLX, NFX, NI, NKE, NLSN, NOC, NOV, NRG, NSC, NTAP, NTRS, NUE, NVDA, NWL, NWSA, OI, OKE, OMC, ORCL, ORLY, OXY, PAYX, PBCT, PBI, PCAR, PCG, PCL, PCLN, PCP, PDCO, PEG, PEP, PETM, PFE, PFG, PG, PGR, PH, PHM, PKI, PLD, PLL, PM, PNC, PNR, PNW, POM, PPG, PPL, PRGO, PRU, PSA, PSX, PVH, PWR, PX, PXD, QCOM, QEP, R, RAI, RDC, REGN, RF, RHI, RHT, RIG, RL, ROK, ROP, ROST, RRC, RSG, RTN, SBUX, SCG, SCHW, SE, SEE, SHW, SIAL, SJM, SLB, SNA, SNDK, SNI, SO, SPG, SPLS, SRCL, SRE, STI, STJ, STT, STX, STZ, SWK, SWN, SWY, SYK, SYMC, SYY, T, TAP, TDC, TE, TEG, TEL, TGT, THC, TIF, TJX, TMK, TMO, TRIP, TROW, TRV, TSN, TSO, TSS, TWC, TWX, TXN, TXT, TYC, UNH, UNM, UNP, UPS, URBN, USB, UTX, V, VAR, VFC, VIAB, VLO, VMC, VNO, VRSN, VRTX, VTR, VZ, WAG, WAT, WDC, WEC, WFC, WFM, WHR, WIN, WM, WMB, WMT, WU, WY, WYN, WYNN, XEL, XL, XLNX, XOM, XRAY, XRX, XYL, YHOO, YUM

\subsection{\label{app:ListTickerSymbols_YahooFinance}List of Ticker Symbols for Daily Data set From Yahoo! Finance}

AA, AAPL, ABT, ADBE, ADI, ADM, ADP, ADSK, AEP, AES, AET, AFL, AGN, AIG, ALTR, AMAT, AMD, AMGN, AON, APA, APC, APD, APH, ARG, AVP, AVY, AXP, AZO, BA, BAC, BAX, BBT, BBY, BCR, BDX, BEN, BF.B, BHI, BIG, BIIB, BK, BLL, BMC, BMS, BMY, C, CA, CAG, CAH, CAT, CB, CCE, CCL, CELG, CERN, CI, CINF, CL, CLF, CLX, CMA, CMCSA, CMI, CMS, CNP, COG, COP, COST, CPB, CSC, CSCO, CSX, CTAS, CTL, CVH, CVS, CVX, D, DD, DE, DELL, DHR, DIS, DNB, DOV, DOW, DTE, DUK, ECL, ED, EFX, EIX, EMC, EMR, EOG, EQT, ETN, ETR, EXC, F, FAST, FDO, FDX, FHN, FISV, FITB, FLS, FMC, FRX, GAS, GCI, GD, GE, GIS, GLW, GPC, GPS, GT, GWW, HAL, HAS, HBAN, HCP, HD, HES, HNZ, HOG, HON, HOT, HP, HPQ, HRB, HRL, HRS, HST, HSY, HUM, IBM, IFF, IGT, INTC, IP, IPG, IR, ITW, JCI, JCP, JEC, JNJ, JPM, JWN, K, KEY, KIM, KLAC, KMB, KO, KR, L, LEG, LEN, LH, LLTC, LLY, LM, LMT, LNC, LOW, LSI, LTD, LUK, LUV, MAS, MAT, MCD, MDT, MHP, MKC, MMC, MMM, MO, MOLX, MRK, MRO, MSFT, MSI, MTB, MU, MUR, MWV, MYL, NBL, NBR, NE, NEE, NEM, NI, NKE, NOC, NSC, NTRS, NU, NUE, NWL, OI, OKE, OMC, ORCL, OXY, PAYX, PBCT, PBI, PCAR, PCG, PCL, PCP, PEP, PFE, PG, PGR, PH, PHM, PLL, PNC, PNW, POM, PPG, PPL, PSA, QCOM, R, RDC, RF, ROK, ROST, RRD, RSH, RTN, S, SCG, SCHW, SEE, SHW, SIAL, SLB, SLM, SNA, SO, SPLS, STI, STJ, STT, SUN, SVU, SWK, SWN, SWY, SYK, SYMC, SYY, T, TAP, TE, TEG, TER, TGT, THC, TIF, TJX, TLAB, TMK, TMO, TROW, TRV, TSN, TSO, TXN, TXT, TYC, UNH, UNP, USB, UTX, VAR, VFC, VLO, VMC, VNO, VZ, WAG, WDC, WEC, WFC, WFM, WHR, WM, WMB, WMT, WPO, WY, X, XEL, XL, XLNX, XOM, XRAY, XRX, ZION

\section{\label{Tables}Tables for Secs.~\ref{sec:EpochsIntervals}, \ref{sec:EmpiricalEpochDistributions}}

\subsection{\label{app:IntervalsRange}Overview of the Start and End of Long Intervals for 25 Drading Days}
\begin{table}[H]
	\begin{minipage}{1.0\linewidth}
		\centering
		\caption{\label{tab:IntervalsRange}Overview of the start and end of intervals (month-day) with a length of 25 trading days for 2014.}\vspace{0.3cm}
		\begin{tabular}{cc}
			\toprule
			interval 1 & 01-02 to 02-06 \\
			\midrule
			interval 2 & 02-07 to 03-14 \\
			\midrule
			interval 3 & 03-17 to 04-21 \\
			\midrule
			interval 4 & 04-22 to 05-27 \\
			\midrule
			interval 5 & 05-28 to 07-01 \\
			\midrule
			interval 6 & 07-02 to 08-06 \\
			\midrule
			interval 7 & 08-07 to 09-11 \\
			\midrule
			interval 8 & 09-12 to 10-16 \\
			\midrule
			interval 9 & 10-17 to 11-20 \\
			\midrule
			interval 10 & 11-21 to 12-29 \\
			\bottomrule
		\end{tabular}
	\end{minipage}%
\end{table}%

\subsection{\label{app:FitParameterslaggr_Single}Parameters $l_\mathrm{rot}$, $\chi^2_{\mathrm{ln}}$ and $\chi^2_{\mathrm{lin}}$ for Fits on the Epochs}
\begin{table}[H]
	\begin{minipage}{1.0\linewidth}
		\centering
		\caption{\label{tab:FitParameterslaggr_Single}Parameters $l_\mathrm{rot}$, $\chi^2_{\mathrm{ln}}$ and $\chi^2_{\mathrm{lin}}$ determined by logarithmic and linear fit with return horizon $\Delta t $.}
		\vspace{0.3cm}
		\begin{tabular}{cccccc}
			\toprule
			date & fit &  $\Delta t$ & $ l_\mathrm{rot}$  & $\chi^2_{\mathrm{ln}}$  & $ \chi^2_{\mathrm{lin}} $  \\ \hline
			Oct. 23 & log & 1 s  & 2.769  & 0.027  & --- \\
			Oct. 23 & lin & 1 s  & 2.303  &  ---  & $ 1.334 \cdot 10^{-5}$  \\
			Dec. 8 & log & 1 s  & 2.933  & 0.007  & --- \\
			Dec. 8 & lin & 1 s  & 2.744 &  ---  & $1.215 \cdot 10^{-6}$ \\
			Dec. 17 & log &  1 s & 2.679  & 0.004  & --- \\
			Dec. 17 & lin &  1 s & 2.361  &  --- & $1.047 \cdot 10^{-5}$ \\
			Feb. 20 & log & 10 s  & 3.227  & 0.011  & --- \\
			Feb. 20 & lin & 10 s  & 3.079 &  ---  & $ 9.395 \cdot 10^{-6}$ \\
			Jun. 2 & log & 10 s  & 3.264  & 0.011  & --- \\
			Jun. 2 & lin & 10 s  & 2.984   &  ---  & $ 1.239 \cdot 10^{-5}$  \\
			\bottomrule
		\end{tabular}
	\end{minipage}%
\end{table}%

\subsection{\label{app:FitParameterslaggr_Averaged}Averaged Parameters $\langle l_\mathrm{rot} \rangle$, $\langle \chi^2_{\mathrm{ln}} \rangle$ and $\langle \chi^2_{\mathrm{lin}} \rangle$ for Fits on the Epochs}
\begin{table}[H]
	\begin{minipage}{1.0\linewidth}
		\centering
		\caption{\label{tab:FitParameterslaggr_Averaged}Averaged parameters $\langle l_\mathrm{rot} \rangle$, $\langle \chi^2_{\mathrm{ln}} \rangle$ and $\langle \chi^2_{\mathrm{lin}} \rangle$ determined by logarithmic and linear fit with return horizon $\Delta t$.}
		\vspace{0.3cm}
		\begin{tabular}{ccccc}
			\toprule
			fit &  $\Delta t$ & $\langle l_\mathrm{rot} \rangle$  & $\langle \chi^2_{\mathrm{ln}} \rangle$  & $\langle \chi^2_{\mathrm{lin}} \rangle$  \\ \hline
			log &  1 s & 2.601  & 0.024  & --- \\
			lin &  1 s & 2.301  &  --- & $8.549 \cdot 10^{-6} $ \\
			log &  10 s & 3.523  & 0.055  & --- \\
			lin &  10 s & 3.669  & ---  & $5.845 \cdot 10^{-6} $ \\
			\bottomrule
		\end{tabular}
	\end{minipage}%
\end{table}%

\subsection{\label{app:FitParametersCapital_Single}Fitting Parameters for Fits on the Long Interval}
\begin{table}[H]
	\begin{minipage}{1.0\linewidth}
		\centering
		\caption{\label{tab:FitParametersCapital_Single}Fitting parameters for distributions of the aggregated returns on long intervals in trading days (td) and return horizon $\Delta t$ determined by logarithmic and linear fit.}\vspace{0.3cm}
		\begin{tabular}{cccccccccc}
			\toprule
			interval &  fit  & $\Delta t$ & interval &  $\langle p \rangle_{\mathrm{GG}}$  & $\langle p \rangle_{\mathrm{GA}}$  & $\langle p \rangle_{\mathrm{GA}}$  & $\langle p \rangle_{\mathrm{AG}}$ & $\langle p \rangle_{\mathrm{AA}}$  & $\langle p \rangle_{\mathrm{AA}}$ \\
			number & & & length &  $N$  & $L_{\mathrm{rot}}$ & $N$  & $N$ & $L_{\mathrm{rot}}$ & $N$  \\
			\midrule
			interval \hspace{0.2cm}9 & log & 1 s & 25 td & 0.782  & 3.467  & 2.438  & 2.852 & 99.554 & 2.910 \\
			interval \hspace{0.2cm}9 & lin & 1 s & 25 td & 2.036  & 4.419  & 5.058  & 5.926 & 100.346 & 6.085 \\
			interval \hspace{0.2cm}8 & log & 10 s & 25 td & 1.190  & 5.444  & 5.256  & 4.192 & 10.990 & 9.935  \\
			interval \hspace{0.2cm}8 & lin & 10 s & 25 td & 2.995  & 7.429  & 10.090  & 5.542 & 13.328 & 18.328 \\
			interval \hspace{0.2cm}1 & log & 1 s & 50 td & 0.805  & 3.548 & 2.154 & 3.056 & 99.607 & 3.123 \\
			interval \hspace{0.2cm}1 & lin & 1 s & 50 td & 2.031  & 4.294 & 4.710 & 5.893 & 100.334 & 6.051 \\
			interval \hspace{0.2cm}2 & log & 10 s & 50 td & 1.204  & 5.294 & 4.699 & 4.343 & 11.407 & 10.768 \\
			interval \hspace{0.2cm}2 & lin & 10 s & 50 td & 3.271  & 9.493 & 14.110 & 6.807 & 14.449 & 17.661 \\
			\bottomrule
		\end{tabular}
	\end{minipage}%
\end{table}%

\subsection{\label{app:FitParametersCapital_Single_ChiSquared}Values of $\chi^2$ for Fits on the Long Interval}
\begin{table}[H]
	\begin{minipage}{1.0\linewidth}
		\centering
		\caption{\label{tab:FitParametersCapital_Single_ChiSquared}Values of $\chi^2$ for distributions of the aggregated returns on long intervals in trading days (td) and return horizon $\Delta t$ determined by logarithmic and linear fit.}\vspace{0.3cm}
		\begin{tabular}{cccccccc}
			\toprule
			interval &  fit  & $\Delta t$ & interval &  $\langle p \rangle_{\mathrm{GG}}$ & $\langle p \rangle_{\mathrm{GA}}$  & $\langle p \rangle_{\mathrm{AG}}$ & $\langle p \rangle_{\mathrm{AA}}$  \\
			length & & & number &   \multicolumn{4}{c}{$\chi^2_{\mathrm{ln}}$ / $\chi^2_{\mathrm{lin}}$}         \\
			\midrule
			25 td  & log & 1 s & interval \hspace{0.2cm}9 & 0.054  &   0.002 & 0.002  &  0.002  \\
			25 td  & lin & 1 s & interval \hspace{0.2cm}9 & $1.539 \cdot 10^{-4}$  & $5.758 \cdot 10^{-7}$  & $1.609 \cdot 10^{-6}$  &  $1.719 \cdot 10^{-6}$  \\
			25 td  & log & 10 s & interval \hspace{0.2cm}8 & 0.062  & 0.004  & 0.004  &  0.004  \\
			25 td & lin & 10 s & interval \hspace{0.2cm}8 & $ 4.742 \cdot 10^{-5}$  & $ 2.477 \cdot 10^{-7}$  & $ 4.961 \cdot 10^{-6}$  &  $ 2.837 \cdot 10^{-7}$  \\
			50 td  & log & 1 s & interval \hspace{0.2cm}1 & 0.044 & 0.003 & 0.005 &  0.005 \\
			50 td & lin & 1 s & interval \hspace{0.2cm}1  & $1.478 \cdot 10^{-4}$  & $7.913 \cdot 10^{-7}$ & $3.105 \cdot 10^{-6}$ &  $3.287 \cdot 10^{-6}$  \\
			50 td  & log & 10 s & interval \hspace{0.2cm}2 & 0.056  & 0.004 & 0.004 &  0.004  \\
			50 td  & lin & 10 s & interval \hspace{0.2cm}2 &  $ 3.826 \cdot 10^{-5}$  &  $ 2.330 \cdot 10^{-7}$ &  $ 3.477 \cdot 10^{-6}$ &   $ 8.817 \cdot 10^{-7}$  \\
			\bottomrule
		\end{tabular}
	\end{minipage}%
\end{table}%

\subsection{\label{app:FitParametersCapital_Averaged}Averaged Fitting Parameters for Fits on the Long Interval}
\begin{table}[H]
	\begin{minipage}{1.0\linewidth}
		\centering
		\caption{\label{tab:FitParametersCapital_Averaged}Averaged fitting parameters for distributions of the aggregated returns on long intervals in trading days (td) and return horizon $\Delta t$ determined by logarithmic and linear fit.}\vspace{0.3cm}
		\begin{tabular}{ccccccccc}
			\toprule
			fit  & $\Delta t$ & interval &  $\langle p \rangle_{\mathrm{GG}}$  & $\langle p \rangle_{\mathrm{GA}}$  & $\langle p \rangle_{\mathrm{GA}}$  & $\langle p \rangle_{\mathrm{AG}}$ & $\langle p \rangle_{\mathrm{AA}}$  & $\langle p \rangle_{\mathrm{AA}}$ \\
			& & length &  $\langle N \rangle$  & $\langle L_{\mathrm{rot}} \rangle$ & $\langle N \rangle$  & $\langle N \rangle$ & $\langle L_{\mathrm{rot}} \rangle$ & $\langle N \rangle$  \\
			\midrule
			log & 1 s & 25 td & 0.798  & 3.522  & 2.164  & 3.060 & 86.025 & 3.195 \\
			log & 1 s & 50 td & 0.758  & 3.356  & 2.010  & 2.568 & 64.641 & 2.733 \\
			lin & 1 s & 25 td & 2.022  & 4.331  & 4.823  & 5.951 & 91.013 & 6.171 \\
			lin & 1 s & 50 td & 1.891  & 3.923  & 4.074  & 4.695 & 81.665 & 4.968 \\
			log & 10 s & 25 td & 1.219  & 5.479  & 5.051  & 4.696 & 19.483 & 10.836 \\
			log & 10 s & 50 td & 1.134  & 5.179  & 4.814  & 3.683 & 11.081 & 10.074 \\
			lin & 10 s & 25 td & 3.360  & 14.329  & 23.923  & 7.324 & 15.644 & 17.543 \\
			lin & 10 s & 50 td & 3.160  & 12.287  & 20.031  & 6.318 & 14.190 & 17.368 \\
			\bottomrule
		\end{tabular}
	\end{minipage}%
\end{table}%

\clearpage

\section{\label{app:LedoitWolfShrinkage}Distributions of the Aggregated Returns Computed With Covariance Matrices Estimated by Ledoit--Wolf Shrinkage}
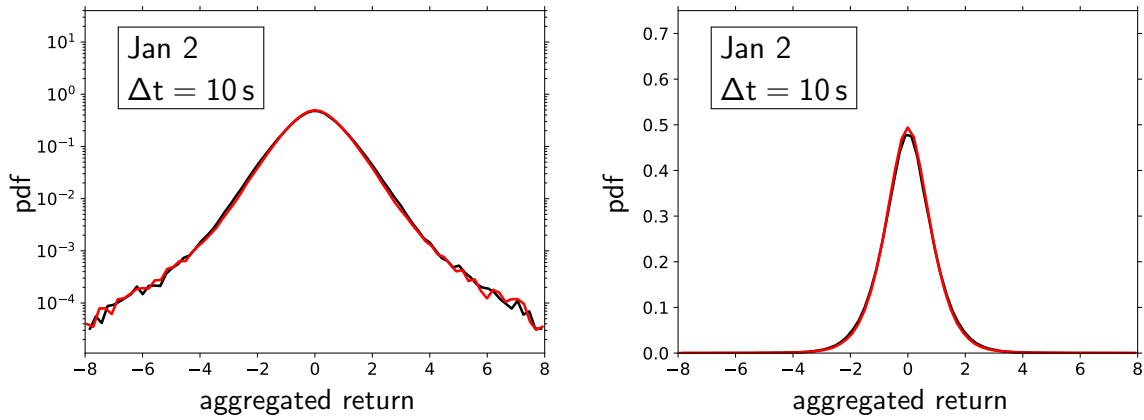
\begin{figure}[htbp]
		\captionsetup[subfigure]{labelformat=empty}
		\centering
		\begin{minipage}{.5\textwidth}
			\centering
			\subfloat[]{\begin{overpic}[width=1.\linewidth]{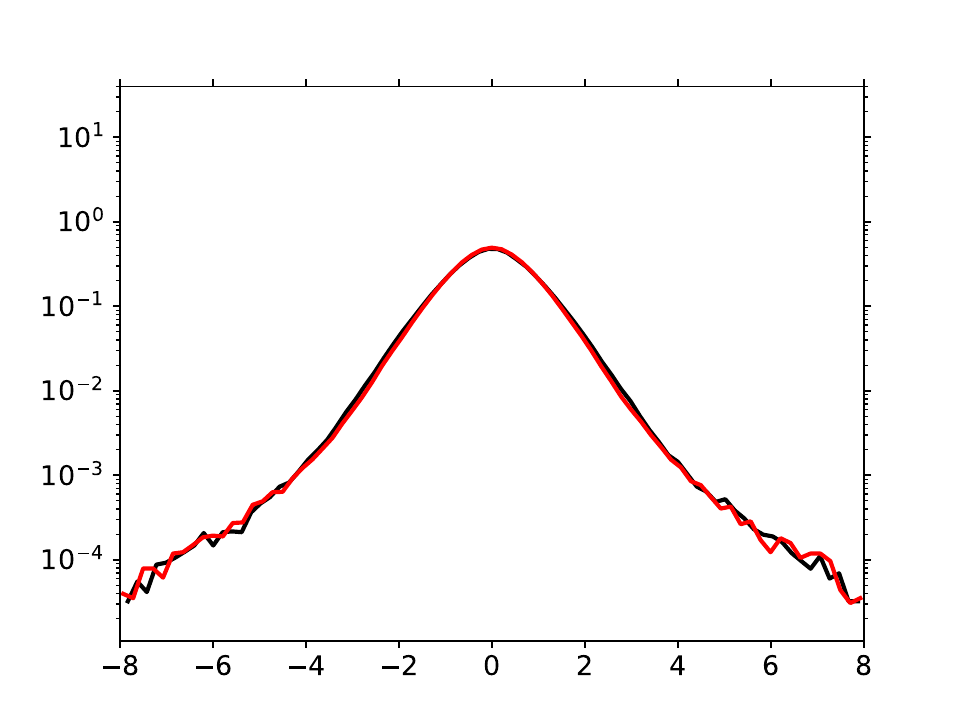}
					\put(18,55){\noindent\fbox{\parbox{1.7cm}{\sffamily Jan 2\\$\mathsf{\Delta t = 10\,s}$}}}
					\put(50,0){\makebox(0,0){\small\sffamily aggregated return}}
					\put(2,35){\makebox(0,0){\rotatebox{90}{\small\sffamily pdf}}}
				\end{overpic}
			}
		\end{minipage}%
		\begin{minipage}{.5\textwidth}
			\centering
			\subfloat[]{\begin{overpic}[width=1.\linewidth]{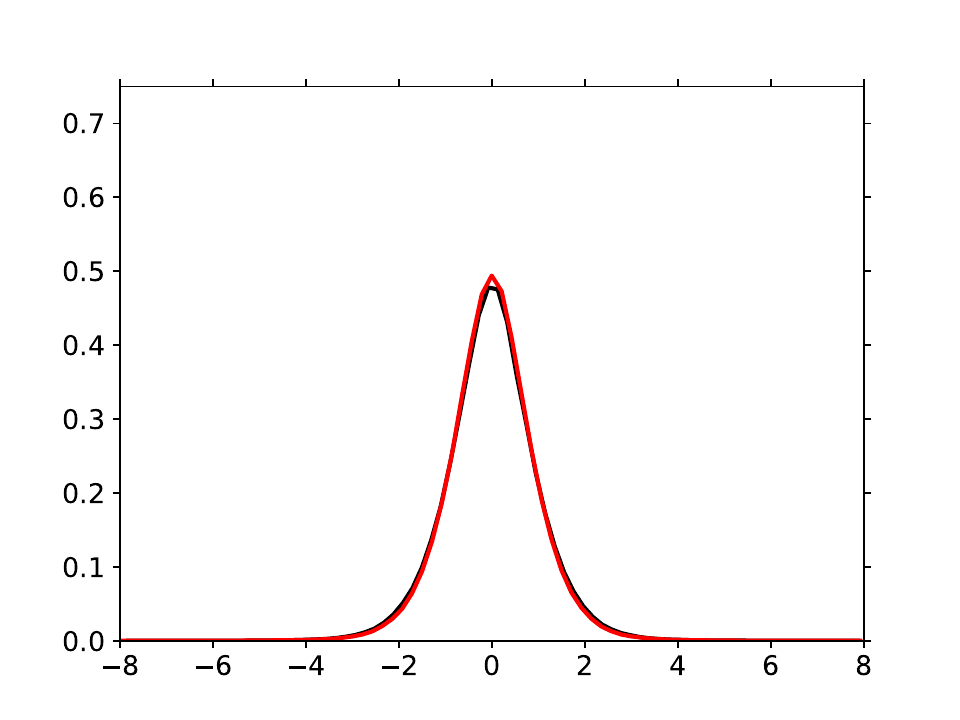}
					\put(18,55){\noindent\fbox{\parbox{1.7cm}{\sffamily Jan 2\\$\mathsf{\Delta t = 10\,s}$}}}
					\put(50,0){\makebox(0,0){\small\sffamily aggregated return}}
					\put(2,35){\makebox(0,0){\rotatebox{90}{\small\sffamily pdf}}}
				\end{overpic}
			}
		\end{minipage}
		\caption{Empirical distributions of aggregated returns for $\Delta t = 10\,\mathrm{s}$ and for the epoch January 02, 2014  without (black) and with Ledoit--Wolf shrinkage (red).}
		\label{fig:LedoitWolf}
	\end{figure}

	The aggregated returns are computed with the normalized, original returns and the correlation matrix over the same long interval or epoch. The normalized, original returns are not noise--dressed.
	Affected by noise are the correlation matrices with a time resolution of $\Delta t = 10$\,s, \textit{i.e} 2220 data points per epoch.
	For the epochs with $\Delta t = 1$\,s, \textit{i.e} 22200 data points per epoch, and for the long intervals, the distributions of the aggregated returns will be even less influenced by noise. %
	Hence, we determine the distributions of the aggregated returns for $\Delta t = 10$\,s over the length of an epoch. We apply Ledoit--Wolf shrinkage to reduce the noise in the covariance matrices, derive from them the correlation matrices and work out the corresponding distributions of the aggregated returns.
	We compare the distributions of the aggregate returns for the correlation matrices, where we do not use a noise reduction method and where we use Ledoit--Wolf shrinkage for the epoch January 02, 2014, see Fig.~\ref{fig:LedoitWolf}.
	Both distributions match almost perfectly on the logarithmic and the linear scale.
	For all epochs, the behavior of both distributions is very similar.

\section{\label{app:FittingParameters}Tables for Sec.~\ref{sec:EmpiricalLongIntervalDistributions}}
\subsection{Intervals with a Length of~25 Drading Days}
\begin{table}[H]
	\begin{minipage}{1.0\linewidth}
		\centering
		\caption{\label{tab:FitParametersCapital_22200_LN}Fitting parameters for distributions of the aggregated returns for long intervals (25 trading days) with $\Delta t = 1\,\mathrm{s}$ determined on a logarithmic scale.}\vspace{0.3cm}
		\begin{tabular}{ccccccc}
			\toprule
			&  $\langle p \rangle_{\mathrm{GG}}$  & $\langle p \rangle_{\mathrm{GA}}$  & $\langle p \rangle_{\mathrm{GA}}$  & $\langle p \rangle_{\mathrm{AG}}$ & $\langle p \rangle_{\mathrm{AA}}$  & $\langle p \rangle_{\mathrm{AA}}$ \\
			&  $N$  & $L_{\mathrm{rot}}$ & $N$  & $N$ & $L_{\mathrm{rot}}$ & $N$    \\
			\midrule
			interval \hspace{0.2cm}1 & 0.838  & 3.664  & 2.287  & 3.527 & 99.729 & 3.613 \\
			interval \hspace{0.2cm}2 & 0.833  & 3.653  & 2.157  & 3.401 & 99.696 & 3.482 \\
			interval \hspace{0.2cm}3 & 0.901  & 3.983  & 2.135  & 4.778 & 100.059 & 4.933 \\
			interval \hspace{0.2cm}4 & 0.767  & 3.392  & 2.086 & 2.607 & 99.490 & 2.659 \\
			interval \hspace{0.2cm}5 & 0.725  & 3.213  & 2.098  & 2.219 & 9.313 & 2.859  \\
			interval \hspace{0.2cm}6 & 0.775  & 3.408  & 2.338  & 2.692 & 53.755 & 2.791 \\
			interval \hspace{0.2cm}7 & 0.787  & 3.468  & 2.210 & 2.846 & 99.552 & 2.904 \\
			interval \hspace{0.2cm}8 & 0.733  & 3.282  & 1.745  & 2.175 & 99.379  & 2.209 \\
			interval \hspace{0.2cm}9 & 0.782  & 3.467  & 2.438  & 2.852 & 99.554 & 2.910  \\
			interval 10 & 0.838  & 3.690  & 2.141  & 3.506 & 99.724 & 3.591 \\
			\bottomrule
		\end{tabular}
	\end{minipage}%
\end{table}%
\begin{table}[H]
	\begin{minipage}{1.0\linewidth}
		\centering
		\caption{\label{tab:FitParametersCapital_22200_linear}Fitting parameters for distributions of the aggregated returns for long intervals (25 trading days) with $\Delta t = 1\,\mathrm{s}$ determined on a linear scale.}\vspace{0.3cm}
		\begin{tabular}{ccccccc}
			\toprule
			&  $\langle p \rangle_{\mathrm{GG}}$  & $\langle p \rangle_{\mathrm{GA}}$  & $\langle p \rangle_{\mathrm{GA}}$  & $\langle p \rangle_{\mathrm{AG}}$ & $\langle p \rangle_{\mathrm{AA}}$  & $\langle p \rangle_{\mathrm{AA}}$ \\
			&  $N$  & $L_{\mathrm{rot}}$ & $N$  & $N$ & $L_{\mathrm{rot}}$ & $N$    \\
			\midrule
			interval \hspace{0.2cm}1 & 2.129  & 4.679  & 5.489  & 7.165 & 100.650  & 7.409 \\
			interval \hspace{0.2cm}2 & 2.152  & 4.804  & 5.750  & 7.540 & 100.061 & 7.822 \\
			interval \hspace{0.2cm}3 & 2.112  & 4.350  & 4.544  & 7.019 & 94.685 & 7.269 \\
			interval \hspace{0.2cm}4 & 2.030  & 4.415  & 5.063  & 5.863 & 100.330 & 6.019 \\
			interval \hspace{0.2cm}5 & 2.017  & 4.514  & 5.345  & 5.776 & 100.307 & 5.928 \\
			interval \hspace{0.2cm}6 & 2.030  & 4.348  & 4.877  & 5.857 & 100.328 & 6.014 \\
			interval \hspace{0.2cm}7 & 2.008  & 4.157  & 4.345  & 5.684 & 100.283 & 5.831 \\
			interval \hspace{0.2cm}8 & 1.753  & 3.606  & 3.679  & 3.535 & 13.003 & 4.073 \\
			interval \hspace{0.2cm}9 & 2.036  & 4.419  & 5.058  & 5.926 & 100.346 & 6.085 \\
			interval 10 & 1.958  & 4.015  & 4.080  & 5.146 & 100.141 & 5.264 \\
			\bottomrule
		\end{tabular}
	\end{minipage}%
\end{table}%
\begin{table}[H]
	\begin{minipage}{1.0\linewidth}
		\centering
		\caption{\label{tab:FitParametersCapital_2220_lN}Fitting parameters for distributions of the aggregated returns for long intervals (25 trading days) with $\Delta t = 10\,\mathrm{s}$ determined on a logarithmic scale.}\vspace{0.3cm}
		\begin{tabular}{ccccccc}
			\toprule
			&  $\langle p \rangle_{\mathrm{GG}}$  & $\langle p \rangle_{\mathrm{GA}}$  & $\langle p \rangle_{\mathrm{GA}}$  & $\langle p \rangle_{\mathrm{AG}}$ & $\langle p \rangle_{\mathrm{AA}}$  & $\langle p \rangle_{\mathrm{AA}}$ \\
			&  $N$  & $L_{\mathrm{rot}}$ & $N$  & $N$ & $L_{\mathrm{rot}}$ & $N$    \\
			\midrule
			interval \hspace{0.2cm}1 & 1.263  & 5.296 & 4.124  & 4.979 & 77.535 & 5.218 \\
			interval \hspace{0.2cm}2 & 1.267  & 5.284  & 3.976  & 5.033 & 18.300  & 6.562 \\
			interval \hspace{0.2cm}3 & 1.432  & 5.873  & 4.581  & 8.226 & 17.615 & 11.219 \\
			interval \hspace{0.2cm}4 & 1.128  & 5.055  & 4.587  & 3.570 & 9.595 & 8.349 \\
			interval \hspace{0.2cm}5 & 1.100  & 4.875  & 4.252  & 3.341 & 9.066 & 7.825 \\
			interval \hspace{0.2cm}6 & 1.151  & 5.594  & 5.889  & 3.818 & 11.940 & 14.802 \\
			interval \hspace{0.2cm}7 & 1.276  & 5.639  & 5.227  & 5.244 & 13.227 & 11.985 \\
			interval \hspace{0.2cm}8 & 1.190  & 5.444  & 5.256  & 4.192 & 10.990 & 9.935 \\
			interval \hspace{0.2cm}9 & 1.122  & 5.967  & 6.909  & 3.576 & 13.541  & 19.038 \\
			interval 10 & 1.255  & 5.764  & 5.714  & 4.981 & 13.026 & 13.433 \\
			\bottomrule
		\end{tabular}
	\end{minipage}%
\end{table}%
\begin{table}[H]
	\begin{minipage}{1.0\linewidth}
		\centering
		\caption{\label{tab:FitParametersCapital_2220_linear}Fitting parameters for distributions of the aggregated returns for long intervals (25 trading days) with $\Delta t = 10\,\mathrm{s}$ determined on a linear scale.}\vspace{0.3cm}
		\begin{tabular}{ccccccc}
			\toprule
			&  $\langle p \rangle_{\mathrm{GG}}$  & $\langle p \rangle_{\mathrm{GA}}$  & $\langle p \rangle_{\mathrm{GA}}$  & $\langle p \rangle_{\mathrm{AG}}$ & $\langle p \rangle_{\mathrm{AA}}$  & $\langle p \rangle_{\mathrm{AA}}$ \\
			&  $N$  & $L_{\mathrm{rot}}$ & $N$  & $N$ & $L_{\mathrm{rot}}$ & $N$    \\
			\midrule
			interval \hspace{0.2cm}1 & 3.427  & 16.967  & 29.292  & 7.624 & 16.684 & 17.016 \\
			interval \hspace{0.2cm}2 & 3.523  & 16.850  & 28.952  & 8.190 & 16.766 & 17.182 \\
			interval \hspace{0.2cm}3 & 3.621  & 10.384  & 15.432  & 8.939 & 17.463 & 16.690 \\
			interval \hspace{0.2cm}4 & 3.195  & 11.636  & 18.715  & 6.418 & 14.163 & 18.301 \\
			interval \hspace{0.2cm}5 & 3.273  & 17.829  & 31.234  & 6.790 & 14.456 & 18.158 \\
			interval \hspace{0.2cm}6 & 3.315  & 13.602  & 22.591  & 7.009 & 15.021 & 17.692 \\
			interval \hspace{0.2cm}7 & 3.447  & 12.108  & 19.331  & 7.770 & 15.944 & 17.512 \\
			interval \hspace{0.2cm}8 & 2.995  & 7.429  & 10.090  & 5.542 & 13.328 & 18.328 \\
			interval \hspace{0.2cm}9 & 3.428  & 25.756  & 47.019  & 7.602 & 16.657 & 16.971 \\
			interval 10 & 3.379  & 10.732  & 16.575  & 7.357 & 15.962 & 17.579 \\
			\bottomrule
		\end{tabular}
	\end{minipage}%
\end{table}%
\begin{table}[H]
	\begin{minipage}{1.0\linewidth}
		\centering
		\caption{\label{tab:ChiSquaredAveraged_25TD}Averaged $\chi^2$ for logarithmic and linear scale ($\langle \chi^2_{\mathrm{ln}} \rangle$ and $\langle \chi^2_{\mathrm{lin}} \rangle$)  with $\Delta t = 1\,\mathrm{s}$ and $\Delta t = 10\,\mathrm{s}$ for the model distributions on the long interval.}\vspace{0.3cm}
		\begin{tabular}{ccccc}
			\toprule
			& \multicolumn{2}{c}{$\Delta t = 1\,\mathrm{s}$}  & \multicolumn{2}{c}{$\Delta t = 10\,\mathrm{s}$} \\
			\midrule
			& $\langle \chi^2_{\mathrm{ln}} \rangle$   & $\langle \chi^2_{\mathrm{lin}} \rangle$   & $\langle \chi^2_{\mathrm{ln}} \rangle$   & $\langle \chi^2_{\mathrm{lin}} \rangle$  \\
			\midrule
			GG & 0.047  & $1.516 \cdot 10^{-4} $  & 0.064  & $5.041 \cdot 10^{-5}$ \\
			\midrule
			GA & 0.003  & $1.125 \cdot 10^{-6} $ & 0.010  & $2.283 \cdot 10^{-7} $ \\
			\midrule
			AG & 0.006  & $3.377 \cdot 10^{-6} $  & 0.012  & $5.219 \cdot 10^{-6} $ \\
			\midrule
			AA & 0.006  & $3.440 \cdot 10^{-6} $  &  0.011 & $2.514 \cdot 10^{-6} $\\
			\bottomrule
		\end{tabular}
	\end{minipage}%
\end{table}%

\subsection{Intervals with a Length of~50 Trading Days}

\begin{table}[H]
	\begin{minipage}{1.0\linewidth}
		\centering
		\caption{\label{tab:FitParametersCapital_22200_LN_50TD}Fitting parameters for distributions of the aggregated returns for long intervals (50 trading days) with $\Delta t = 1\,\mathrm{s}$ determined on a logarithmic scale.}\vspace{0.3cm}
		\begin{tabular}{ccccccc}
			\toprule
			&  $\langle p \rangle_{\mathrm{GG}}$  & $\langle p \rangle_{\mathrm{GA}}$  & $\langle p \rangle_{\mathrm{GA}}$  & $\langle p \rangle_{\mathrm{AG}}$ & $\langle p \rangle_{\mathrm{AA}}$  & $\langle p \rangle_{\mathrm{AA}}$ \\
			&  $N$  & $L_{\mathrm{rot}}$ & $N$  & $N$ & $L_{\mathrm{rot}}$ & $N$    \\
			\midrule
			interval \hspace{0.2cm}1 & 0.805  & 3.548  & 2.154 & 3.056 & 99.607 & 3.123 \\
			interval \hspace{0.2cm}2 & 0.808  & 3.577  & 2.053 & 3.068 & 99.610 & 3.136 \\
			interval \hspace{0.2cm}3 & 0.725  & 3.210  & 2.066 & 2.209 & 12.344 & 2.611 \\
			interval \hspace{0.2cm}4 & 0.662  & 2.980  & 1.608 & 1.686 & 12.099 & 1.919 \\
			interval \hspace{0.2cm}5 & 0.789  & 3.464  & 2.170 & 2.819 & 99.545 & 2.876 \\
			\bottomrule
		\end{tabular}
	\end{minipage}%
\end{table}%
\begin{table}[H]
	\begin{minipage}{1.0\linewidth}
		\centering
		\caption{\label{tab:FitParametersCapital_22200_linear_50TD}Fitting parameters for distributions of the aggregated returns for long intervals (50 trading days) with $\Delta t = 1\,\mathrm{s}$ determined on a linear scale.}\vspace{0.3cm}
		\begin{tabular}{ccccccc}
			\toprule
			&  $\langle p \rangle_{\mathrm{GG}}$  & $\langle p \rangle_{\mathrm{GA}}$  & $\langle p \rangle_{\mathrm{GA}}$  & $\langle p \rangle_{\mathrm{AG}}$ & $\langle p \rangle_{\mathrm{AA}}$  & $\langle p \rangle_{\mathrm{AA}}$ \\
			&  $N$  & $L_{\mathrm{rot}}$ & $N$  & $N$ & $L_{\mathrm{rot}}$ & $N$    \\
			\midrule
			interval \hspace{0.2cm}1 & 2.031  & 4.294 & 4.710 & 5.893 & 100.334 & 6.051  \\
			interval \hspace{0.2cm}2 & 1.958  & 3.997 & 3.984 & 5.187 & 100.152 & 5.307  \\
			interval \hspace{0.2cm}3 & 1.901  & 3.971 & 4.182 & 4.670 & 100.017  & 4.764 \\
			interval \hspace{0.2cm}4 & 1.654  & 3.435 & 3.486 & 3.086 & 7.815 & 3.986 \\
			interval \hspace{0.2cm}5 & 1.909  & 3.918 & 4.010 & 4.642 & 100.009 & 4.734 \\
			\bottomrule
		\end{tabular}
	\end{minipage}%
\end{table}%
\begin{table}[H]
	\begin{minipage}{1.0\linewidth}
		\centering
		\caption{\label{tab:FitParametersCapital_2220_lN_50TD}Fitting parameters for distributions of the aggregated returns for long intervals (50 trading days) with $\Delta t = 10\,\mathrm{s}$ determined on a logarithmic scale.}\vspace{0.3cm}
		\begin{tabular}{ccccccc}
			\toprule
			&  $\langle p \rangle_{\mathrm{GG}}$  & $\langle p \rangle_{\mathrm{GA}}$  & $\langle p \rangle_{\mathrm{GA}}$  & $\langle p \rangle_{\mathrm{AG}}$ & $\langle p \rangle_{\mathrm{AA}}$  & $\langle p \rangle_{\mathrm{AA}}$ \\
			&  $N$  & $L_{\mathrm{rot}}$ & $N$  & $N$ & $L_{\mathrm{rot}}$ & $N$    \\
			\midrule
			interval \hspace{0.2cm}1 & 1.203 & 5.075 & 3.896 & 4.250 & 14.025 & 6.050 \\
			interval \hspace{0.2cm}2 & 1.204 & 5.294  & 4.699 & 4.343 & 11.407 & 10.768 \\
			interval \hspace{0.2cm}3 & 1.064 & 5.011  & 4.911  & 3.075 & 9.900 & 11.890 \\
			interval \hspace{0.2cm}4 & 1.073 & 4.842  & 4.357  & 3.140 & 8.663 & 7.539 \\
			interval \hspace{0.2cm}5 & 1.126 & 5.674  & 6.208  & 3.610 & 11.410 & 14.124 \\
			\bottomrule
		\end{tabular}
	\end{minipage}%
\end{table}%
\begin{table}[H]
	\begin{minipage}{1.0\linewidth}
		\centering
		\caption{\label{tab:FitParametersCapital_2220_linear_50TD}Fitting parameters for distributions of the aggregated returns for long intervals (50 trading days) with $\Delta t = 10\,\mathrm{s}$ determined on a linear scale.}\vspace{0.3cm}
		\begin{tabular}{ccccccc}
			\toprule
			&  $\langle p \rangle_{\mathrm{GG}}$  & $\langle p \rangle_{\mathrm{GA}}$  & $\langle p \rangle_{\mathrm{GA}}$  & $\langle p \rangle_{\mathrm{AG}}$ & $\langle p \rangle_{\mathrm{AA}}$  & $\langle p \rangle_{\mathrm{AA}}$ \\
			&  $N$  & $L_{\mathrm{rot}}$ & $N$  & $N$ & $L_{\mathrm{rot}}$ & $N$    \\
			\midrule
			interval \hspace{0.2cm}1 & 3.332  & 16.756  & 28.984 & 7.088 & 16.119 & 15.897 \\
			interval \hspace{0.2cm}2 & 3.271  & 9.493  & 14.110 & 6.807 & 14.449 & 17.661 \\
			interval \hspace{0.2cm}3 & 3.128  & 14.086  & 23.815 & 6.110 & 13.631 & 17.840 \\
			interval \hspace{0.2cm}4 & 2.793  & 7.296 & 10.185 & 4.794 & 12.406 & 17.758 \\
			interval \hspace{0.2cm}5 & 3.275  & 13.806 & 23.067 & 6.790 & 14.341 & 17.683 \\
			\bottomrule
		\end{tabular}
	\end{minipage}%
\end{table}%
\begin{table}[H]
	\begin{minipage}{1.0\linewidth}
		\centering
		\caption{\label{tab:ChiSquaredAveraged_50TD}Averaged $\chi^2$ for logarithmic and linear scale ($\langle \chi^2_{\mathrm{ln}} \rangle$ and $\langle \chi^2_{\mathrm{lin}} \rangle$)  with $\Delta t = 1\,\mathrm{s}$ and $\Delta t = 10\,\mathrm{s}$ for the model distributions on the long interval.}\vspace{0.3cm}
		\begin{tabular}{ccccc}
			\toprule
			& \multicolumn{2}{c}{$\Delta t = 1\,\mathrm{s}$}  & \multicolumn{2}{c}{$\Delta t = 10\,\mathrm{s}$} \\
			\midrule
			& $\langle \chi^2_{\mathrm{ln}} \rangle$   & $\langle \chi^2_{\mathrm{lin}} \rangle$   & $\langle \chi^2_{\mathrm{ln}} \rangle$   & $\langle \chi^2_{\mathrm{lin}} \rangle$  \\
			\midrule
			GG & 0.047  & $1.790 \cdot 10^{-4} $  & 0.061  & $4.858 \cdot 10^{-5} $ \\
			\midrule
			GA & 0.002  & $1.351 \cdot 10^{-6} $ & 0.005  & $2.223 \cdot 10^{-7} $ \\
			\midrule
			AG & 0.004  & $3.958 \cdot 10^{-6} $ & 0.006 & $7.057 \cdot 10^{-6} $ \\
			\midrule
			AA & 0.004  & $3.287 \cdot 10^{-6} $ & 0.004  & $2.489 \cdot 10^{-6} $ \\
			\bottomrule
		\end{tabular}
	\end{minipage}%
\end{table}%

\end{appendices}

\end{document}